\documentclass[preprint, 5p, times, twocolumn]{elsarticle}
\usepackage{amssymb}
\usepackage{graphicx}
\usepackage{fixltx2e}
\usepackage{pdfpages}

\journal{Acta Materialia}

\begin{document}

\begin{frontmatter}

\title{Migration barriers for diffusion of As and P atoms in InP and InAs via vacancies and interstitial atoms}

\author{Ivan~A.~Aleksandrov}
\ead{aleksandrov@isp.nsc.ru}
\author{Konstantin~S.~Zhuravlev}
\address{Rzhanov Institute of Semiconductor Physics, Siberian Branch of Russian Academy of Sciences, Novosibirsk, Russia}

\begin{abstract}

Processes of diffusion of As and P atoms in InP and InAs, and atomic and energy structure of group-V vacancies and interstitial P and As atoms in InP and InAs have been investigated using density functional theory. Formation energies of group-V vacancies in InP and InAs and P and As interstitial atoms in InP and InAs have been calculated with hybrid functional. The main types of migration jumps have been determined, and the energy favorable migration paths and migration barriers of As and P atoms diffusion in InP and InAs via vacancies and interstitial atoms have been calculated using climbing image nudged elastic band method. In the case of diffusion of As and P atoms in InP and InAs via interstitial atoms the diffusion process occurs via indirect interstitial mechanism. The migration energy barriers for the vacancy diffusion mechanism are 1.5--2.0 eV, the migration energy barriers for the interstitialcy mechanism are 0.3--0.6 eV. The interstitial atoms have higher formation energies compared to the formation energies of the vacancies, and total activation energies of the diffusion are comparable for the vacancy and interstitialcy mechanisms. The obtained results will be useful for modeling of the diffusion processes in semiconductor structures based on InP and InAs.

\end{abstract}

\begin{keyword}

diffusion \sep InP \sep InAs \sep density functional theory \sep hybrid functional \sep vacancy \sep interstitial atom \sep vacancy mechanism \sep interstitialcy mechanism \sep dumbbell interstitial \sep split-interstitial \sep correlation factor \sep zinc-blende structure

\end{keyword}

\end{frontmatter}

\section{Introduction}
\label{sec:introduction}
Heterostructures based on A\textsubscript{3}B\textsubscript{5} compounds on InP substrates are promising for creating of optoelectronic devices in the spectral range of 1.3--1.6~$\mu$m: semiconductor lasers~\cite{Teng_2002, Blokhin_2022}, photodetectors~\cite{Smiri_2020, Dmitriev_2019}, single photon sources~\cite{Benyoucef_2013}, as well as for the creation of field-effect and bipolar transistors~\cite{Ajayan_2015} and photonic integrated circuits~\cite{Smit_2019}. 

Atomic diffusion is a fundamentally important process for the manufacturing technology of the semiconductor structures. Diffusion of atoms at the material interfaces and diffusion of dopants is one of the limiting factors for reducing the size of nanoelectronic devices. The diffusion of atoms in heterostructures affects the sharpness of the composition profile at the heterointerfaces, so it must be taken into account during the calculations of the parameters of the device structures and when choosing modes for the growth of heterostructures and for the manufacture of devices. Modification of the composition profile at heterointerfaces affects the energies of optical transitions and radiative lifetimes in heterostructures~\cite{Bauer_2019, Shamirzaev_2011}. Controlled annealing of structures with quantum wells can be used to intentionally change the energies of optical transitions~\cite{Teng_2002}.

To theoretically describe the diffusion processes, various continuum models are used~\cite{Bracht_2007, Bursik_2002, Zimmermann_1993}, as well as kinetic Monte Carlo models~\cite{Martin_Bragado_2018, Chen_2011} and molecular dynamics~\cite{Bloechl_1993, Posselt_2008}. The key parameters of continuum and Monte Carlo models are the energy barriers of the atomic diffusion, and the formation energies of defects involved in the diffusion process, which determine their concentration for the equilibrium case. To determine these parameters, fitting of experimental results and ab initio calculations~\cite{Mantina_2009} are used. Calculation of these parameters using density functional theory is useful in that it allows one to obtain a theoretically based estimate of these parameters, and obtain information about the most probable types of atomic transitions that determine the microscopic diffusion mechanism. Among the A\textsubscript{3}B\textsubscript{5} compounds with sphalerite structure, diffusion processes have been most studied in gallium arsenide~\cite{El-Mellouhi_2006, Levasseur-Smith_2008, Zollo_2012, Wright_2016}. Calculations of the migration energy barriers are known for Ga and As vacancies in GaAs~\cite{El-Mellouhi_2006, El-Mellouhi_2006a}, interstitial Ga atoms~\cite{Levasseur-Smith_2008, Schick_2011} and interstitial As atoms~\cite{Zollo_2012, Wright_2016} in GaAs, and silicon atoms in GaAs~\cite{Reveil_2020}. Diffusion processes in InP and InAs have been less studied theoretically. There are known calculations of the energy structure of defects in InP~\cite{Seitsonen_1994, Mishra_2012, Tahini_2013, Chroneos_2014}, calculations of migration barriers of Zn diffusion in InP~\cite{Hoeglund_2008}, neutral intrinsic defects and silicon in InAs~\cite{Reveil_2017}, and silicon atoms in InGaAs~\cite{Reveil_2017a}. We conclude that for such fundamentally important processes as the self-diffusion of P in InP and As in InAs and the diffusion of As in InP and P in InAs, the theoretical calculation of energy barriers of diffusion and migration paths is an actual task.

In this work, calculations of the migration energy barriers and migration paths for the diffusion of As and P atoms in InP and InAs via vacancies and interstitial atoms, formation energies of phosphorus vacancies in InP, arsenic vacancies in InAs and interstitial As and P atoms in InP and InAs have been carried out. Activation energies and pre-exponential factors for the diffusion coefficients of As and P in InP and InAs have been estimated. The calculation results are compared with the experimental data available in the literature.

\section{Calculation details}
\label{sec:calculation details}
\subsection{Details of DFT calculations of atomic structure and formation energies of defects and migration energy barriers for As and P in InAs and InP}

Calculations of the atomic structure and formation energies of vacancies and interstitial atoms in InP and InAs were carried out using density functional theory in the Quantum ESPRESSO software package~\cite{Giannozzi_2009, Giannozzi_2017} for supercells corresponding to 64 and 216 lattice sites of bulk InP and InAs. Calculations of defect formation energies were carried out using the hybrid HSE functional~\cite{Heyd_2003, Heyd_2006} with 64-site supercells and the generalized gradient approximation with the PBE functional~\cite{Perdew_1996} with 64- and 216-site supercells for convergence checking. Calculations with the hybrid functional were carried out with a standard value of the Hartree-Fock exchange fraction $\alpha$=0.25, which gives the values of the band gap of InP $E_g$=1.419~eV and InAs $E_g$=0.507~eV, which are consistent with the experiment for InP $E_g$=1.424~eV~\cite{Vurgaftman_2001} and InAs $E_g$=0.417~eV~\cite{Vurgaftman_2001}. For the calculations with the PBE functional we used optimized norm-conserving Vanderbilt pseudopotentials~\cite{Hamann_2013, van_Setten_2018}. For calculations with the hybrid HSE functional, pseudopotentials of the same type were used, but generated without nonlinear core correction using the ONCVPSP code~\cite{Hamann_2013}. The cutoff energy for the wave functions was 80~Ry, which gives total energy convergence of 0.3~meV/atom in the case of the PBE functional and 4~meV/atom in the case of the HSE functional. For defect calculations, a $2\times2\times2$ k-point mesh was used, shifted by half a grid step relative to the Gamma point along each coordinate, which gives total energy convergence of 1~meV/atom. The following convergence criteria were used for relaxation of the atomic structure: $10^{-5}$~Ry for total energy and $10^{-4}$~eV/\AA{} for forces. Spin polarization was taken into account in the collinear approximation. To take into account the interaction of supercells with charged defects, the total energy was corrected using the method proposed in~\cite{Freysoldt_2009, Freysoldt_2010}, with the experimental values of the dielectric constants of InP $\varepsilon$=12.55~\cite{Neidert_1982} and InAs $\varepsilon$=15.15~\cite{Hass_1962}.
Formation energies of the defects were calculated using the expression:~\cite{Van_de_Walle_2004, Freysoldt_2014}: 
\begin{equation}
\label{eq:eq1}
 E_{form}^{def}(q)=E_{tot}^{def}(q)-E_{tot}^{bulk}-\sum_{i}{n_i\mu_i}+q(E_f+E_V),
\end{equation}
where $E_{form}^{def}(q)$ is the formation energy of the defect in the charge state $q$, $E_{tot}^{def}(q)$ is total energy of a supercell with a defect, $E_{tot}^{bulk}$ is total energy of a perfect bulk supercell, $n_i$ is number of atoms of i-th element added ($n_i>0$) or removed ($n_i<0$) from the supercell with the defect, $\mu_i$ is a chemical potential of the corresponding element, $E_V$ is the valence band edge of the bulk material, and $E_F$ is the Fermi level counted from the valence band edge. The limiting phases for the chemical potentials in the calculation of formation energies were: bulk metallic indium with bcc tetragonal structure, bulk black phosphorous, bulk gray arsenic, bulk InP for chemical potential of P impurity in InAs and bulk InAs for chemical potential of As impurity in InP. 

Migration energy barriers and migration paths were calculated using climbing image nudged elastic band method (CI~NEB)~\cite{Henkelman_2000} with the PBE functional for 64-site supercells. We have checked convergence of the migration barrier value with the supercell size for neutral P\textsubscript{i} in InP using 217-atom supercell. The migration barrier increases by 0.016 eV from 0.396~eV for 65-atom supercell to 0.412~eV for 217-atom supercell. Spin polarization was supposed to be unchanged during migration transitions.

InP lattice parameter, calculated with the PBE and HSE functionals, is a=5.959~\AA{} and a=5.854~\AA{}, respectively, which differs by 1.5\% and 0.3\% from the experimental value a=5.870~\AA{}~\cite{Vurgaftman_2001}. For InAs, the calculated lattice parameter is a=6.171~\AA{} and a=6.032~\AA{}, for the PBE and HSE functionals, respectively, which differs by 1.9\% and 0.4\% from the experimental value a=6.058~\AA{}~\cite{Vurgaftman_2001}. Enthalpy of InP formation relative to bcc tetragonal indium and black phosphorous calculated with the PBE functional is $-$0.477~eV, and with the HSE functional is $-$0.653~eV. Enthalpy of InAs formation is $-$0.461~eV for the PBE functional and $-$0.769~eV for the HSE functional. These values are consistent within 0.17~eV with the experimental values $-$0.499~eV for InP~\cite{Vasilev_2006} and $-$0.622~eV for InAs~\cite{Yamaguchi_1994}.

Concentration of defects can be calculated from the expression~\cite{Freysoldt_2014}:
\begin{equation}
\label{eq:eq2}
 N_{def}=N_{sites} \exp\left(-\frac{G_{form}}{k_B T}\right),
\end{equation}
where $N_{def}$ is the number of defects per unit volume, $N_{sites}$ is the number of the lattice sites per unit volume, $G_{form}$ is the Gibbs energy of defect formation: $G_{form}=H_{form}-TS_{form}$, where $H_{form}=E_{form}+PV_{form}$ is the enthalpy of  defect formation, $V_{form}$ is the volume of defect formation, and $S_{form}$ is the entropy of defect formation.
Expression~\ref{eq:eq1} for the formation energy is valid in the limit of low temperatures. More accurate methods are known for calculation of defect concentrations at finite temperatures, which take into account vibrational and electronic excited states~\cite{Freysoldt_2014, Grabowski_2009, Zhang_2018}. Vibrational excitations can be considered in the quasi-harmonic approximation or taking into account anharmonicity. The assumption of a close to the equilibrium concentration of defects is fulfilled only for certain experimental conditions (sufficiently high temperatures and close to equilibrium vapor pressures). At relatively low temperatures, a ``frozen'' concentration of defects can often be observed, which is higher than the equilibrium concentration~\cite{Khreis_1997}.

\subsection{Calculation of diffusion coefficients of As and P atoms in InP and InAs}
Diffusion coefficient of atoms for one type of the jumps can be calculated as~\cite{Manning_1962, Peterson_1978}:
\begin{equation}
\label{eq:eq3}
 D=\frac{1}{6}r^2\Gamma f,
\end{equation}
where $r$ is the jump distance, $\Gamma$ is the number of jumps per unit time, $f$ is the correlation coefficient. For vacancy mechanism of diffusion the jump frequency $\Gamma$ can be found as~\cite{Peterson_1978}: $\Gamma=Z\frac{N_{def}}{N_{sites}}w$, where $Z$ is the number of the nearest neighbor places for the jump, $N_{def}$ is the number of defects per unit volume, $N_{sites}$ is the number of lattice sites per unit volume, $w$ is jump probability per unit time in one direction. Diffusion coefficient of vacancies is expressed as:
\begin{equation}
\label{eq:eq4}
 D=\frac{1}{6}r^2 Z w.
\end{equation}
Jump frequency $w$ can be found as~\cite{Mehrer_2007, Vineyard_1957} $w=\tilde{\nu}\exp\left(\frac{S_m}{k_B}\right)\exp\left(-\frac{H_m}{k_B T}\right)=\nu^*\exp\left(-\frac{H_m}{k_B T}\right)$, where $\tilde{\nu}$ is the frequency of vibrational mode in the jump direction, $H_m$ is the enthalpy of migration, $S_m$ is the entropy of migration. In further calculations we do not take into account the migration entropy and migration volume and assume that $w=\nu\exp\left(-\frac{E_m}{k_B T}\right)$, where $E_m$ is migration barrier energy of the atom and $\nu$ is attempt frequency. The frequency of vibrational mode in the jump direction can be approximately calculated as the Einstein frequency of the moving atom in the jump direction~\cite{Vineyard_1957}. We used this method  in our earlier work~\cite{Aleksandrov_2020}. In this work depending on the transition type we took into account one, two or three atoms with largest displacements. The frequency $\nu$ was calculated as $\frac{1}{2\pi}\sqrt{\frac{K}{M}}$, the effective mass $M$ of the vibrational mode was calculated as $M=\frac{\sum_a{r_a^2 m_a}}{\sum_a{r_a^2}}$, where $r_a$ and $m_a$ are displacement and mass of the atom $a$, force constant $K$ was determined from the expression $E=\frac{K}{2}\sum_a{r_a^2}$ for several points of the dependence of the energy on the displacements of the considered atoms along the migration path near the equilibrium position.

For self-diffusion of atoms in group-V elements sublattice by vacancy mechanism we have used the correlation coefficient, calculated by Compaan and Haven~\cite{Compaan_1958} for fcc lattice $f$=0.78146. For diffusion of As atoms in InP and P atoms in InAs by vacancy mechanism the correlation factors were calculated by Manning expression~\cite{Manning_1964}, obtained in the five-frequency model for fcc lattice:
\begin{equation}
\label{eq:eq5}
 f=\frac{w_1+7F_3w_3/2}{w_2+w_1+7F_3w_3/2},
\end{equation}
\begin{equation}
\label{eq:eq6}
 7F_3(\alpha)=7-\frac{10\alpha^4+180.5\alpha^3+927\alpha^2+1341\alpha}{2\alpha^4+40.2\alpha^3+254\alpha^2+597\alpha+436},
\end{equation}
where $w_1$ is the jump rate of the vacancy between first-nearest neighbor sites of the impurity atom in the group-V sublattice, $w_2$ is the jump rate of the vacancy to the impurity atom site, $w_3$ is the jump rate of the vacancy from first coordination shell of the impurity atom to second coordination shell in the group-V sublattice, $w_4$ is the jump rate of the reverse of a $w_3$ jump, $w_0$ is the jump rate of all other jumps in the group-V sublattice, $\alpha=\frac{w_4}{w_0}$. We assume that $w_1=w_0=\nu_{bulk}\exp\left(-\frac{E_{m bulk}}{k_B T}\right)$, $w_2=\nu\exp\left(-\frac{E_m}{k_B T}\right)$, $w_3=\nu_{bulk}\exp\left(-\frac{E_{m bulk}+E_{bind}/2}{k_B T}\right)$, $w_4=\nu_{bulk}\exp\left(-\frac{E_{m bulk}-E_{bind}/2}{k_B T}\right)$, where $\nu_{bulk}$ and $E_{m bulk}$ are attempt frequency and migration energy barrier for self-diffusion in the group-V sublattice. The binding energy $E_{bind}$ for the complex of phosphorous vacancy with the arsenic substitution atom As\textsubscript{P}V\textsubscript{P} was calculated as:
$E_{bind}=E_{form}(V_P)+E_{form}(As_P)-E_{form}(As_PV_P)$.

In the case of non-zero binding energy between the impurity atom and vacancy, the probability of presence of vacancy near the impurity atom is $\frac{N_{def}}{N_{sites}}\exp\left(\frac{E_{bind}}{k_B T}\right)$, and the diffusion coefficient is: 
\begin{equation}
\label{eq:eq7}
 D=\frac{1}{6}r^2 Z f \nu\frac{N_{def}}{N_{sites}}\exp\left(-\frac{E_m}{k_B T}\right)\exp\left(\frac{E_{bind}}{k_B T}\right).
\end{equation}

We have also calculated the diffusion coefficients via vacancies and via interstitial atoms using the KineCluE code~\cite{Schuler_2020}, which implements the self-consistent mean-field theory for clusters of finite size. We used migration energy barriers and attempt frequencies calculated by DFT for calculation of the diffusion coefficients in this model.

\section{Calculation results}
\label{sec:results}
\subsection{Atomic structure and formation energies of group-V vacancies and complexes As\textsubscript{P}V\textsubscript{P} and P\textsubscript{As}V\textsubscript{As} in InP and InAs}

\begin{figure}[h]
\includegraphics[width=0.99\columnwidth]{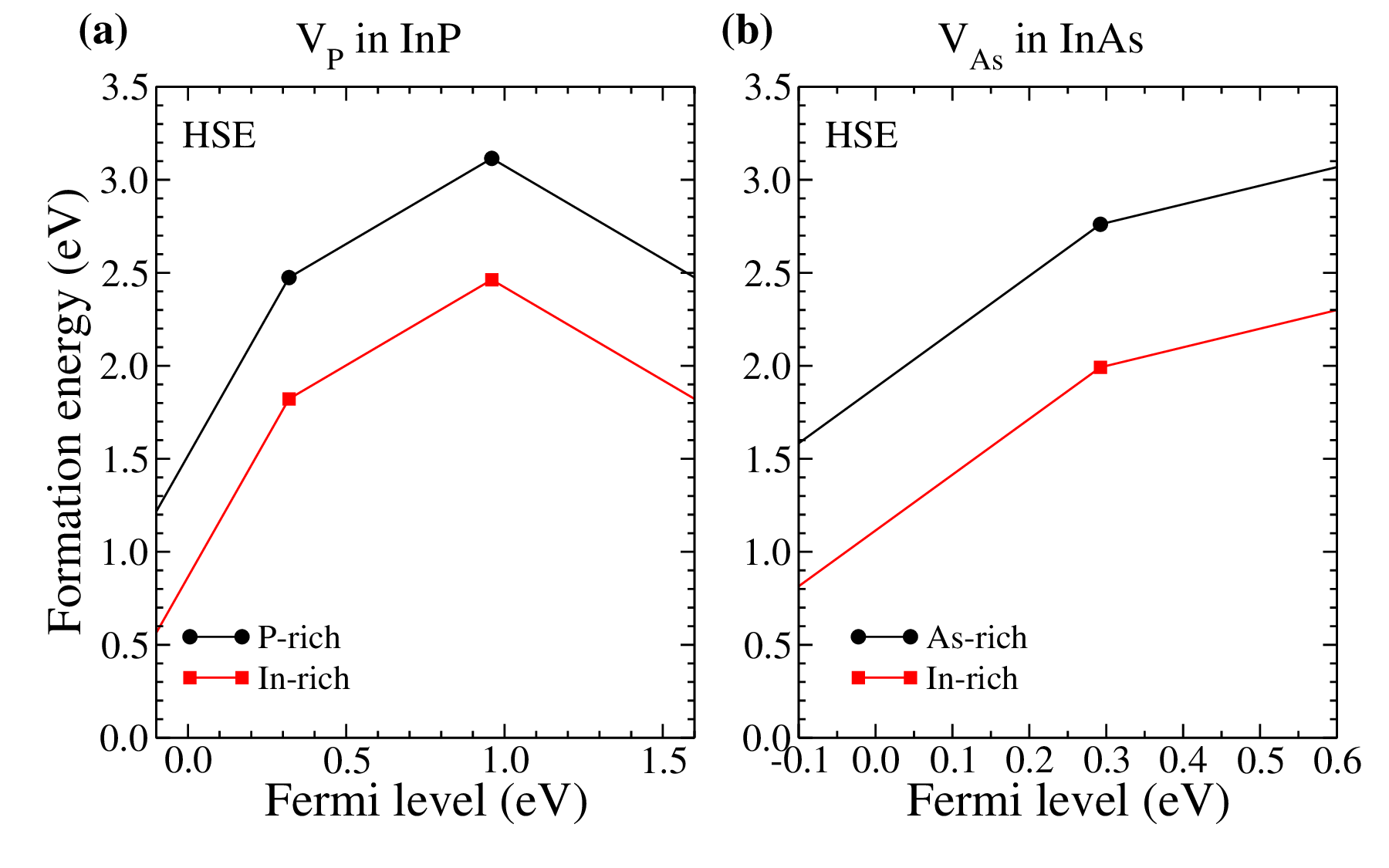}
\includegraphics[width=0.99\columnwidth]{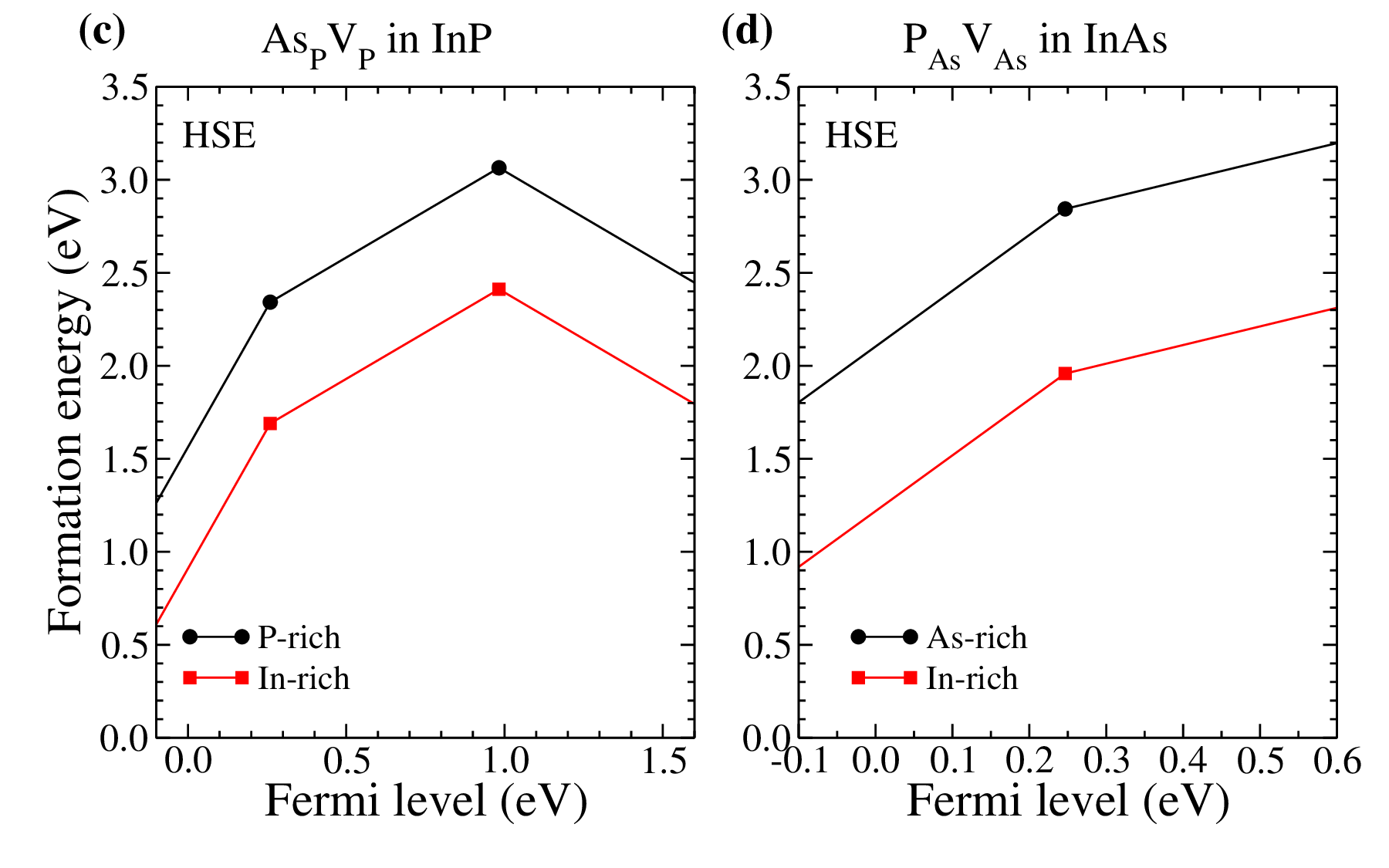}
\caption{\label{fig:figure1} Dependence of the phosphorous vacancy formation energy in InP on the Fermi level for P-rich and In-rich conditions (a), dependence of the arsenic vacancy formation energy in InAs on the Fermi level for As-rich and In-rich conditions (b), and dependences of the formation energies of the As\textsubscript{P}V\textsubscript{P} complex in InP (c) and P\textsubscript{As}V\textsubscript{As} complex in InAs (d) on the Fermi level. Calculation with the hybrid HSE functional for the 63-atom supercell.}
\end{figure}

Figure~\ref{fig:figure1}a shows the dependence of the formation energy of phosphorous vacancies in InP on the Fermi level calculated with the HSE functional for a 63-atom supercell.
Convergence of the formation energies with the supercell size was selectively checked with the PBE functional by increasing the supercell size from 64 to 216 lattice sites. The difference in the formation energies between the 216 and 64-site supercells is less than 0.10 eV for V\textsubscript{P} in InP, 0.12 eV for V\textsubscript{As} in InAs, 0.20 eV for P\textsubscript{i} in InP and 0.16 eV for As\textsubscript{i} in InAs (figures~S1, S3a and S4a of the supplementary material to this article). For the most of the charge states of the considered defects the formation energies are sufficiently converged with the 64-site supercell, the largest difference is observed for the formation energy of the +2 charge state of the interstitials.
The band gap of InP calculated with the PBE functional is 0.438~eV, this value is considerably underestimated compared to the experimental value $E_g$=1.424~eV~\cite{Vurgaftman_2001}. Calculation with the hybrid HSE functional gives a value of the InP band gap $E_g$=1.419~eV which is closer to the experiment. According to calculations with the HSE functional, the charge states +3, +1 and $-$1 are thermodynamically stable for V\textsubscript{P} in InP. The thermodynamic transition level (+3/+1) is 0.319~eV above the valence band edge, the thermodynamic transition level (+1/$-$1) is 0.960~eV above the valence band edge. The ionization energy of a singly negative vacancy is higher than the ionization energy of a neutral vacancy by 0.206~eV, which is typical for ``negative-U'' centers. This effect was previously observed for V\textsubscript{P} in InP by Alatalo et al.~\cite{Alatalo_1993}. In addition, we have observed such an effect for the thermodynamic transition (+3/+1). The values of the thermodynamic transition levels differ from the result of the work of Mishra et al.~\cite{Mishra_2012}, in which the calculation of defect formation energies with the HSE functional was carried out for the atomic configuration relaxed in the generalized gradient approximation in a 63-atom supercell, and the thermodynamic transition level (+1/0) for V\textsubscript{P} in InP was observed near the edge of the conduction band, and the +1 charge state was thermodynamically stable for lower Fermi levels inside the band gap. This difference is probably due to the fact that in the work of Mishra et al.~\cite{Mishra_2012} the energy levels of thermodynamic transitions were estimated using the Kohn-Sham levels for a neutral vacancy.

Table~\ref{tab:table1} shows the symmetry types of the phosphorus vacancy in InP for different charge states and the interatomic distances for the indium atoms nearest to the vacancy.
\begin{table}
\caption{\label{tab:table1} Symmetry of V\textsubscript{P} in InP and interatomic distances for indium atoms nearest to the vacancy in different charge states. Calculation with HSE functional for supercell size of 63 atoms. If the In-In interatomic distances are different, their number is given in parentheses for each distance. If several different configurations with close total energies are found, the difference in total energies $\Delta E$ is shown for them relative to the configuration with the minimum energy in a given charge state.}
\begin{tabular}{c c c c}
\hline
Charge & Symmetry & In-In distances, \AA & $\Delta E$, eV\\ 
\hline
+3 & T\textsubscript{d} & 5.04 & \\
+2 & T\textsubscript{d} & 4.59 & \\
+1 & C\textsubscript{2v} & 4.71 (1), 4.18 (4), 3.39 (1) & 0 \\
+1 & T\textsubscript{d} & 4.08 & 0.045 \\
0 & T\textsubscript{d} & 3.95 & \\
$-$1 & D\textsubscript{2d} & 3.74 (2), 3.62 (4) & \\
\hline
\end{tabular}
\end{table}
If several different configurations with close total energies are found, interatomic distances and the difference in total energies relative to the configuration with the minimum energy in a given charge state are also given for them. According to the calculations with the HSE functional, the P vacancy in InP in the charge states +3, +2 and 0 has T\textsubscript{d} symmetry, in the charge state +1 it has C\textsubscript{2v} symmetry with the (110) symmetry plane, and in the charge state $-$1 it has D\textsubscript{2d} symmetry. The transition from T\textsubscript{d} to D\textsubscript{2d} symmetry with a change in the charge state of V\textsubscript{P} in InP was observed earlier in the work of Seitsonen et al.~\cite{Seitsonen_1994}, where the T\textsubscript{d} symmetry was observed in the charge state +1 and D\textsubscript{2d} symmetry was observed in charge states 0, $-$1, $-$2 in the local density approximation for V\textsubscript{P} in InP.

For the defects considered in our study only 0 and 1/2 spin states are formed. V\textsubscript{P} in InP has 1/2 spin in the charge states +2, 0, $-$2 and zero spin in the charge states +3, +1, $-$1, $-$3. The formation energies calculated with PBE functional for 64-site supercell without spin polarization for +2, 0, and $-$2 charge states of V\textsubscript{P} in InP are higher by 84~meV, 17~meV and 5~meV, respectively, comparing to spin-polarized calculation. Spin-unpolarized calculation leads to 85~meV and 74~meV higher formation energies for +2 and 0 charge states of As\textsubscript{i} in InAs, and to 119~meV and 95~meV higher formation energies for +2 and 0 charge states of P\textsubscript{i} in InP, respectively, comparing to spin-polarized calculation.

Figure~\ref{fig:figure1}b shows the formation energies of V\textsubscript{As} in InAs calculated with the HSE functional in dependence on the Fermi level. The charge states +3 and +1 are thermodynamically stable for V\textsubscript{As} in InAs. The (+3/+1) thermodynamic transition level for V\textsubscript{As} in InAs is 0.292~eV above the valence band edge. Table~\ref{tab:table2} shows the symmetry types of the arsenic vacancy in different charge states.      
\begin{table}
\caption{\label{tab:table2} Symmetry of V\textsubscript{As} in InAs and interatomic distances for indium atoms nearest to the vacancy in different charge states. Calculation with HSE functional for supercell size of 63 atoms. If the In-In interatomic distances are different, their number is given in parentheses for each distance. If several different configurations with close total energies are found, the difference in total energies $\Delta E$ is shown for them relative to the configuration with the minimum energy in a given charge state.}
\begin{tabular}{c c c c}
\hline
Charge & Symmetry & In-In distances, \AA & $\Delta E$, eV\\ 
\hline
+3 & T\textsubscript{d} & 5.22 & \\
+2 & T\textsubscript{d} & 4.67 & \\
+1 & C\textsubscript{2v} & 5.04 (1), 4.38 (4), 3.23 (1) & 0 \\
+1 & T\textsubscript{d} & 4.20 & 0.035 \\
0 & T\textsubscript{d} & 3.95 &  \\
$-$1 & T\textsubscript{d} & 3.63 & 0 \\
$-$1 & D\textsubscript{2d} & 3.74 (2), 3.62 (4) & 0.0095 \\
\hline
\end{tabular}
\end{table}
Calculation with the HSE functional gives T\textsubscript{d} symmetry in the charge states +3, +2, 0 and $-$1, and C\textsubscript{2v} symmetry in the charge state +1.

To consider diffusion of As atoms in InP and P atoms in InAs by vacancy mechanism we have calculated atomic structure and formation energies of the defect complexes As\textsubscript{P}V\textsubscript{P} in InP and P\textsubscript{As}V\textsubscript{As} in InAs. Dependences of the formation energies of the complexes As\textsubscript{P}V\textsubscript{P} in InP and P\textsubscript{As}V\textsubscript{As} in InAs on the Fermi level, calculated with HSE functional for 63-atom supercell are shown in the figure~\ref{fig:figure1}(c,d). Thermodynamic transition levels of As\textsubscript{P}V\textsubscript{P} in InP and P\textsubscript{As}V\textsubscript{As} in InAs are close to the levels of P vacancy in InP and As vacancy in InAs, respectively. Atomic configurations of the complexes As\textsubscript{P}V\textsubscript{P} in InP and P\textsubscript{As}V\textsubscript{As} in InAs have C\textsubscript{s} symmetry with the symmetry plane (110) passing through the lattice sites with the vacancy and the substitutional atom. Binding energies of the As\textsubscript{P}V\textsubscript{P} and P\textsubscript{As}V\textsubscript{As} complexes are shown in the table~\ref{tab:table3}.
\begin{table}
\caption{\label{tab:table3} Binding energies of the complexes As\textsubscript{P}V\textsubscript{P} in InP and P\textsubscript{As}V\textsubscript{As} in InAs. Calculation with the PBE functional for the 63-atom supercell.}
\begin{center}
\begin{tabular}{c c c}
\hline
Charge & $E_{bind}$(As\textsubscript{P}V\textsubscript{P}), meV & $E_{bind}$(P\textsubscript{As}V\textsubscript{As}), meV \\
\hline
+3 & 56.9 & $-$45.3 \\
+1 & 128.7 & $-$10.5 \\
0 & 58.2 & $-$9.6 \\
$-$1 & 70.4 & $-$21.3 \\
\hline
\end{tabular}
\end{center}
\end{table}

\subsection{Atomic structure and formation energies of P and As interstitial atoms in InP and InAs}
\begin{figure}[h]
\includegraphics[width=0.99\columnwidth]{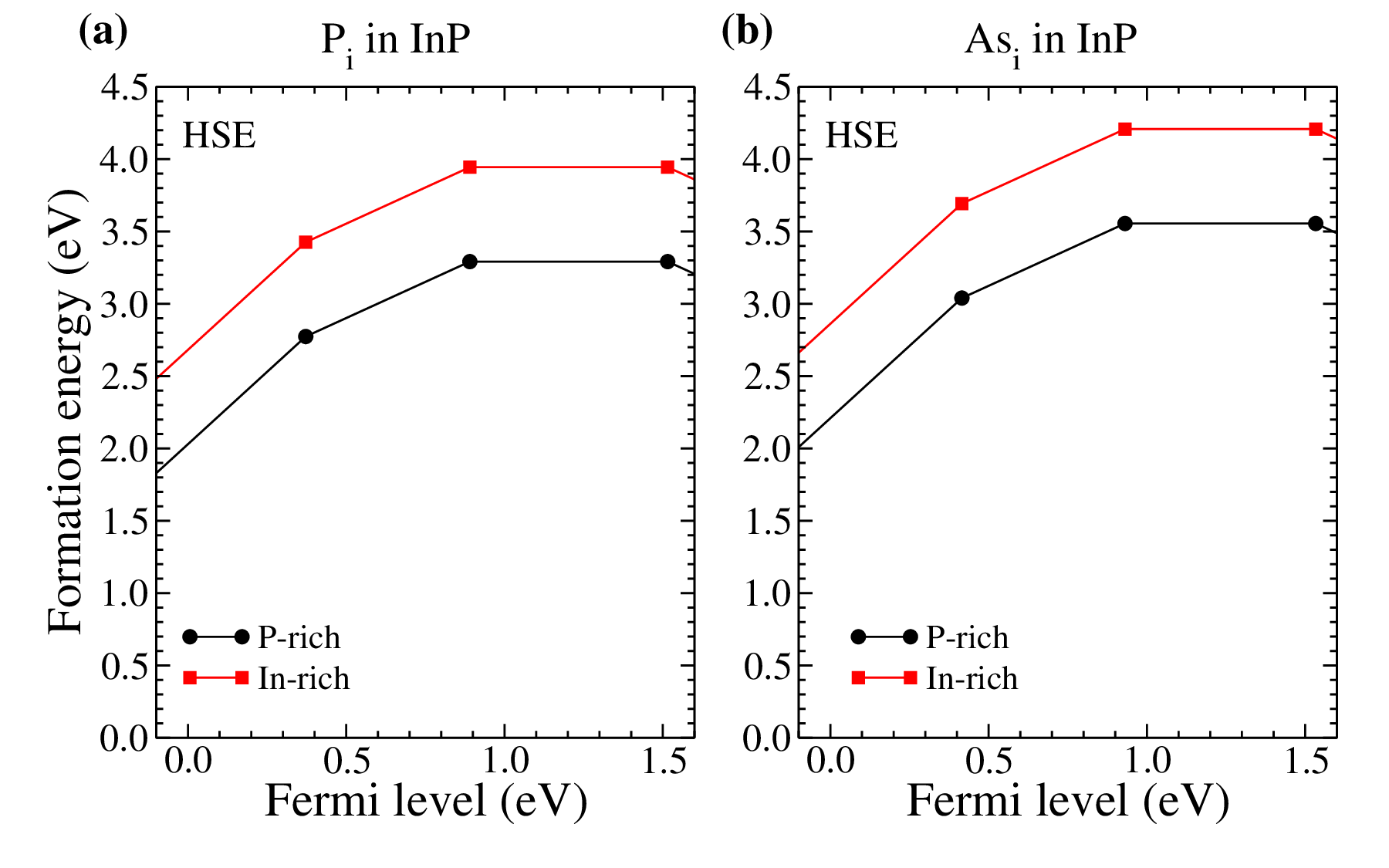}
\includegraphics[width=0.99\columnwidth]{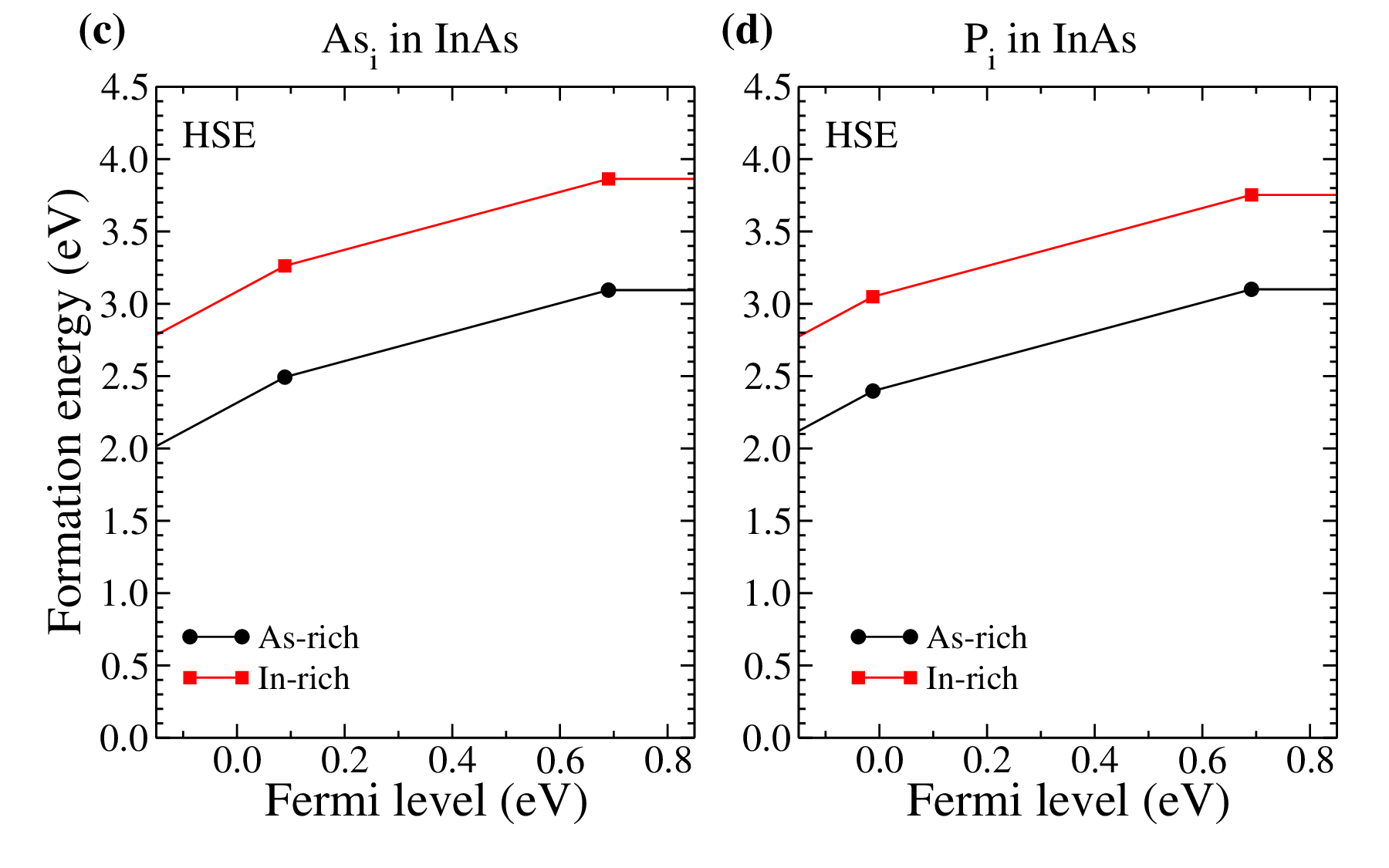}
\caption{\label{fig:figure2} Dependence of the formation energy of interstitial phosphorous (a) and arsenic (b) atoms in InP on the Fermi level for P-rich and In-rich conditions, and dependence of the formation energy of interstitial As atoms (c) and interstitial P atoms (d) in InAs on the Fermi level for As-rich and In-rich conditions. Calculation with the HSE functional for the 65-atom supercell.}
\end{figure}

Figure~\ref{fig:figure2}(a, b) shows formation energies of interstitial P and As atoms in InP on the Fermi level, calculated with the HSE functional. According to the calculation, the charge states +2, +1 and 0 are thermodynamically stable for P\textsubscript{i} in InP and for As\textsubscript{i} in InP.  

\begin{figure}
\includegraphics[width=0.99\columnwidth]{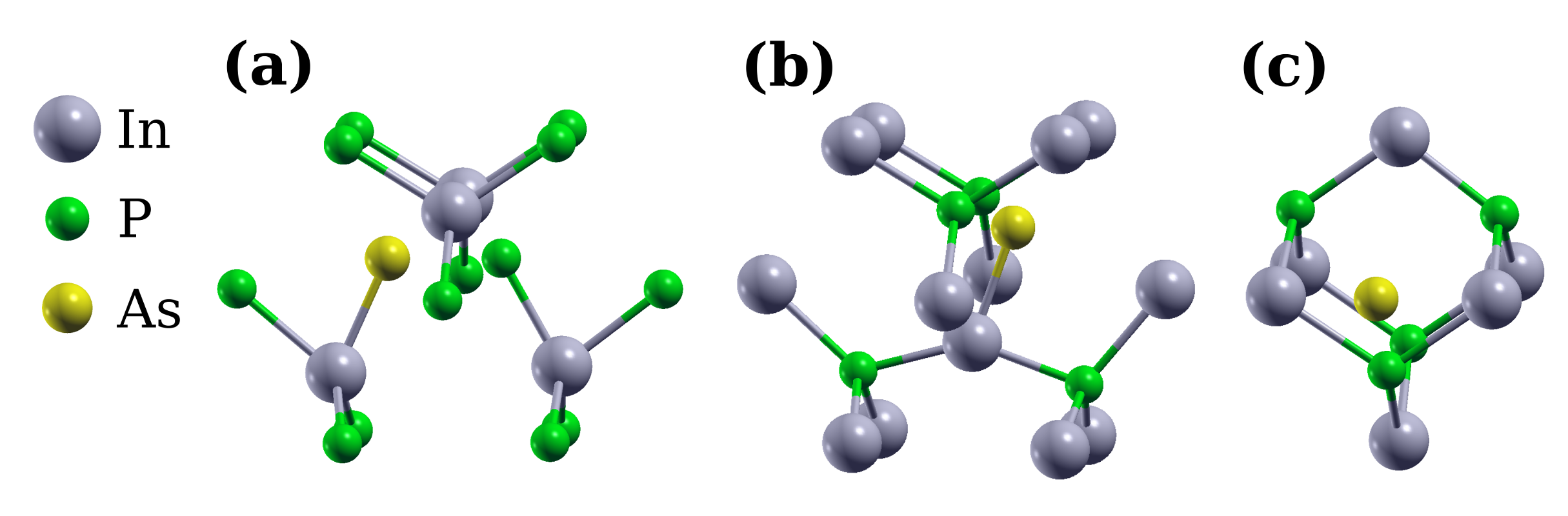}
\caption{\label{fig:figure3} Energy favorable atomic configurations of the interstitial As in InP for neutral charge state (a), +1 charge state (b) and +2 charge state (c).}
\end{figure}

Most energy favorable atomic configurations of As\textsubscript{i} in InP for different charge states are shown in the figure~\ref{fig:figure3}. Split-interstitial configuration in the P atoms sublattice is energy favorable for neutral interstitial As atom (figure~\ref{fig:figure3}a). It has C\textsubscript{s} symmetry and direction of the bond between P and As atoms of the split-interstitial close to the [110] direction. In the charge state +1 the most energy favorable configuration is a split-interstitial configuration in the In sublattice with C\textsubscript{s} symmetry and direction of the bond between In and As atoms of the split-interstitial at an angle of about 20$^\circ$ to the [001] direction (figure~\ref{fig:figure3}b). In the charge state +2 the most energy favorable configuration has C\textsubscript{3v} symmetry (figure~\ref{fig:figure3}c). In this configuration the interstitial As atom is shifted in the [111] direction from the tetrahedral interstitial site in the P atoms sublattice.

For P\textsubscript{i} in InP the energy favorable configurations are similar to the shown in the figure~\ref{fig:figure3} configurations of As\textsubscript{i} in InP. In the neutral charge state the split-interstitial consist of two P atoms and has C\textsubscript{s} symmetry with small deviation from C\textsubscript{2v} symmetry and the direction of the bond between two P atoms close to the [110] direction. For As\textsubscript{i} in InAs in the neutral charge state the split-interstitial consist of two As atoms and has C\textsubscript{2v} symmetry and the [110] direction of the bond between two As atoms. This configuration is similar to the 110a split-interstitial As\textsubscript{i} in GaAs in the notation of the work of Schultz et al.~\cite{Schultz_2009}. The split-interstitial configuration of P\textsubscript{i} in InP in the +1 charge state is similar to the p-001g configuration of As\textsubscript{i} in GaAs in the notation of the work of Schultz et al.~\cite{Schultz_2009}.   

Figure~\ref{fig:figure2}(c, d) shows the dependences of the formation energies of interstitial As and interstitial P atoms in InAs on the Fermi level, calculated with the HSE functional. According to the calculation, the charge states +2 and +1 are thermodynamically stable for As\textsubscript{i} in InAs at the Fermi levels inside the band gap, and the charge state +1 is thermodynamically stable for P\textsubscript{i} in InAs at the Fermi levels inside the band gap. As\textsubscript{i} and P\textsubscript{i} in InAs in the charge states 0, +1, and +2 have the most energy favorable configurations similar to P\textsubscript{i} and As\textsubscript{i} in InP with the same symmetry types excluding the mentioned above slight difference for the neutral As\textsubscript{i} in InAs (C\textsubscript{2v} symmetry) and P\textsubscript{i} in InP (C\textsubscript{s} symmetry). 

\subsection{Migration barriers for diffusion of As and P atoms in InP by vacancy mechanism}

\begin{figure*}[t]
\begin{center}
\includegraphics[width=0.50\columnwidth]{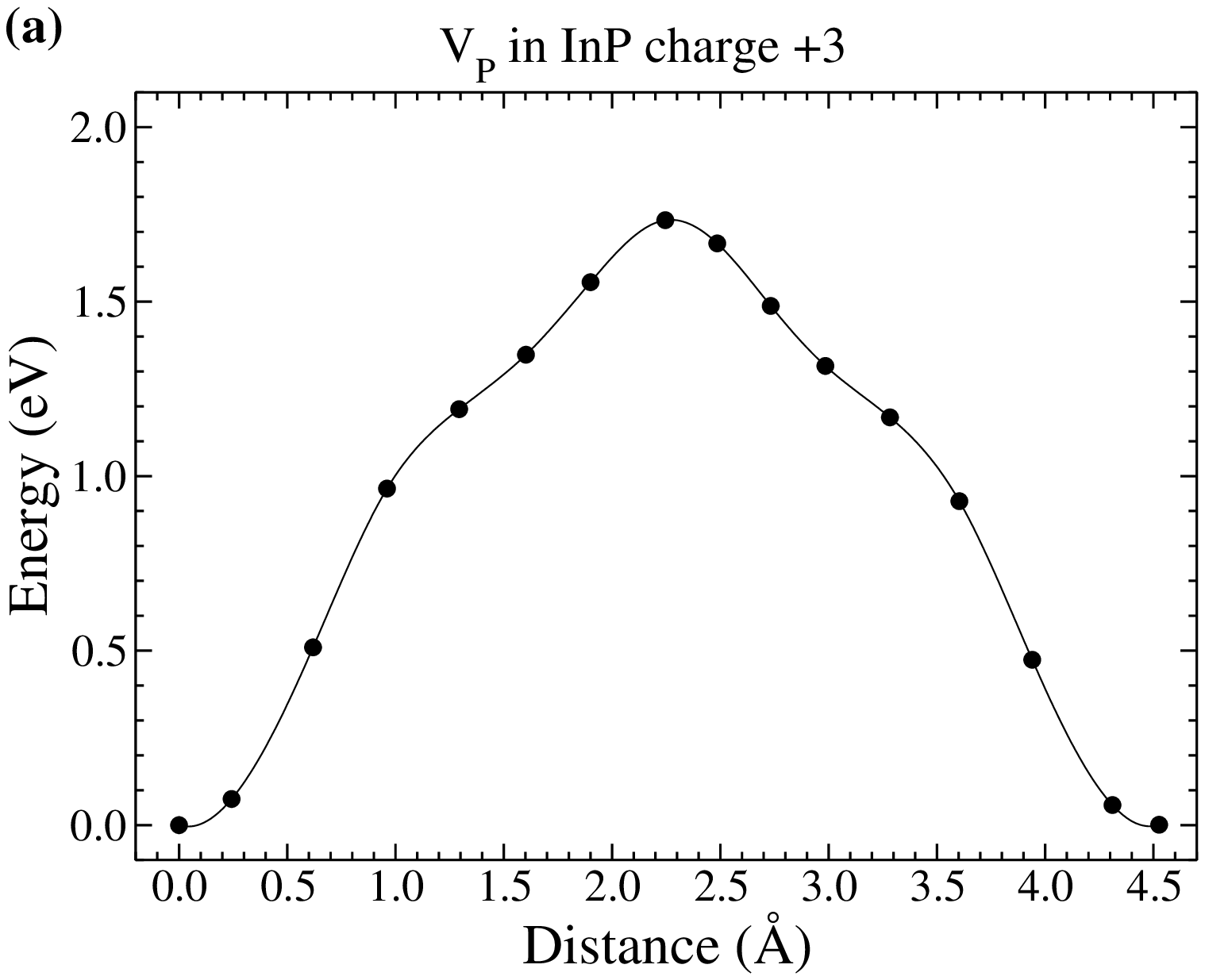}
\includegraphics[width=0.50\columnwidth]{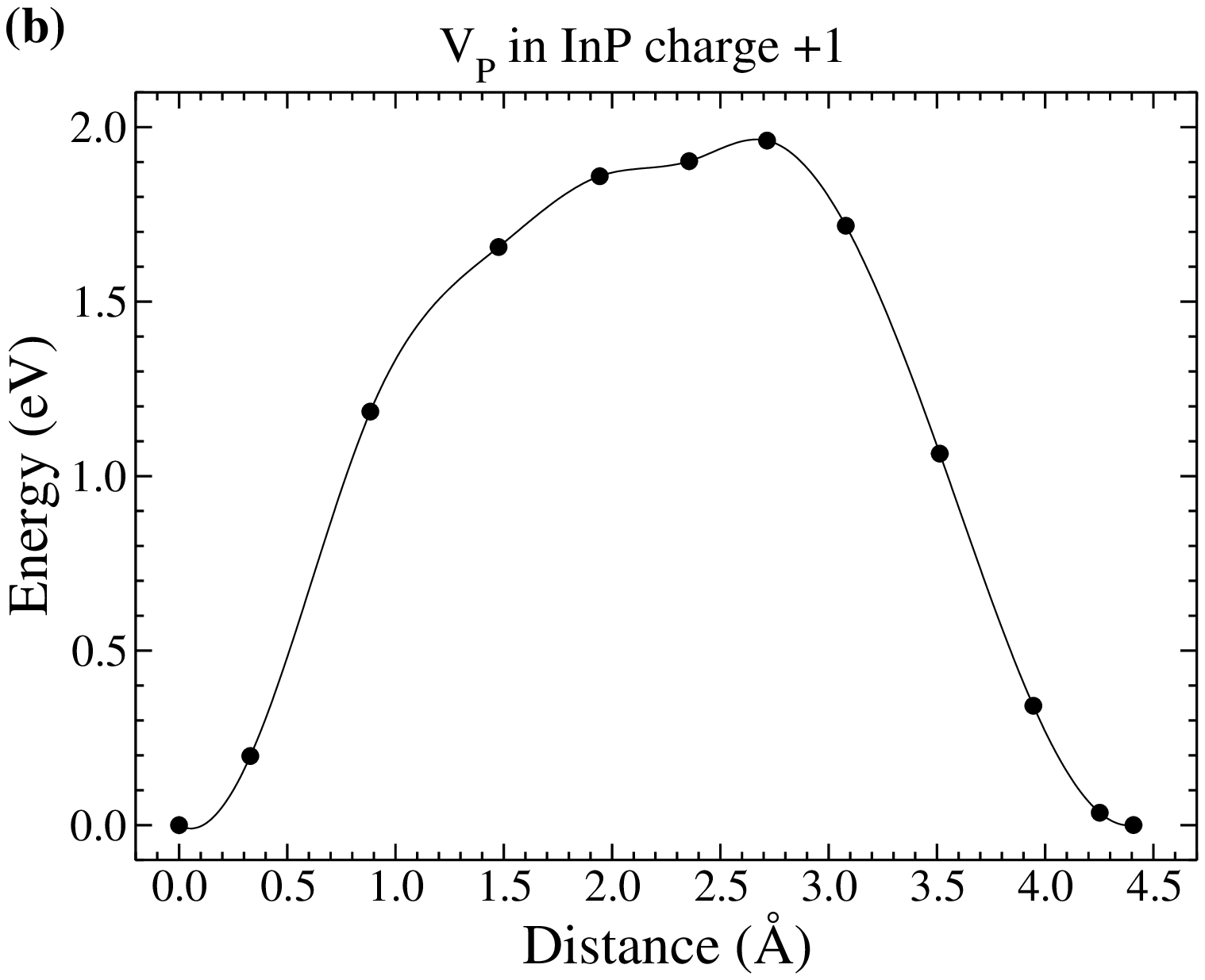}
\includegraphics[width=0.50\columnwidth]{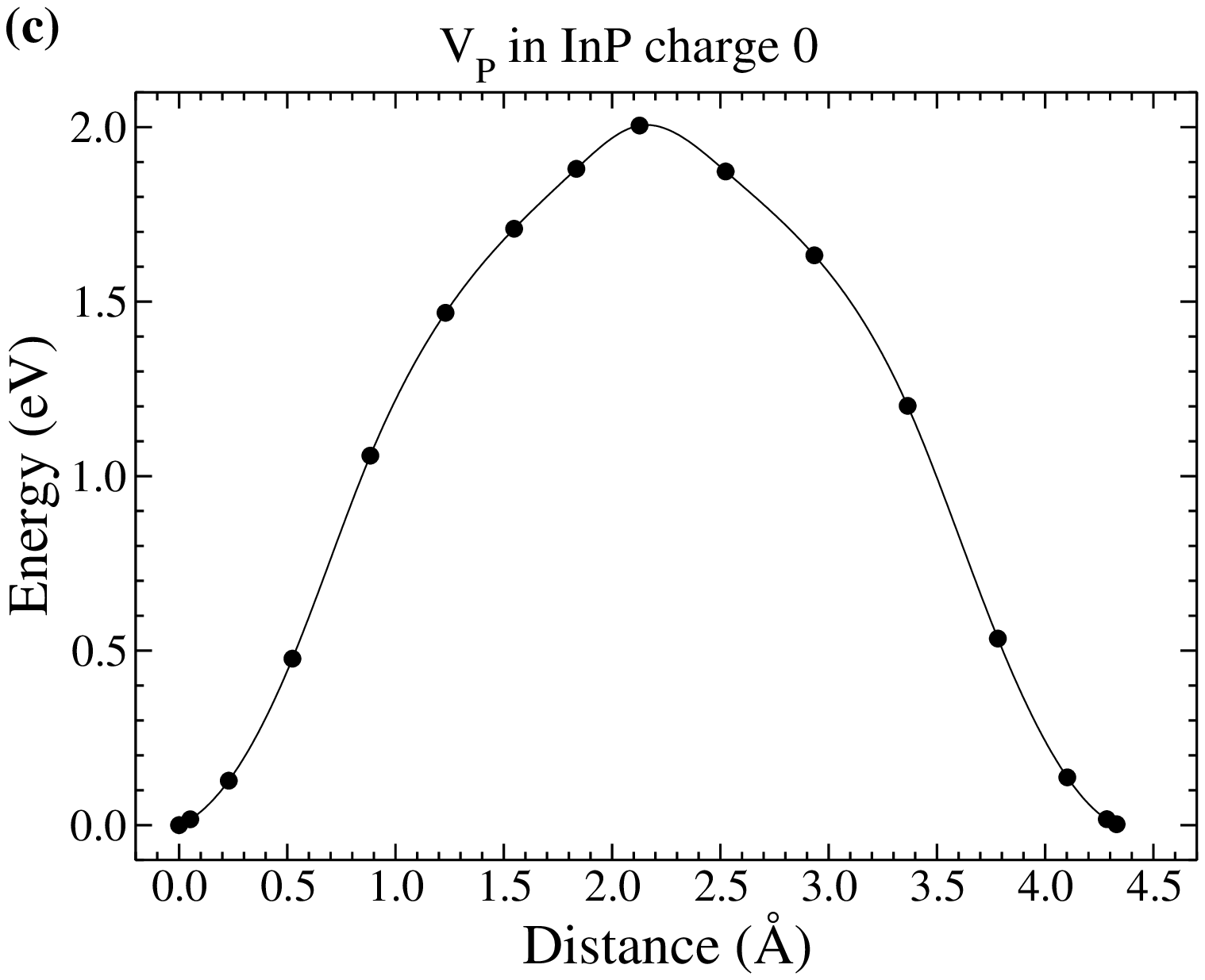}
\includegraphics[width=0.50\columnwidth]{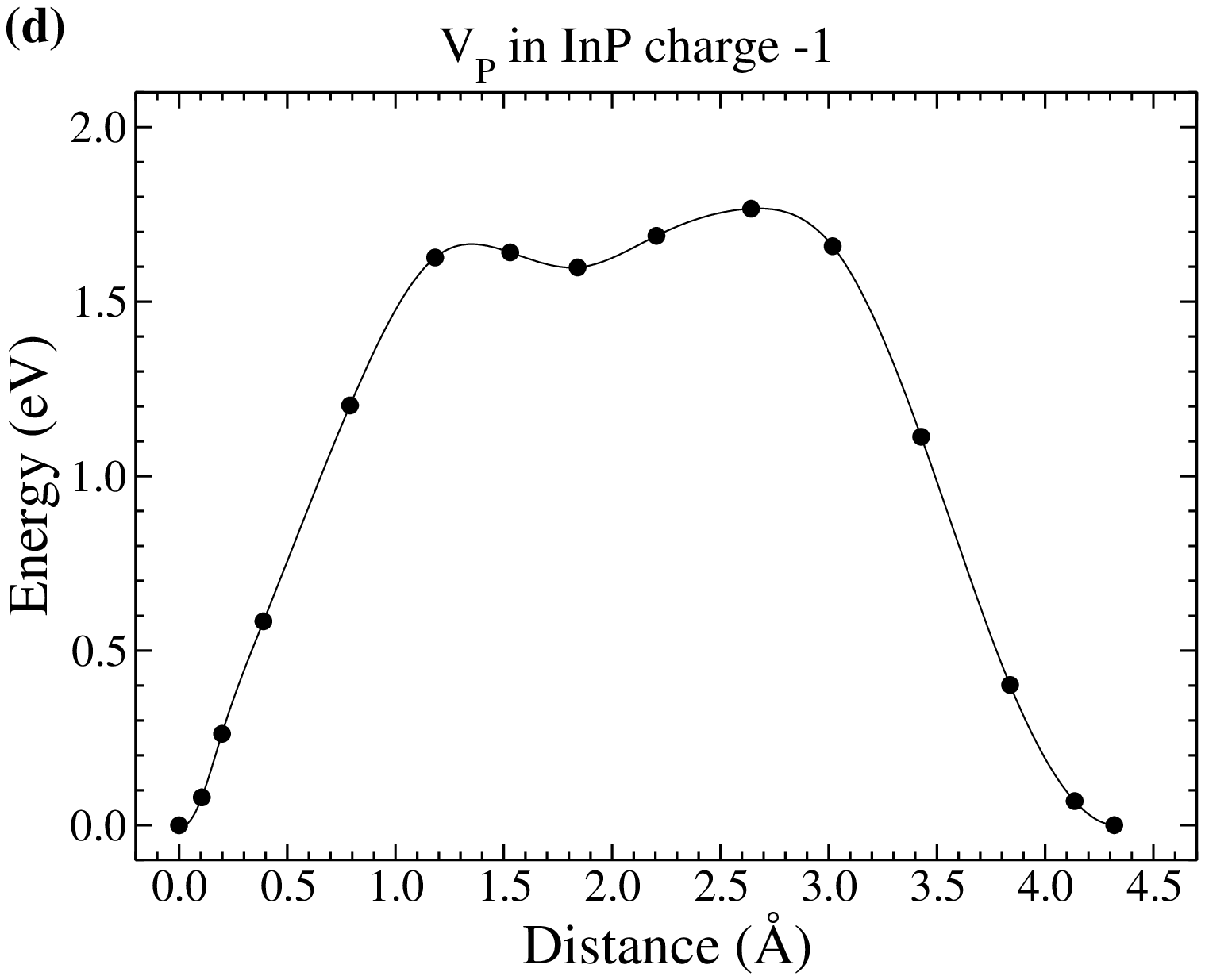}\\
\includegraphics[width=0.50\columnwidth]{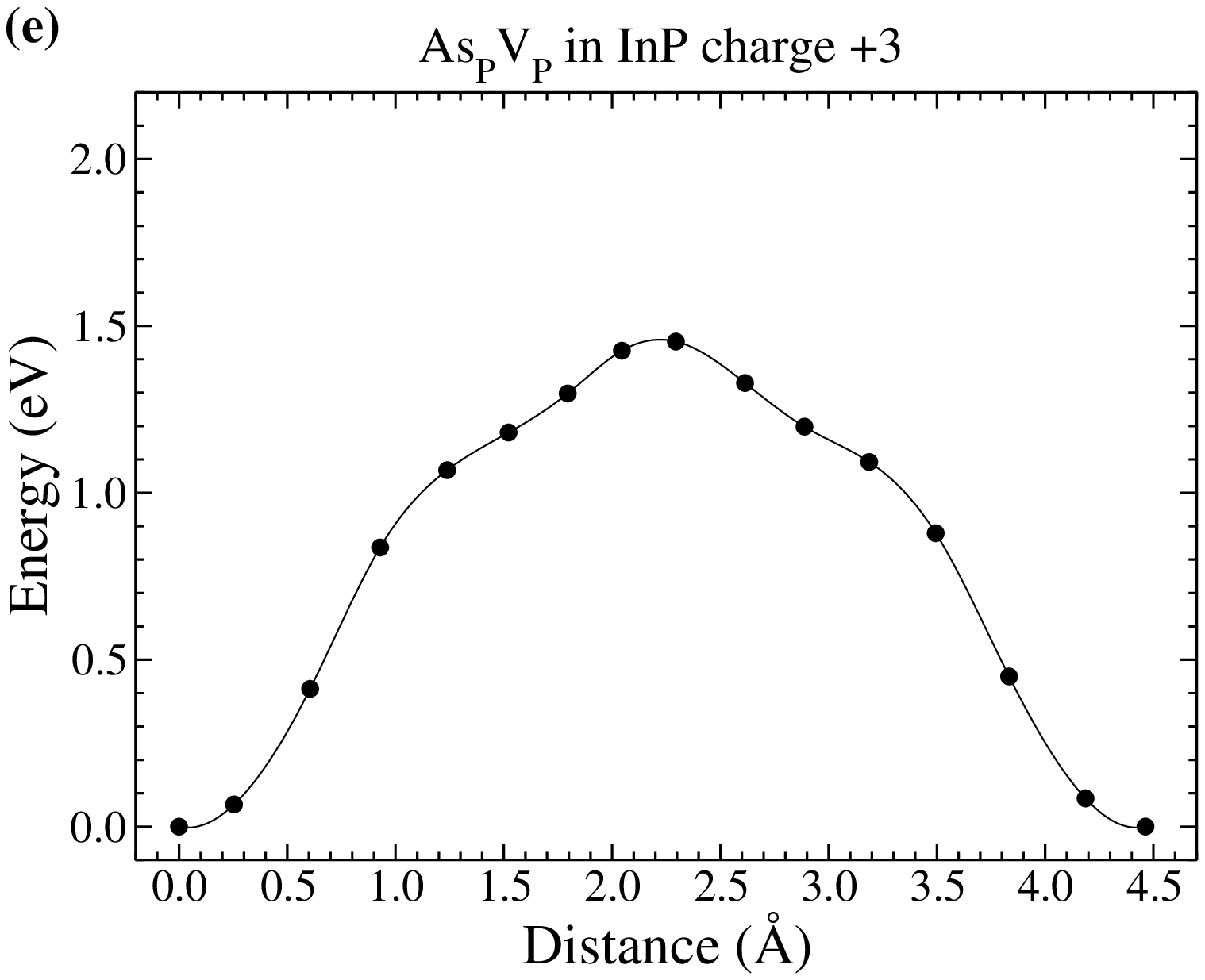}
\includegraphics[width=0.50\columnwidth]{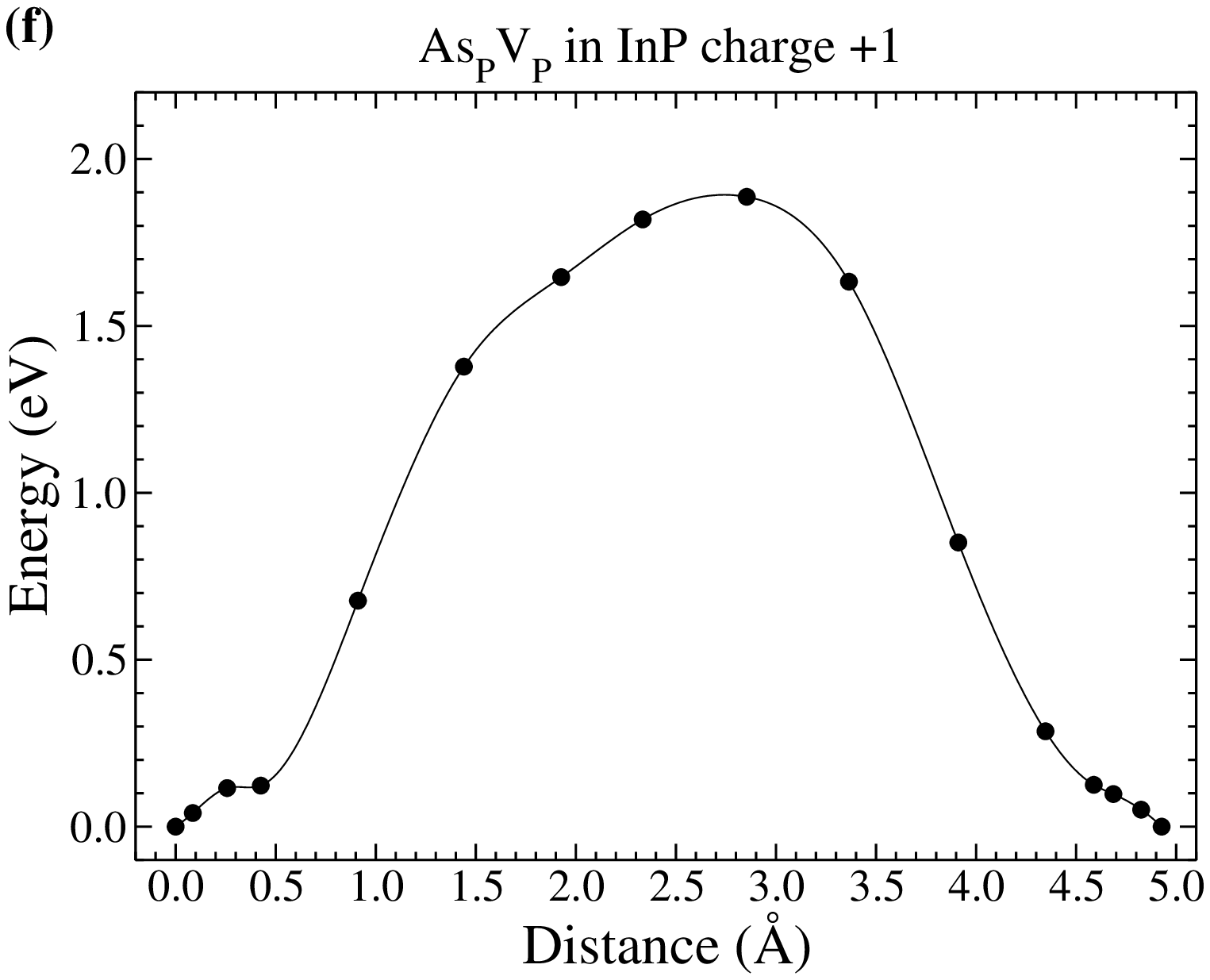}
\includegraphics[width=0.50\columnwidth]{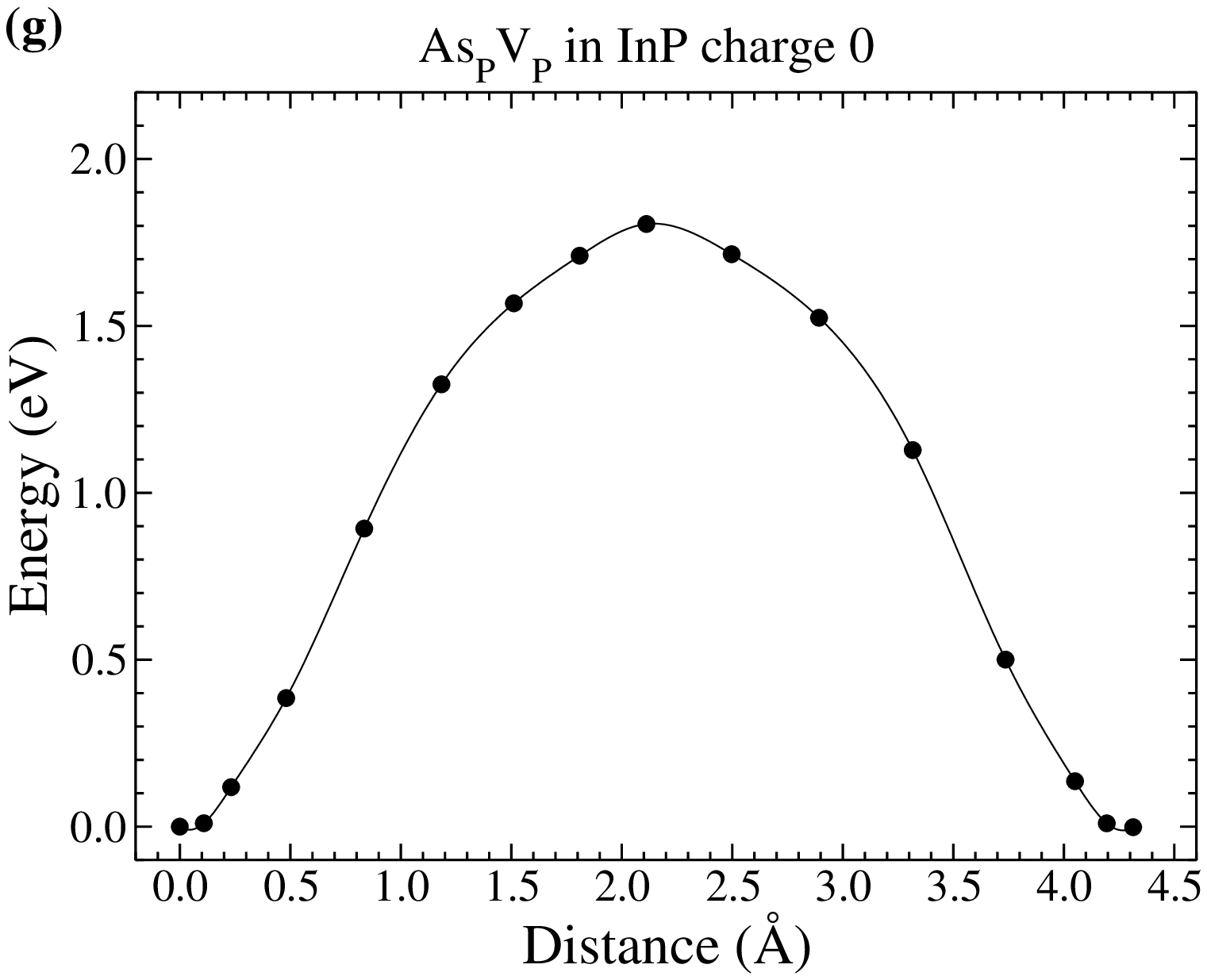}
\includegraphics[width=0.50\columnwidth]{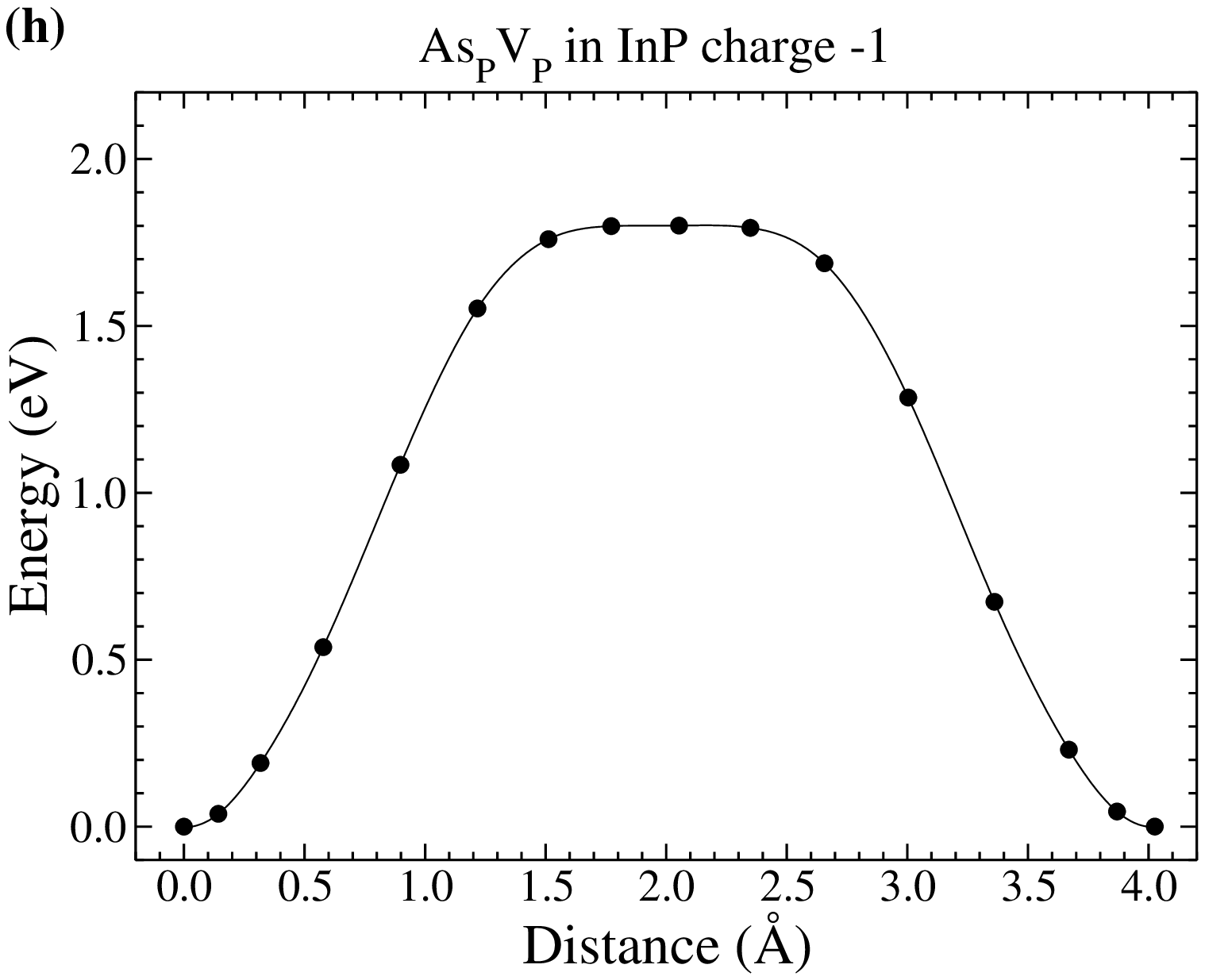}
\caption{\label{fig:figure4} Energy profiles for the process of a P atom migration to the phosphorous vacancy site (a, b, c, d) and a substitutional As atom migration to the phosphorous vacancy site (e, f, g, h) in InP for different charge states of the vacancy.}
\end{center}
\end{figure*}

Figure~\ref{fig:figure4} shows calculated energy profiles for the process of a P atom migration to the phosphorous vacancy site and a substitutional As atom migration to the phosphorous vacancy site in InP for different charge states of the vacancy.
When a P atom moves to the P vacancy site in InP, the symmetry relative to the (1$-$10) plane, in which the initial and final lattice sites are located, is generally preserved. For the P vacancy in InP in the charge state $-$1 the path with a slight deviation from this plane has approximately the same energy barrier. Reflection relative to the (110) plane, which is perpendicular to the (1$-$10) plane, transforms the initial lattice site into the final one. The (110) plane in some cases is a plane of symmetry for the migration trajectory. For a P vacancy in InP in the charge states +1 and $-$1 this symmetry is broken. This asymmetry was previously observed for GaAs~\cite{El-Mellouhi_2006, El-Mellouhi_2006a, Du_2013}.
Table~\ref{tab:table4} shows the calculated migration energy barriers and the attempt frequencies for As atoms diffusion in InP and self-diffusion of P atoms in InP by vacancy mechanism, and the pre-exponential factors for diffusion coefficient of phosphorous vacancies in InP.
\begin{table}
\caption{\label{tab:table4} Migration barriers and attempt frequencies for As atoms diffusion in InP and self-diffusion of P atoms in InP by vacancy mechanism, and pre-exponential factors for the diffusion coefficient of P vacancies in InP.}
\begin{center}
\begin{tabular}{p{0.13\columnwidth} c c c c p{0.16\columnwidth}}
\hline
 &\multicolumn{2}{c}{As in InP} & \multicolumn{2}{c}{P in InP} & \\
\hline
Vacancy charge & $E_m$, eV & $\nu$, THz & $E_m$, eV & $\nu$, THz & D\textsubscript{0} for vacancies, cm\textsuperscript{2}/s  \\
\hline
+3 & 1.453 & 2.53 & 1.734 & 4.62 & 1.64$\cdot10^{-2}$\\
+1 & 1.887 & 2.77 & 1.962 & 5.44 & 1.93$\cdot10^{-2}$\\
0 & 1.807 & 3.24 & 2.005 & 5.30 & 1.88$\cdot10^{-2}$\\
$-$1 & 1.801 & 3.22 & 1.766 & 6.75 & 2.40$\cdot10^{-2}$\\
\hline
\end{tabular}
\end{center}
\end{table}

\subsection{Migration barriers for diffusion of As and P atoms in InAs by vacancy mechanism}
\begin{figure*}
\begin{center}
\includegraphics[width=0.50\columnwidth]{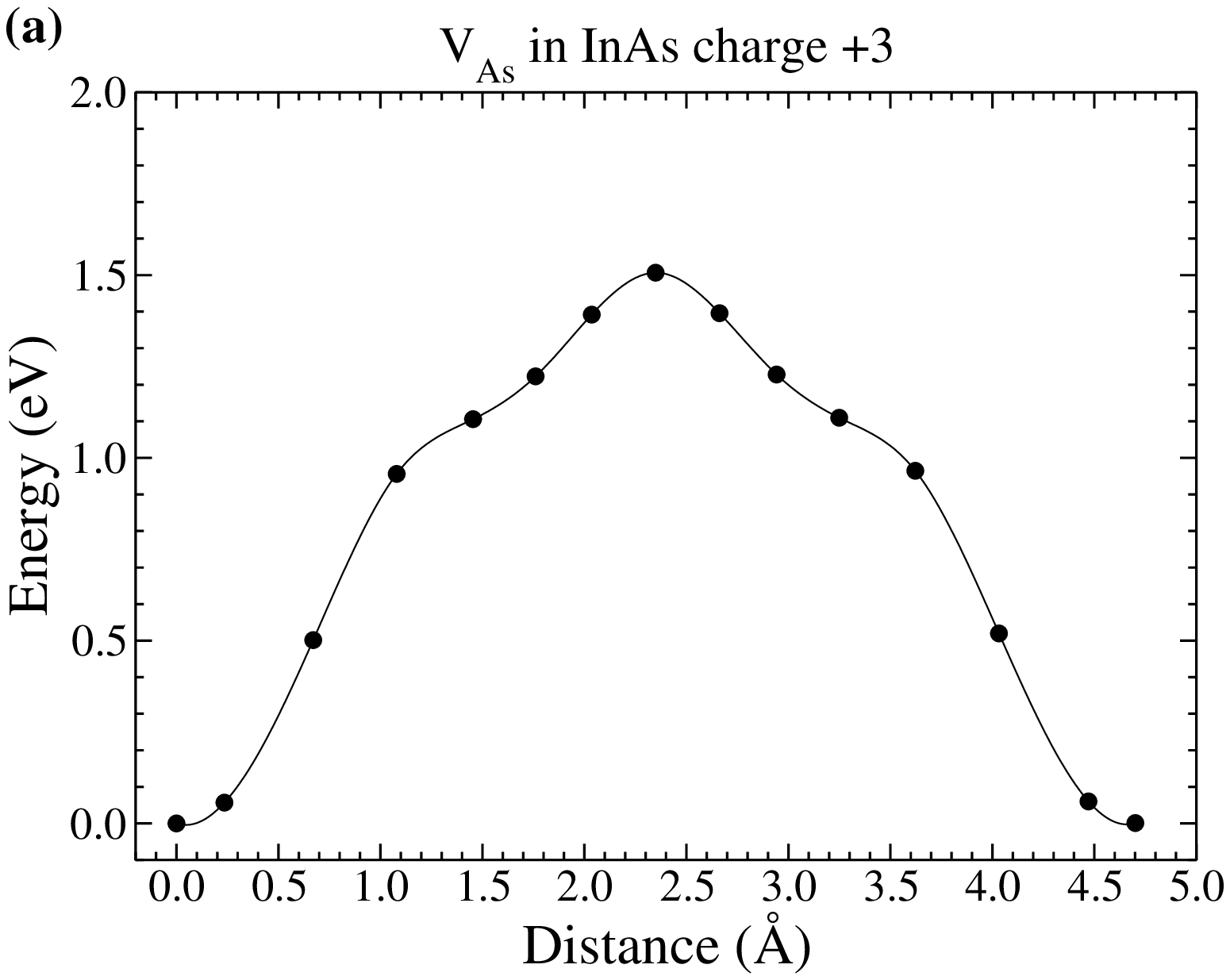}
\includegraphics[width=0.50\columnwidth]{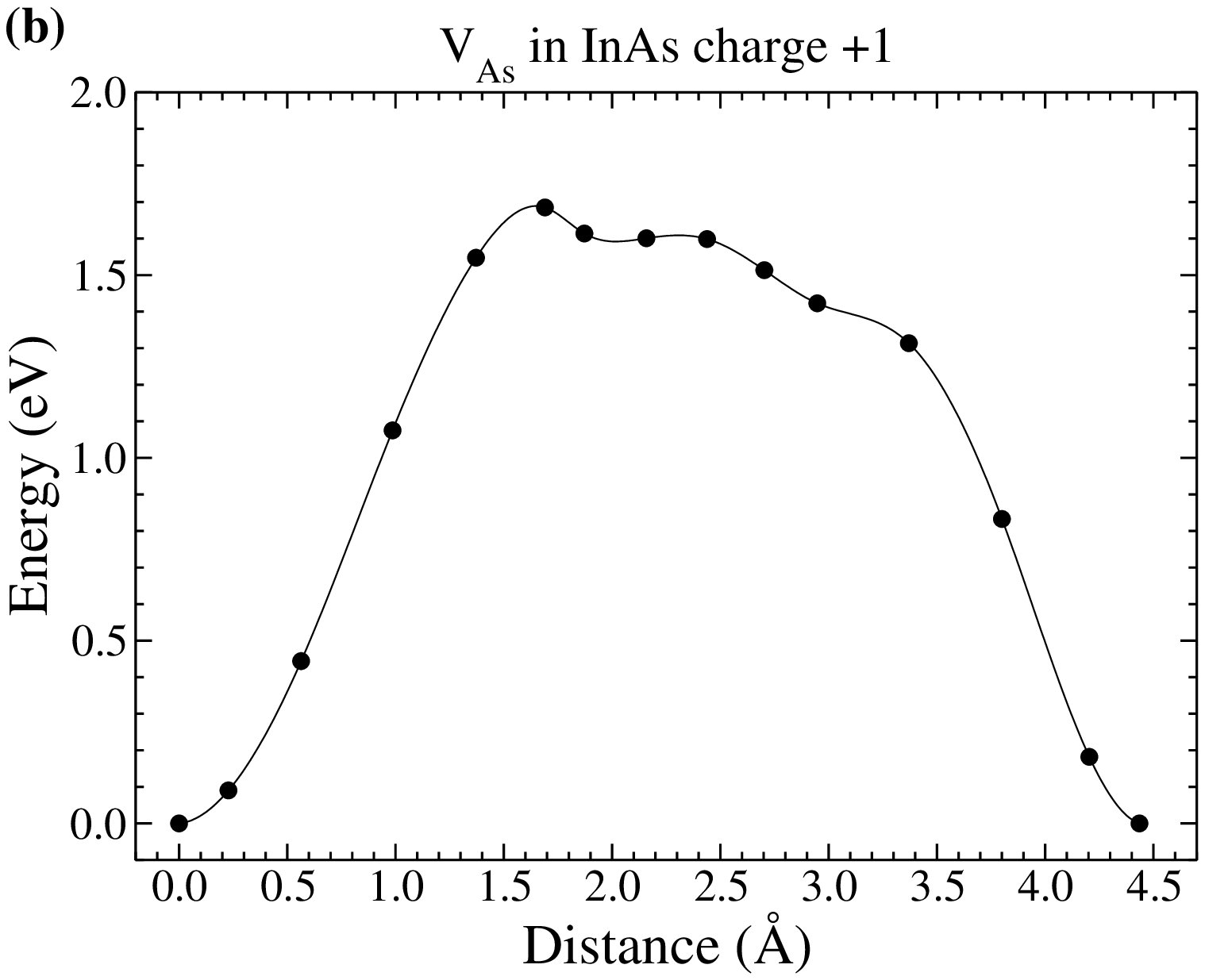}
\includegraphics[width=0.50\columnwidth]{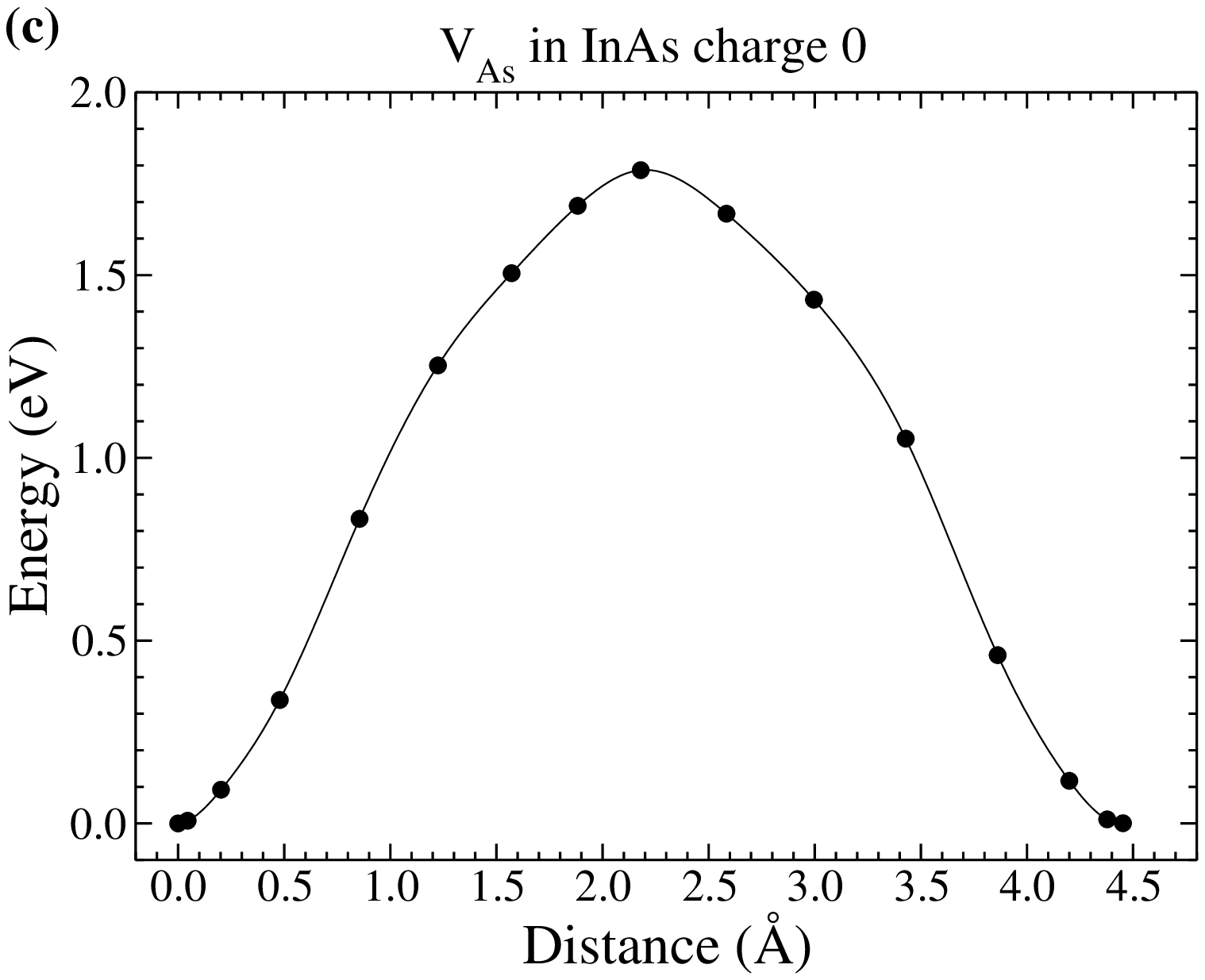}
\includegraphics[width=0.50\columnwidth]{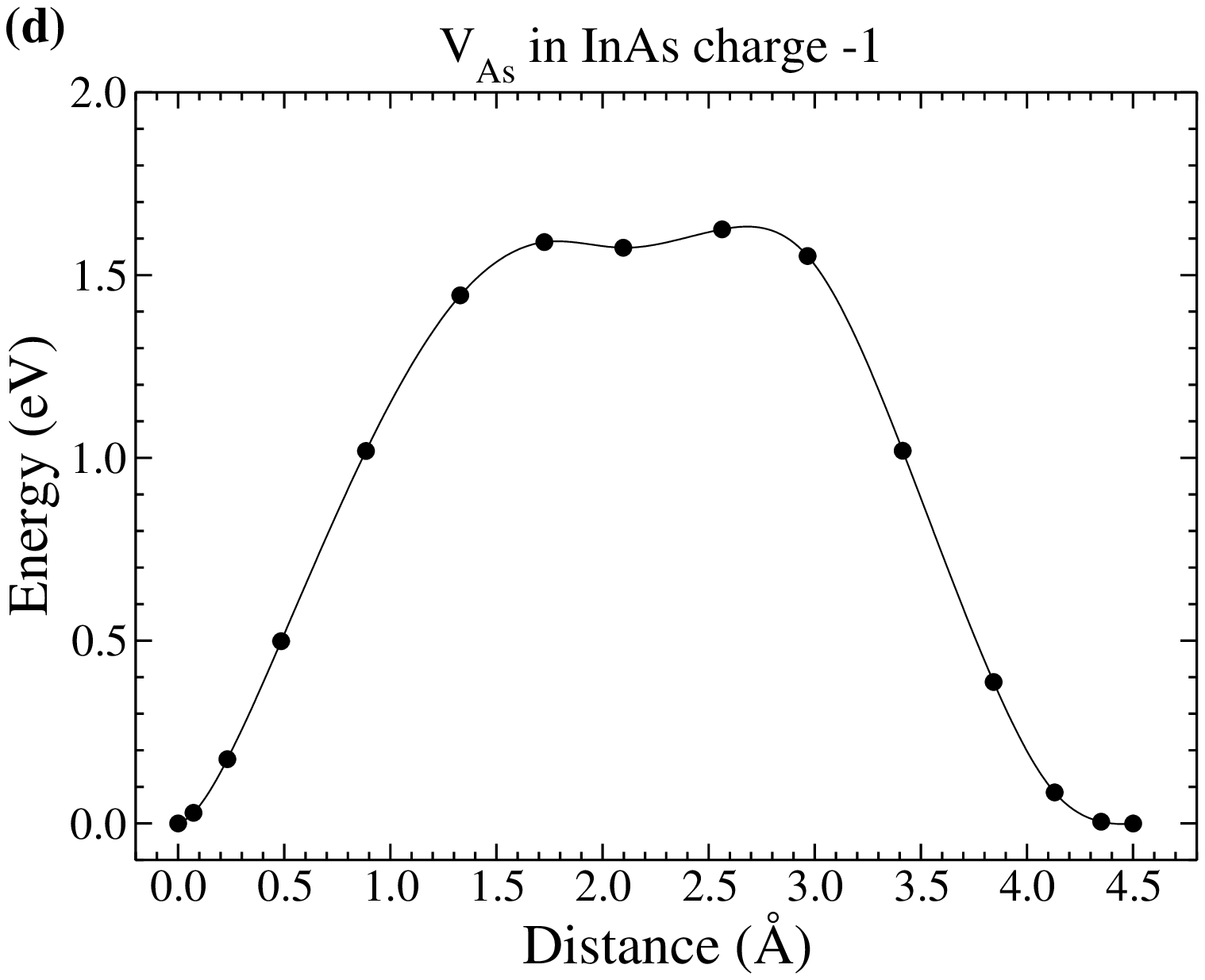}\\
\includegraphics[width=0.50\columnwidth]{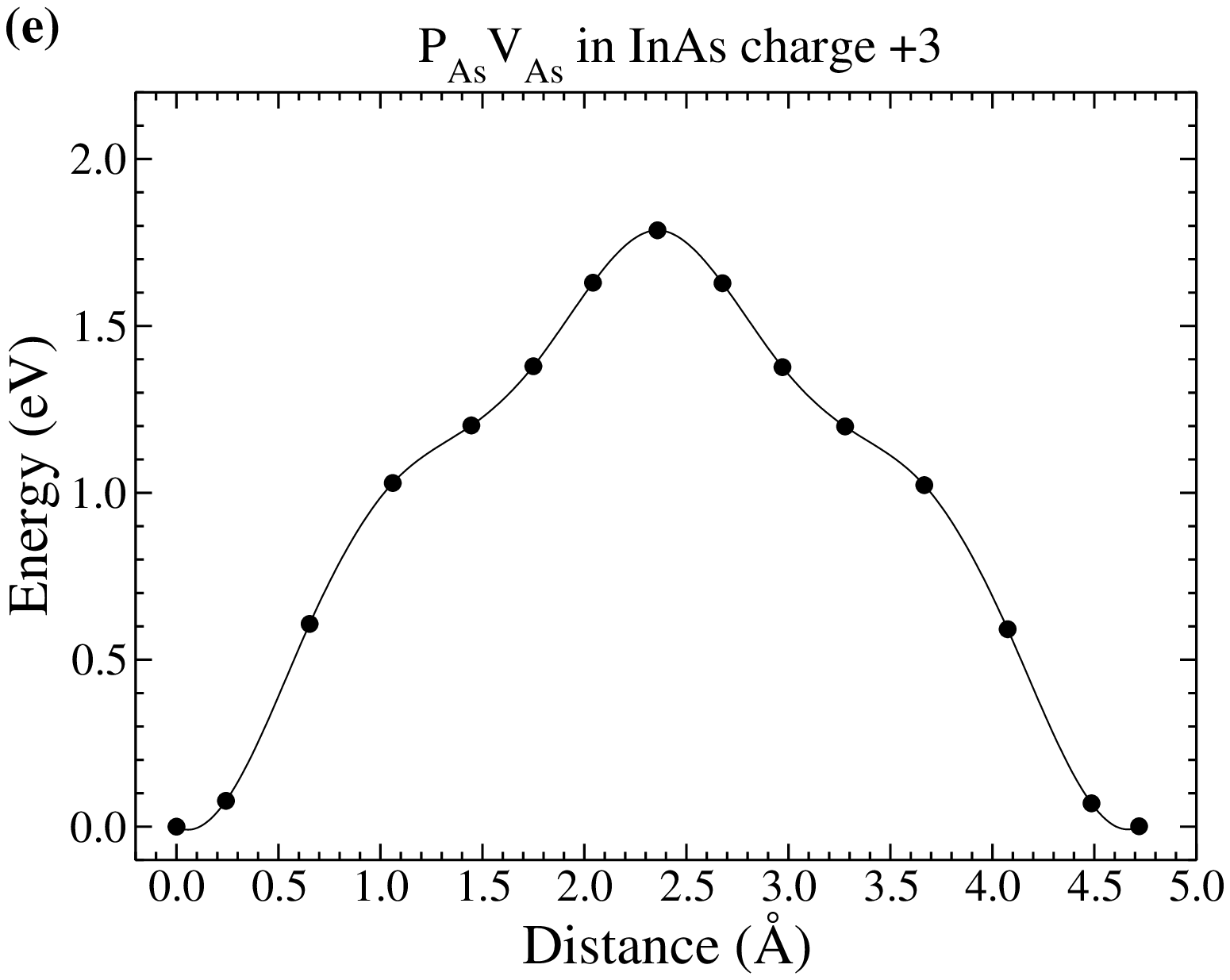}
\includegraphics[width=0.50\columnwidth]{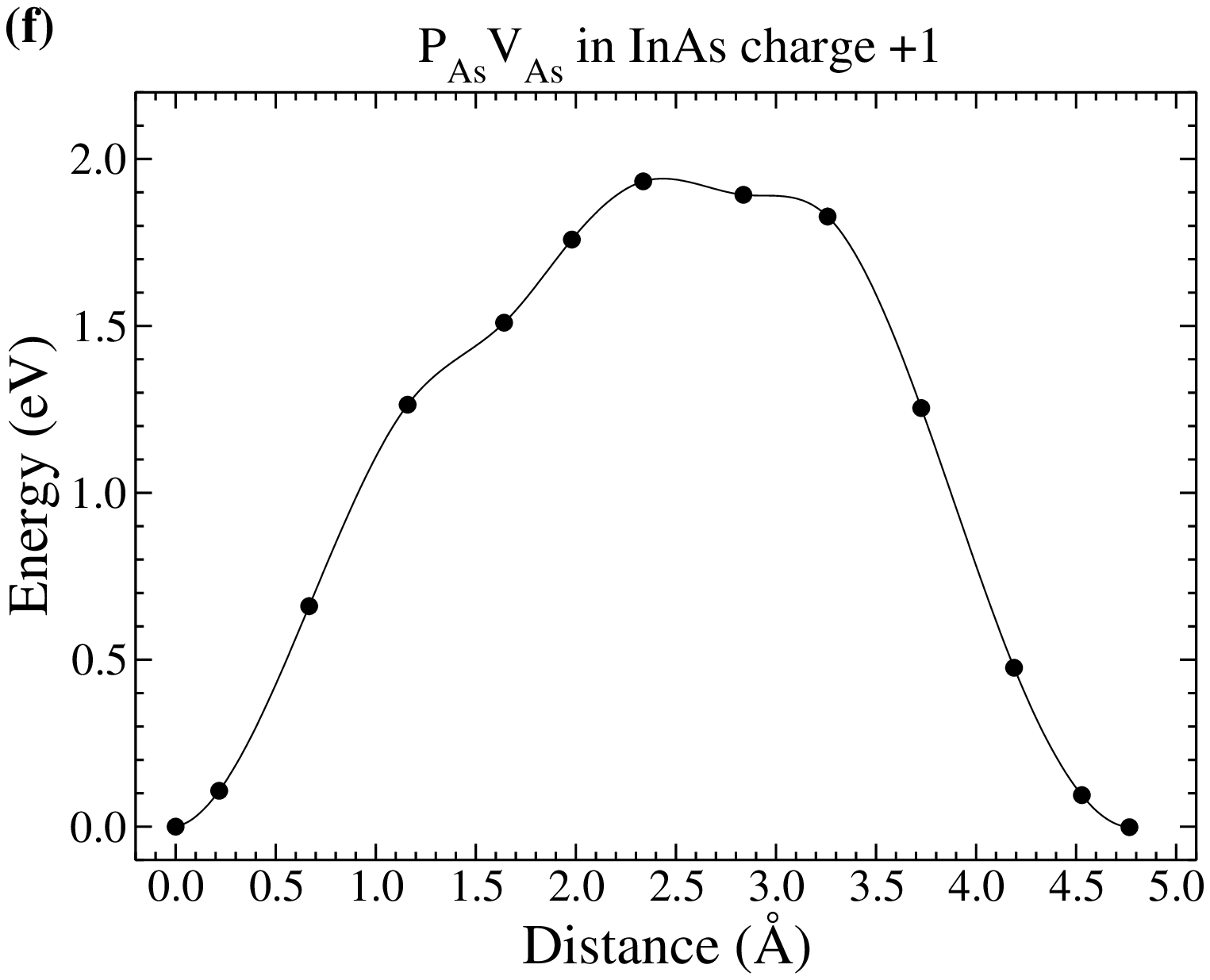}
\includegraphics[width=0.50\columnwidth]{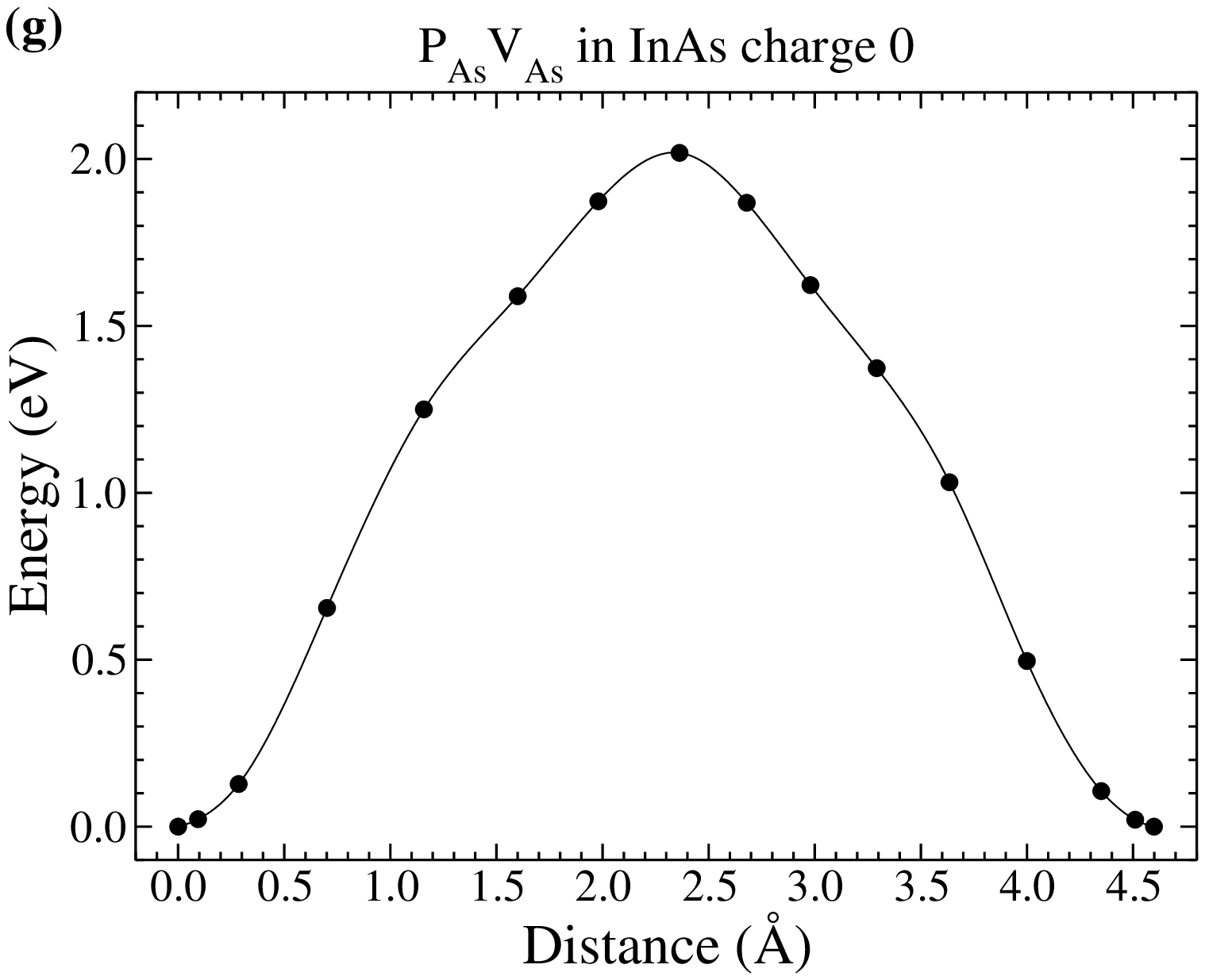}
\includegraphics[width=0.50\columnwidth]{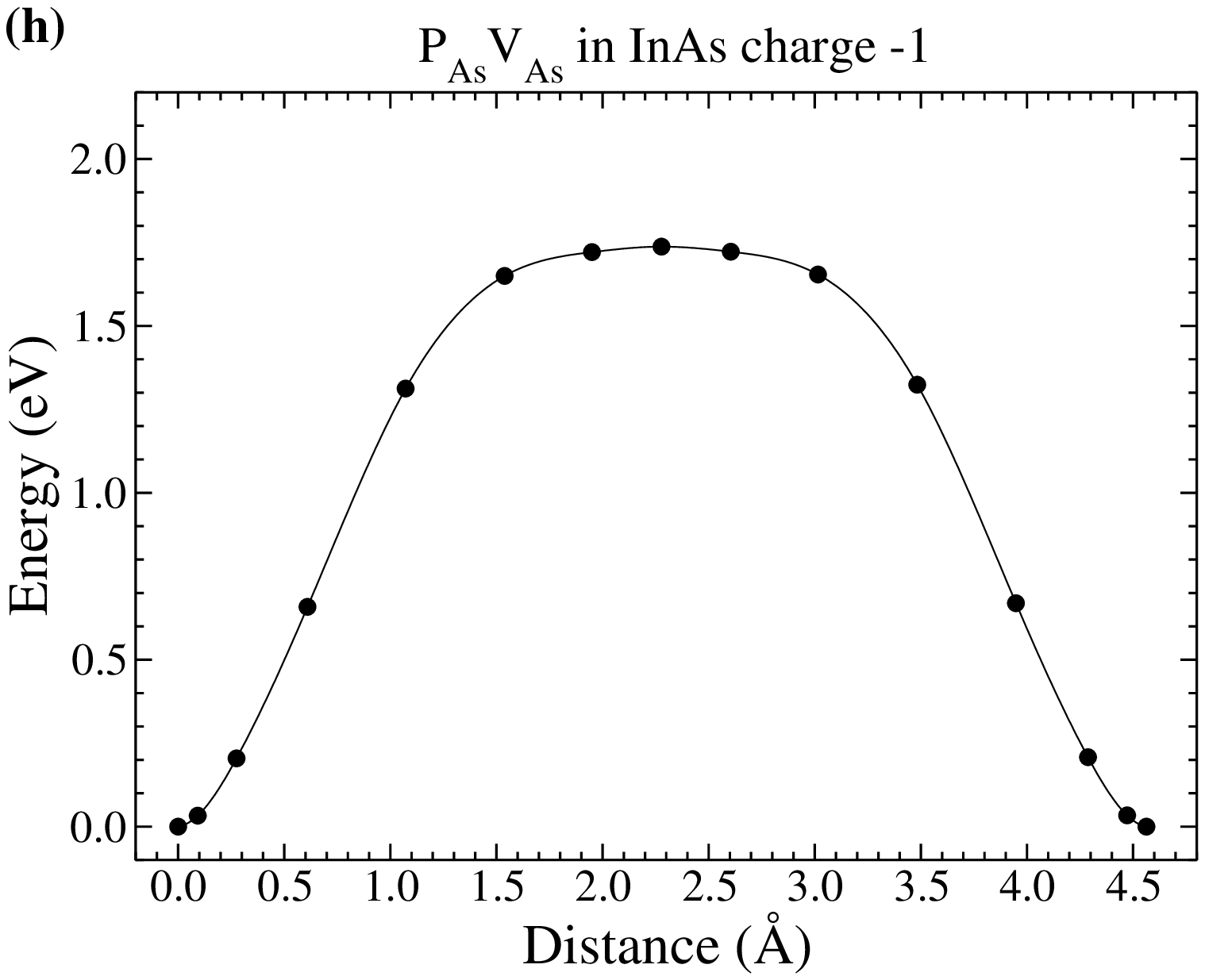}
\caption{\label{fig:figure5} Energy profiles for the process of an As atom migration to the arsenic vacancy site (a, b, c, d) and a substitutional P atom migration to the arsenic vacancy site (e, f, g, h) in InAs for different charge states of the vacancy.}
\end{center}
\end{figure*}

Figure~\ref{fig:figure5} shows calculated energy profiles for the process of As and P atoms migration to the arsenic vacancy site in InAs for different charge states of the vacancy.
An asymmetric diffusion trajectory is observed for an arsenic vacancy in InAs in the charge states +1 and $-$1, similarly to the case of a P vacancy in InP. Table~\ref{tab:table5} shows the calculation results of the migration energy barriers and attempt frequencies for the diffusion of As and P atoms in InAs by vacancy mechanism. 
\begin{table}
\caption{\label{tab:table5} Migration barriers and attempt frequencies for P atoms diffusion in InAs and self-diffusion of As atoms in InAs by vacancy mechanism, and pre-exponential factors for the diffusion coefficient of As vacancies in InAs.}
\begin{center}
\begin{tabular}{p{0.13\columnwidth} c c c c p{0.16\columnwidth}}
\hline
 &\multicolumn{2}{c}{P in InAs} & \multicolumn{2}{c}{As in InAs} & \\
\hline
\hline
Vacancy charge & $E_m$, eV & $\nu$, THz & $E_m$, eV & $\nu$, THz & D\textsubscript{0} for vacancies, cm\textsuperscript{2}/s  \\
\hline
+3 & 1.787 & 4.74 & 1.506 & 2.72 & 1.04$\cdot10^{-2}$\\
+1 & 1.935 & 4.82 & 1.683 & 2.97 & 1.13$\cdot10^{-2}$\\
0 & 2.019 & 5.12 & 1.787 & 3.29 & 1.25$\cdot10^{-2}$\\
$-$1 & 1.738 & 6.58 & 1.625 & 3.32 & 1.27$\cdot10^{-2}$\\
\hline
\end{tabular}
\end{center}
\end{table}
The migration barrier for a neutral As vacancy in InAs calculated in the generalized gradient approximation is $E_m$=1.79~eV. This result is lower by 0.21~eV than the calculation in the local density approximation in the work of Reveil~et~al.~\cite{Reveil_2017}~$E_m$=2.0~eV.

\subsection{Migration barriers for diffusion of interstitial P and As atoms in InP.}
\begin{figure*}
\begin{minipage}{0.33\linewidth}
\begin{center}
\includegraphics[width=0.80\columnwidth]{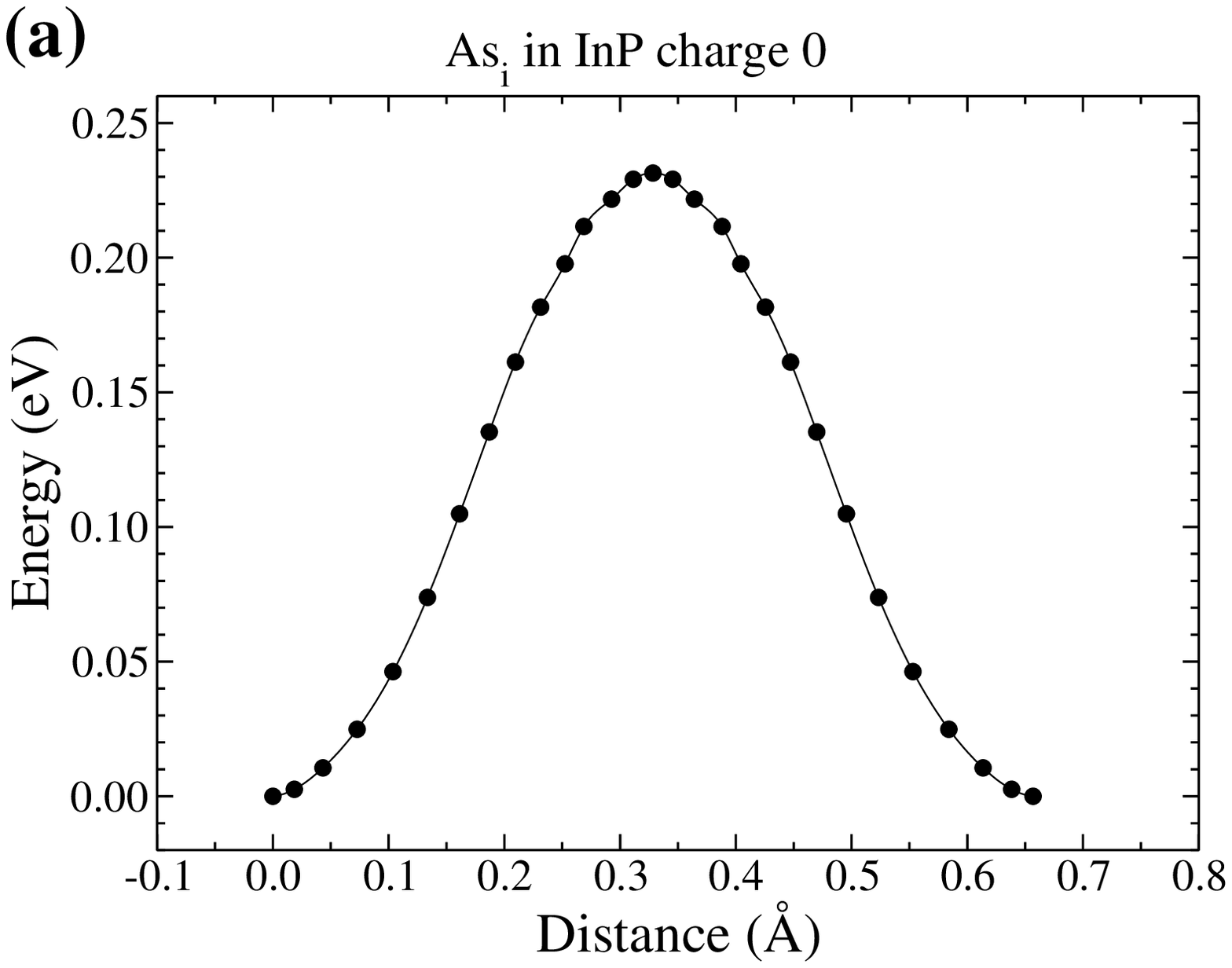}
\includegraphics[width=0.85\columnwidth]{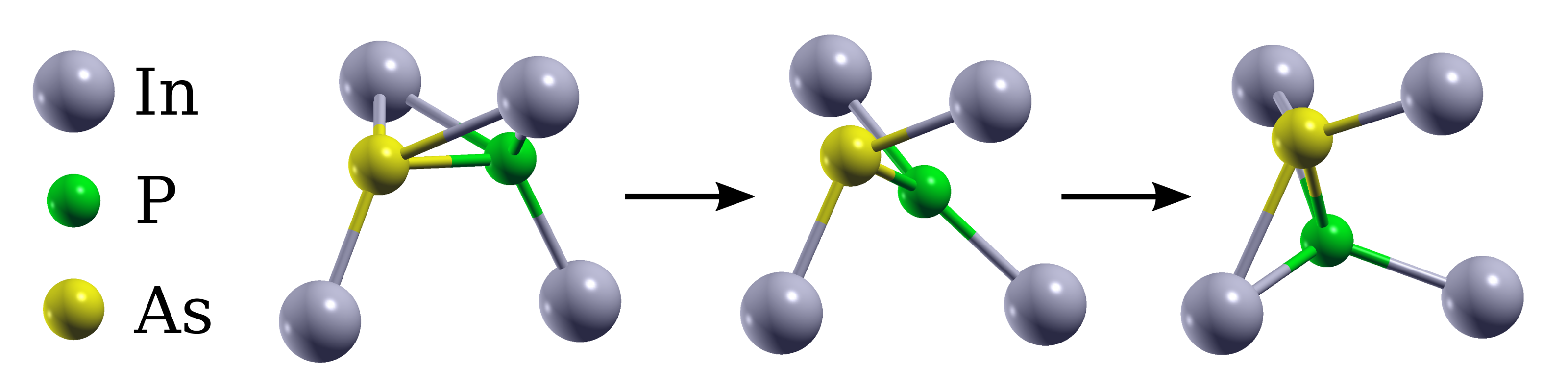}
\end{center}
\end{minipage}
\begin{minipage}{0.33\linewidth}
\begin{center}
\includegraphics[width=0.80\columnwidth]{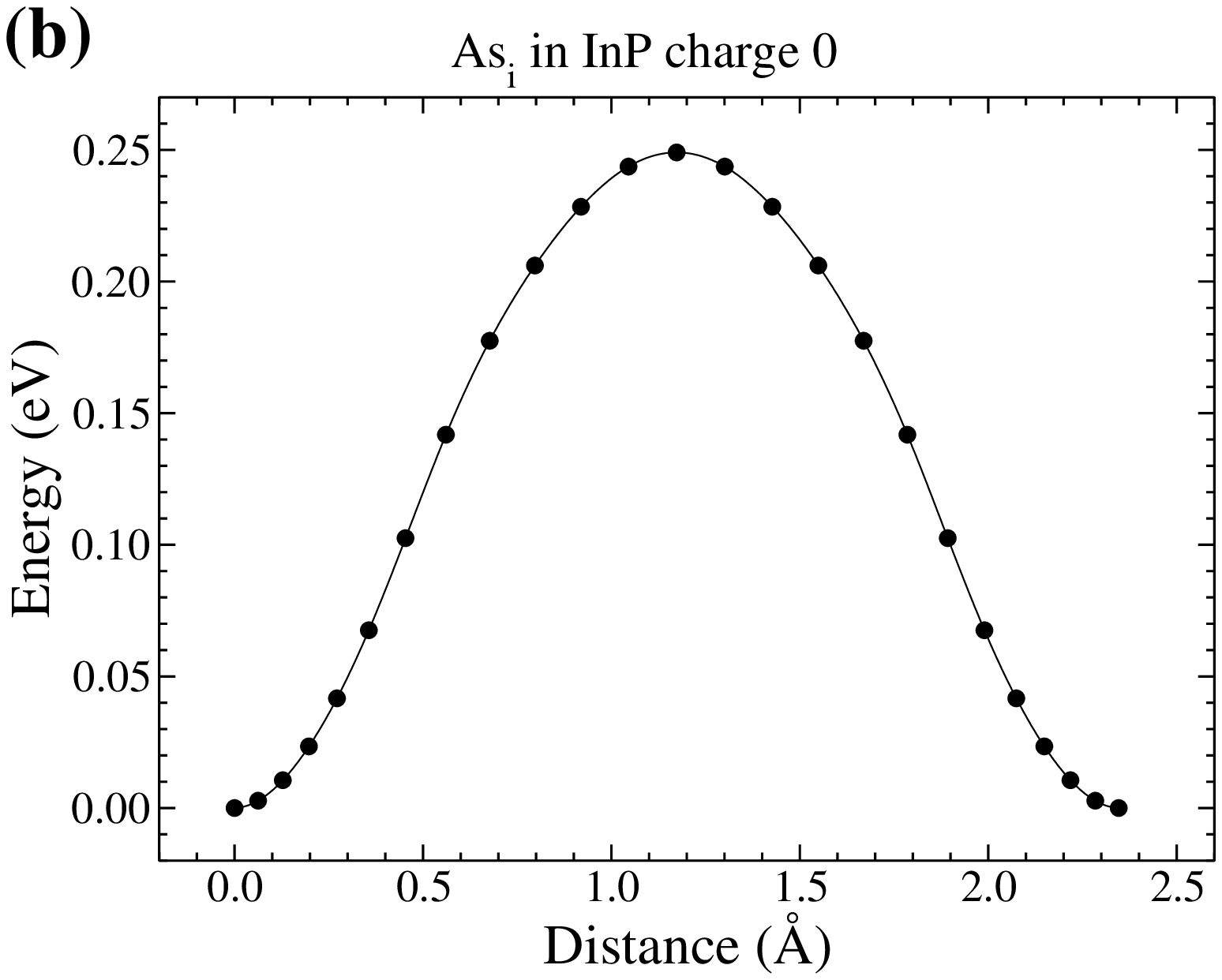}
\includegraphics[width=0.85\columnwidth]{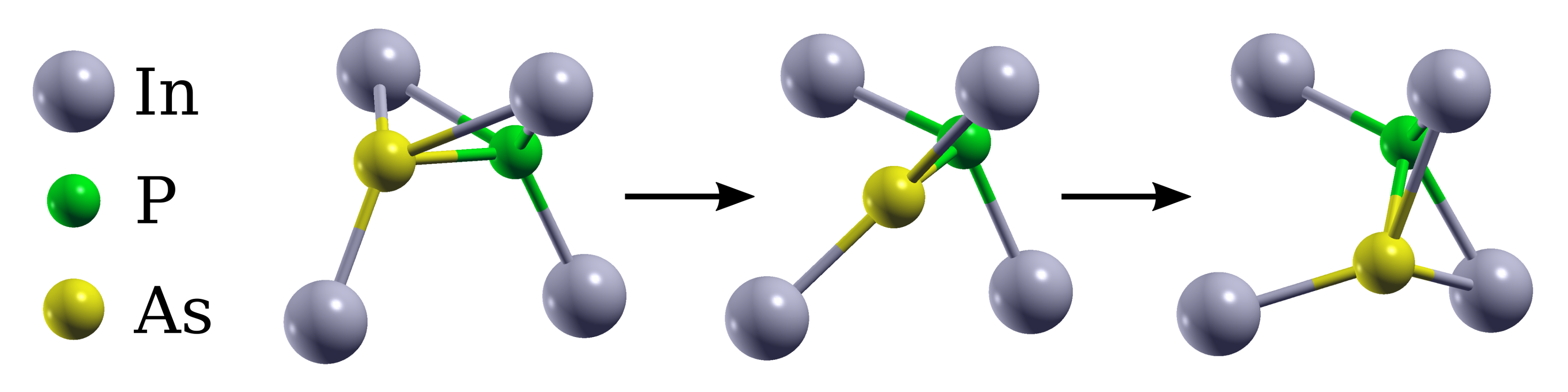}
\end{center}
\end{minipage}
\begin{minipage}{0.33\linewidth}
\begin{center}
\includegraphics[width=0.80\columnwidth]{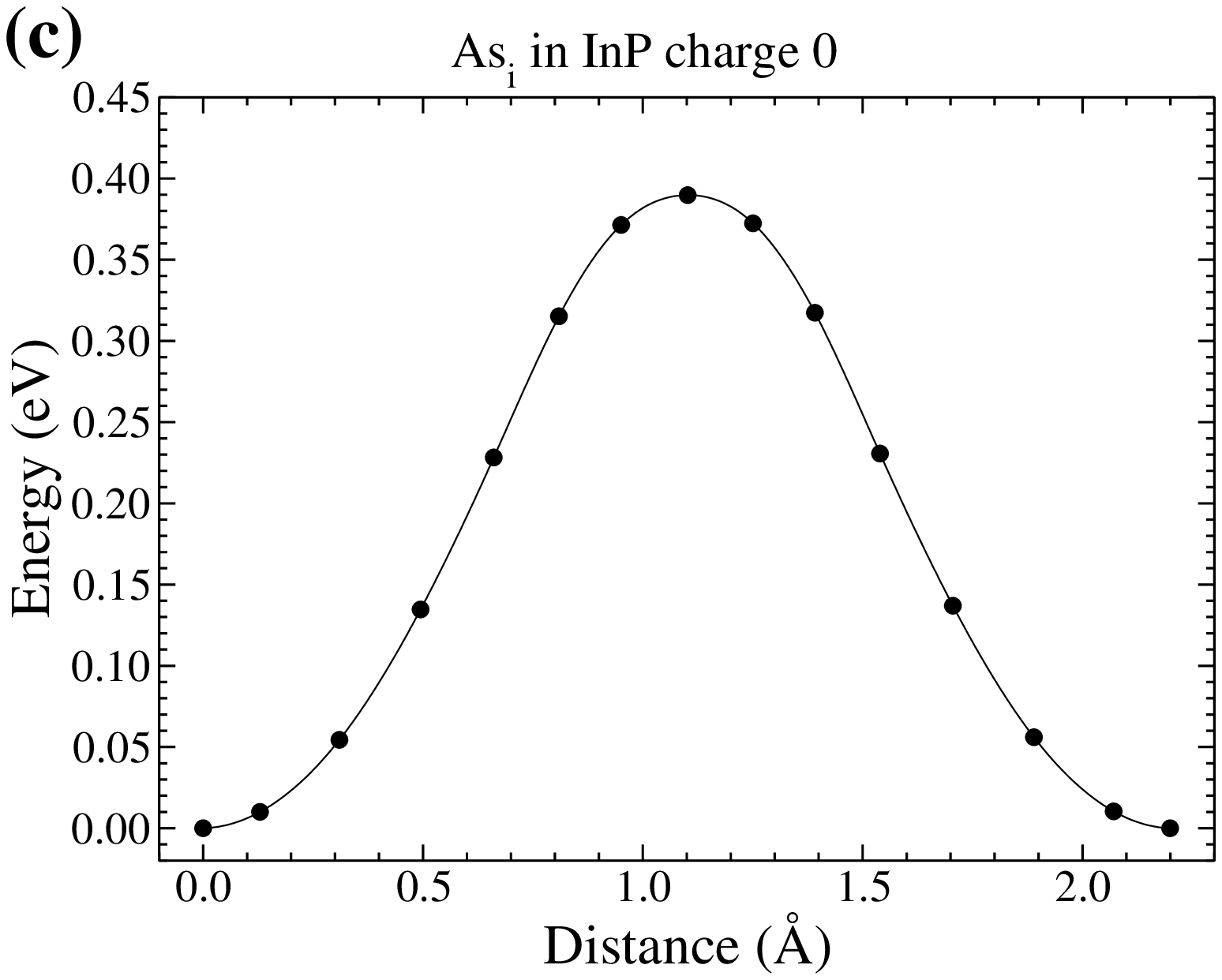}
\includegraphics[width=0.99\columnwidth]{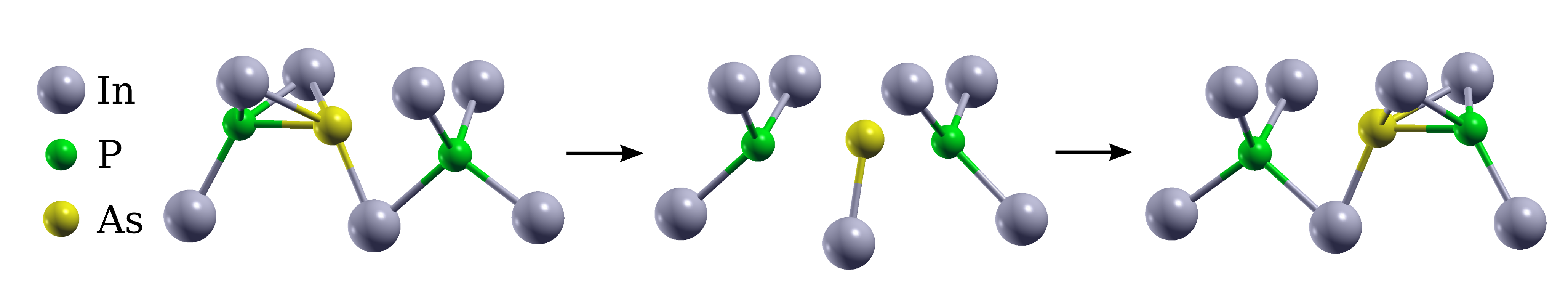}
\end{center}
\end{minipage}
\begin{minipage}{0.33\linewidth}
\begin{center}
\includegraphics[width=0.80\columnwidth]{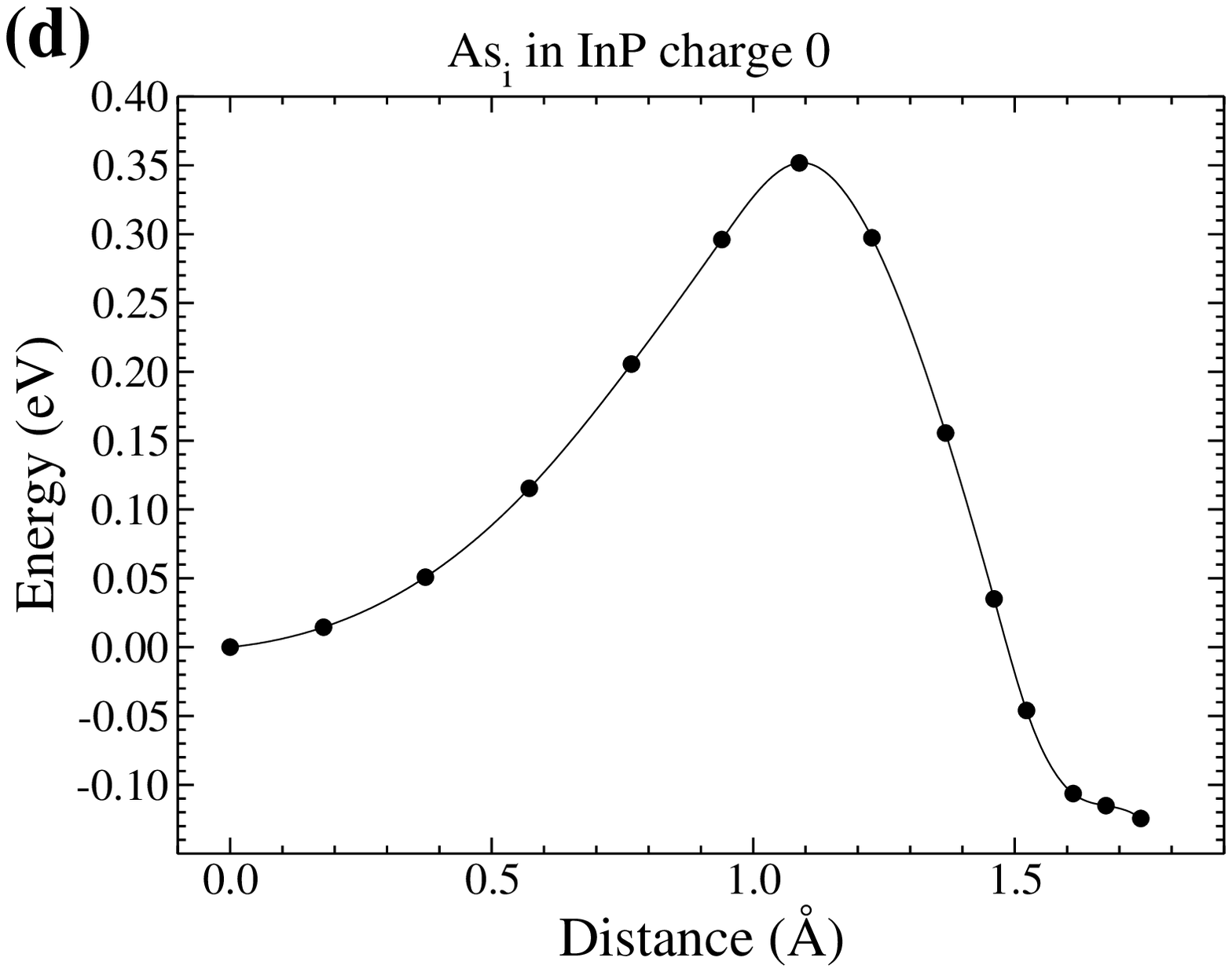}
\includegraphics[width=0.99\columnwidth]{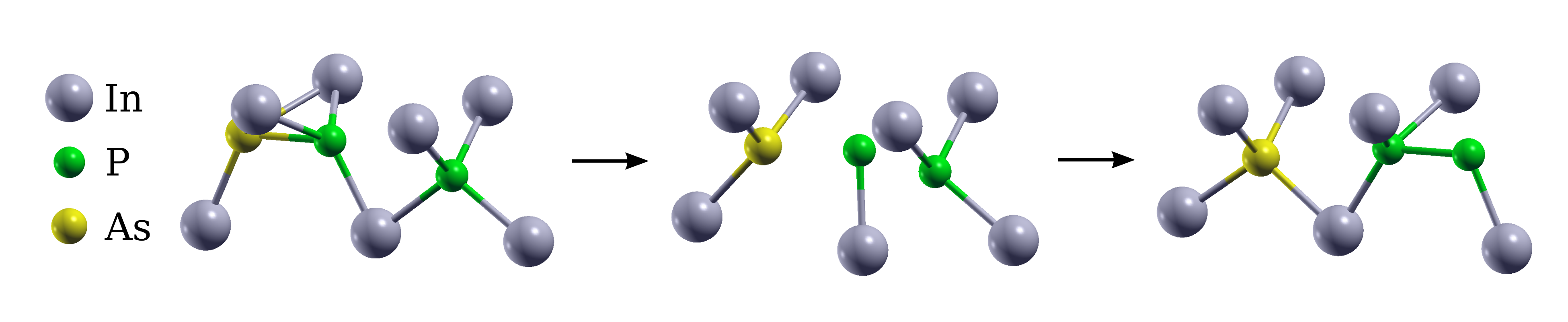}
\end{center}
\end{minipage}
\begin{minipage}{0.33\linewidth}
\begin{center}
\includegraphics[width=0.80\columnwidth]{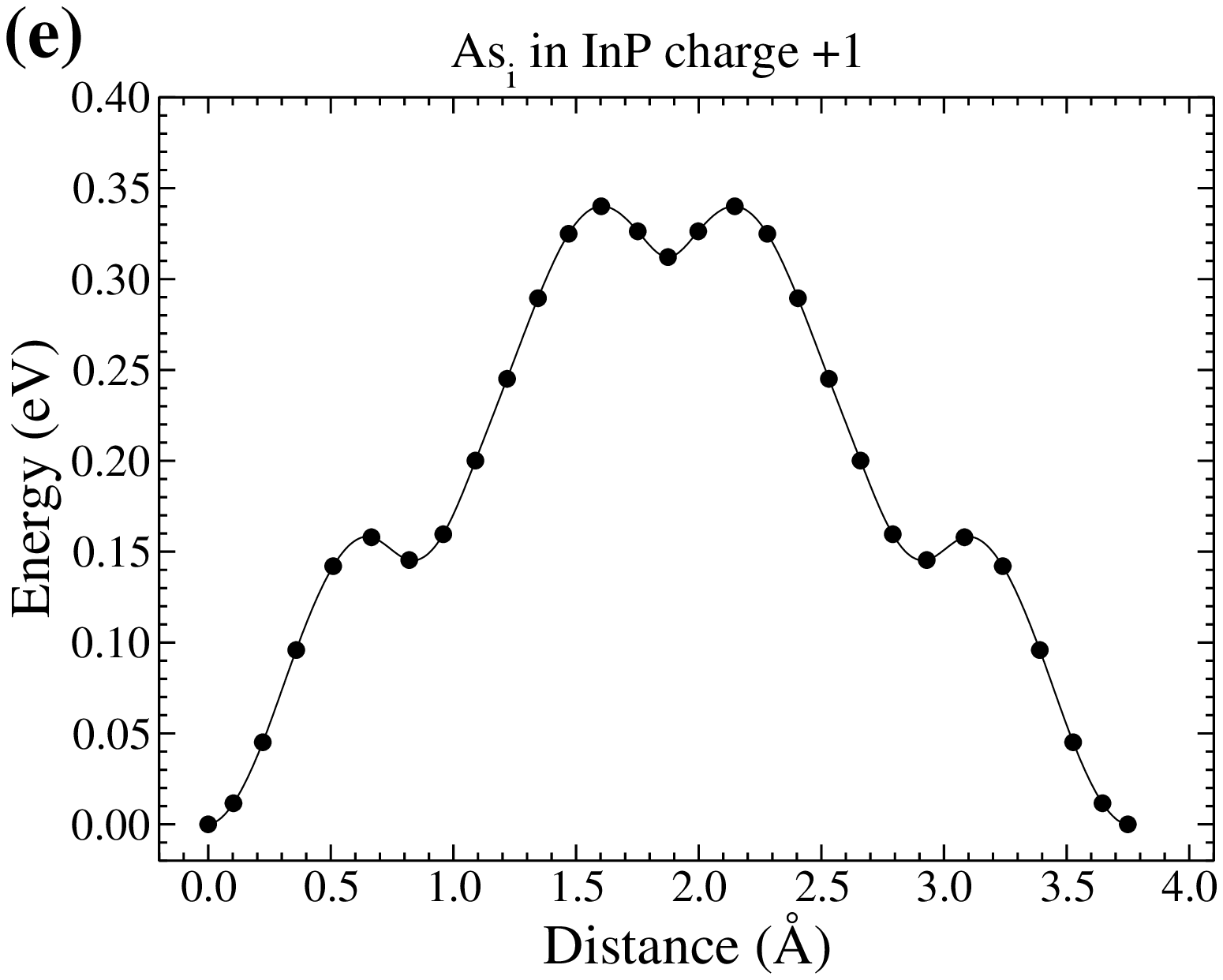}
\includegraphics[width=0.99\columnwidth]{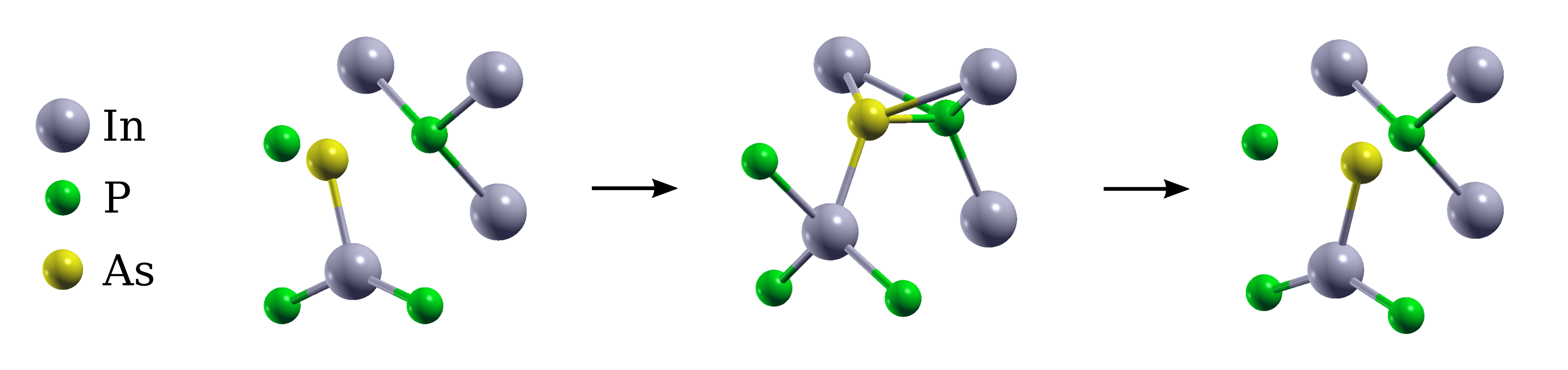}
\end{center}
\end{minipage}
\begin{minipage}{0.33\linewidth}
\begin{center}
\includegraphics[width=0.80\columnwidth]{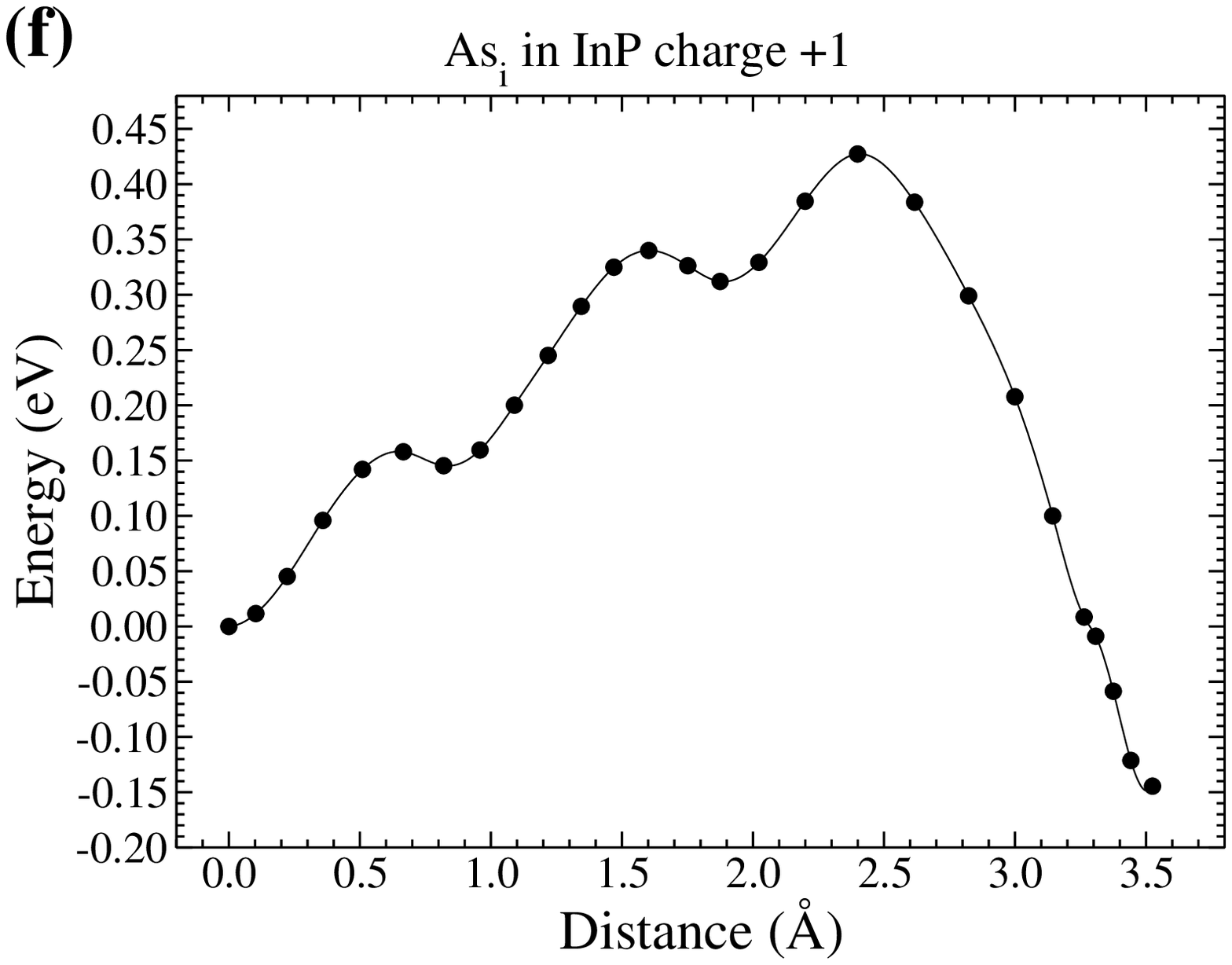}
\includegraphics[width=0.99\columnwidth]{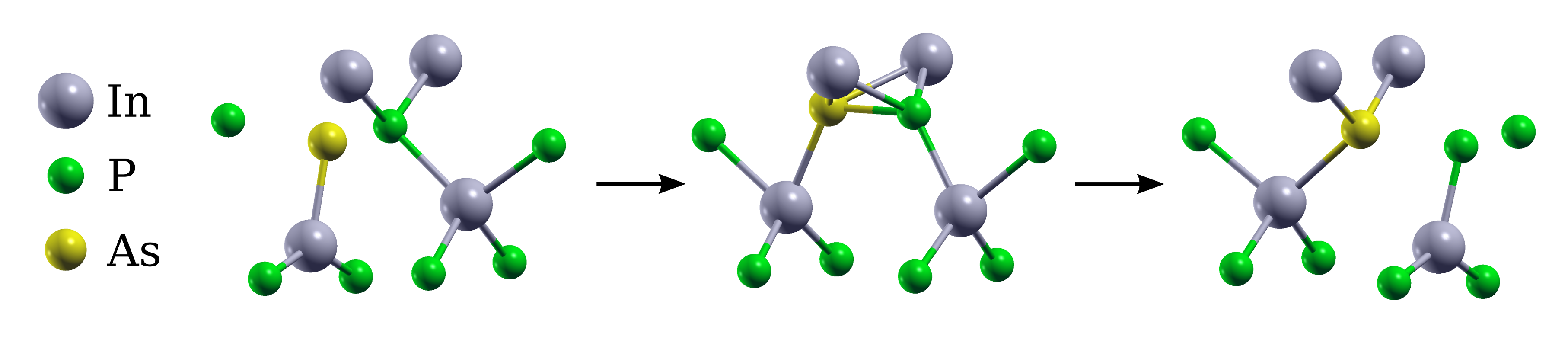}
\end{center}
\end{minipage}
\begin{minipage}{0.33\linewidth}
\begin{center}
\includegraphics[width=0.80\columnwidth]{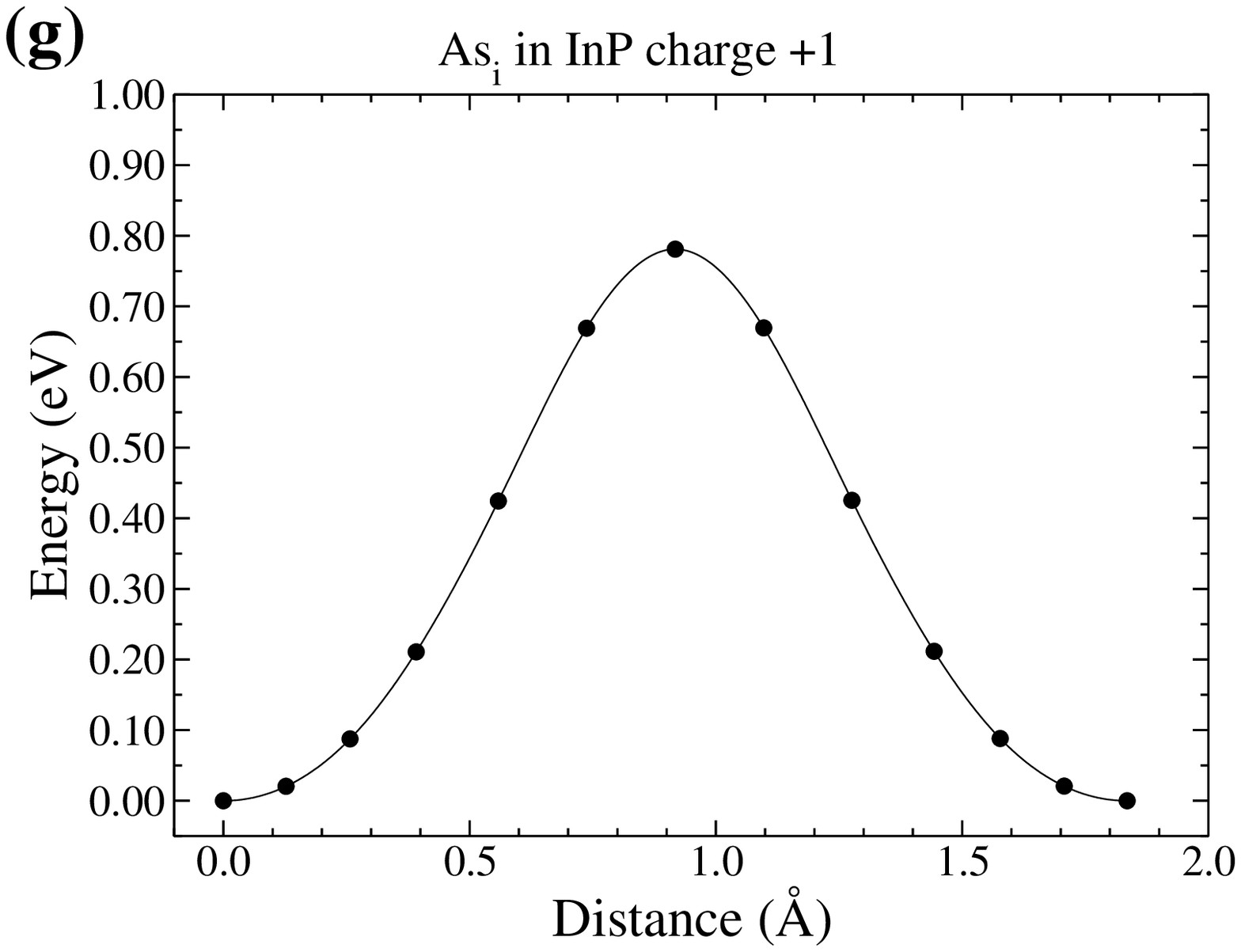}
\includegraphics[width=0.99\columnwidth]{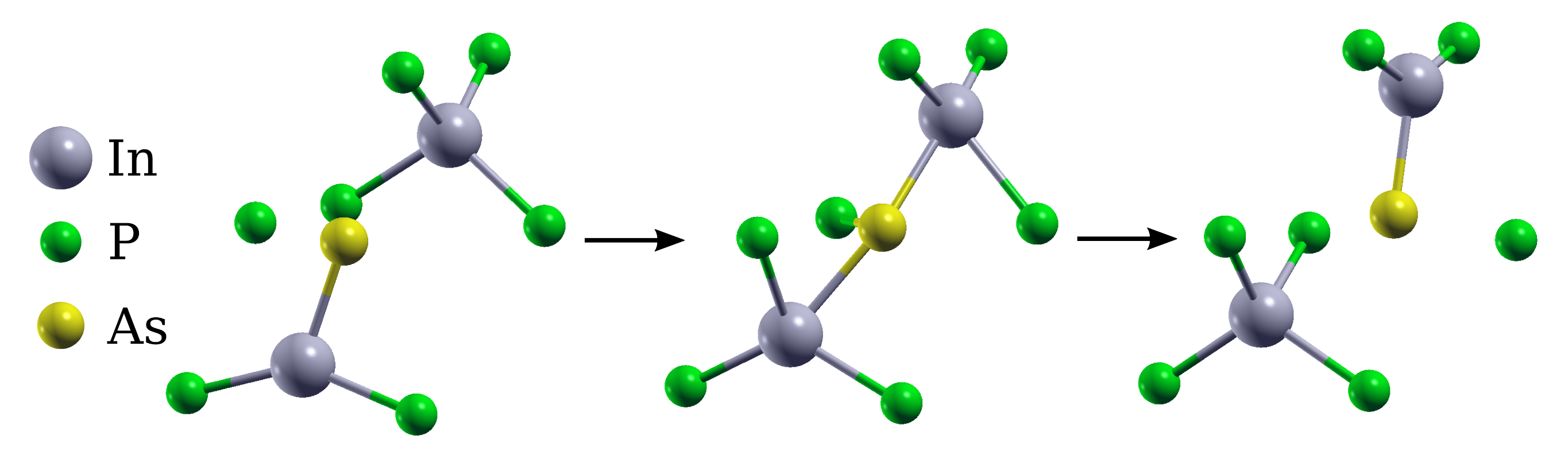}
\end{center}
\end{minipage}
\begin{minipage}{0.33\linewidth}
\begin{center}
\includegraphics[width=0.80\columnwidth]{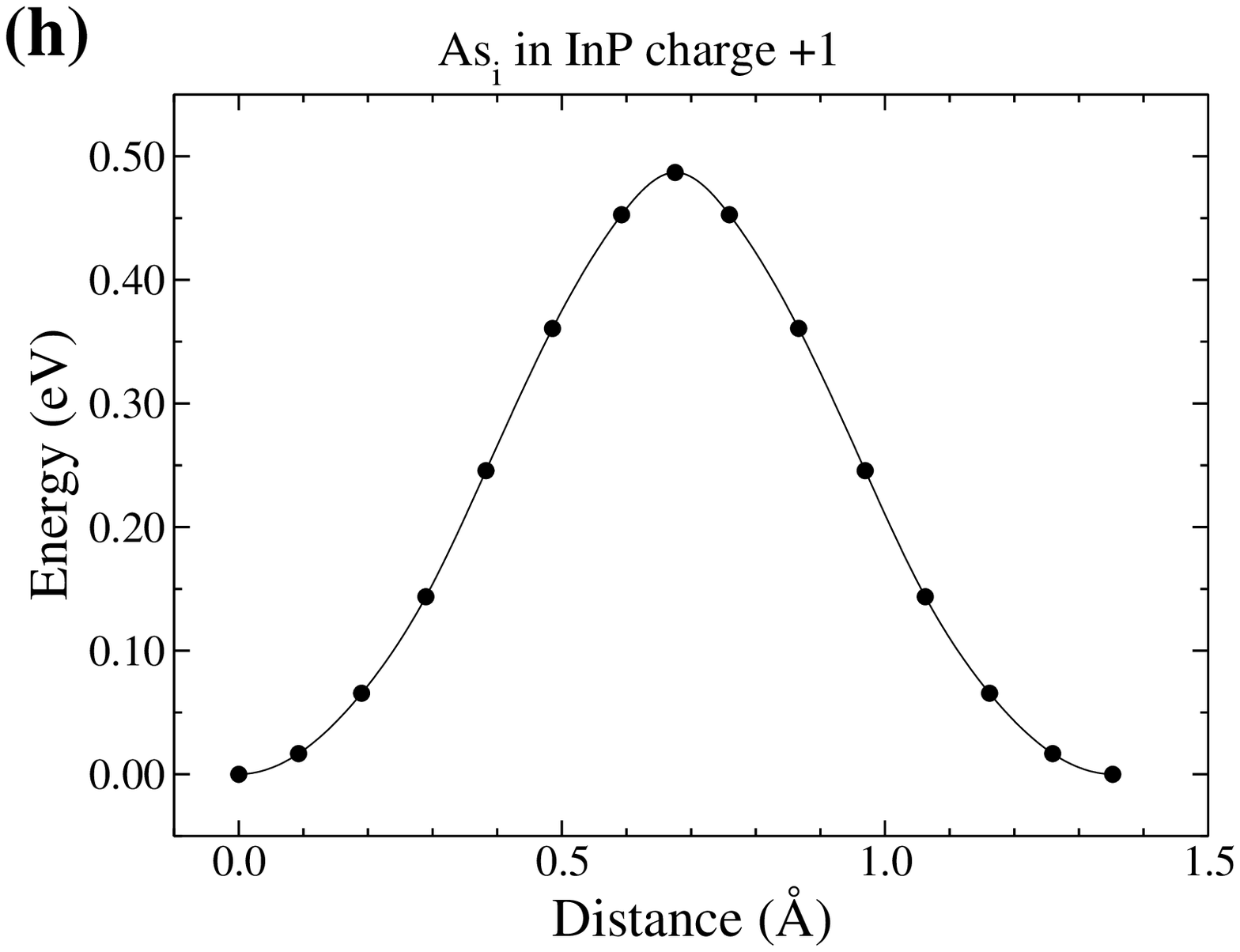}
\includegraphics[width=0.99\columnwidth]{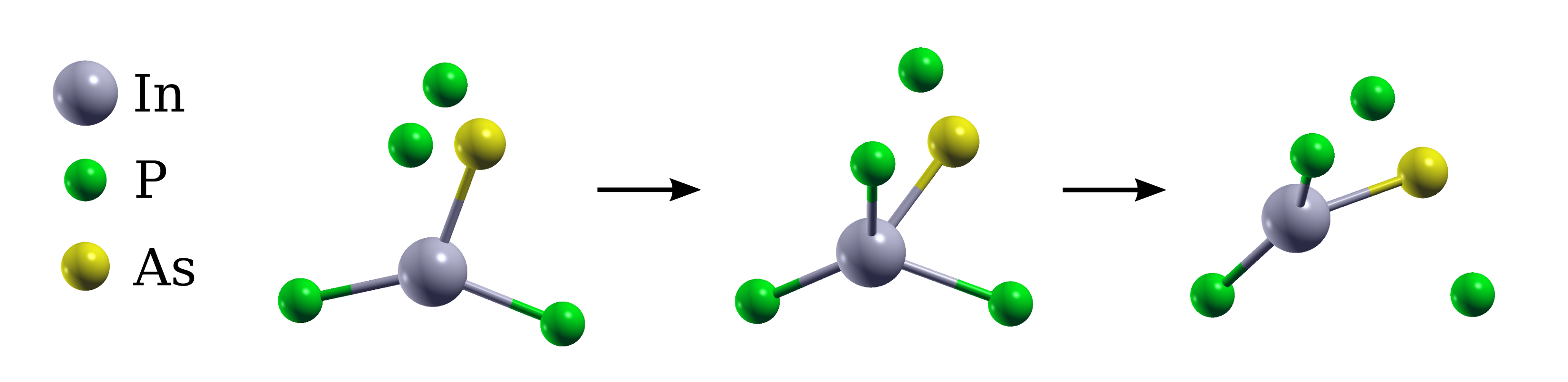}
\end{center}
\end{minipage}
\begin{minipage}{0.33\linewidth}
\begin{center}
\includegraphics[width=0.80\columnwidth]{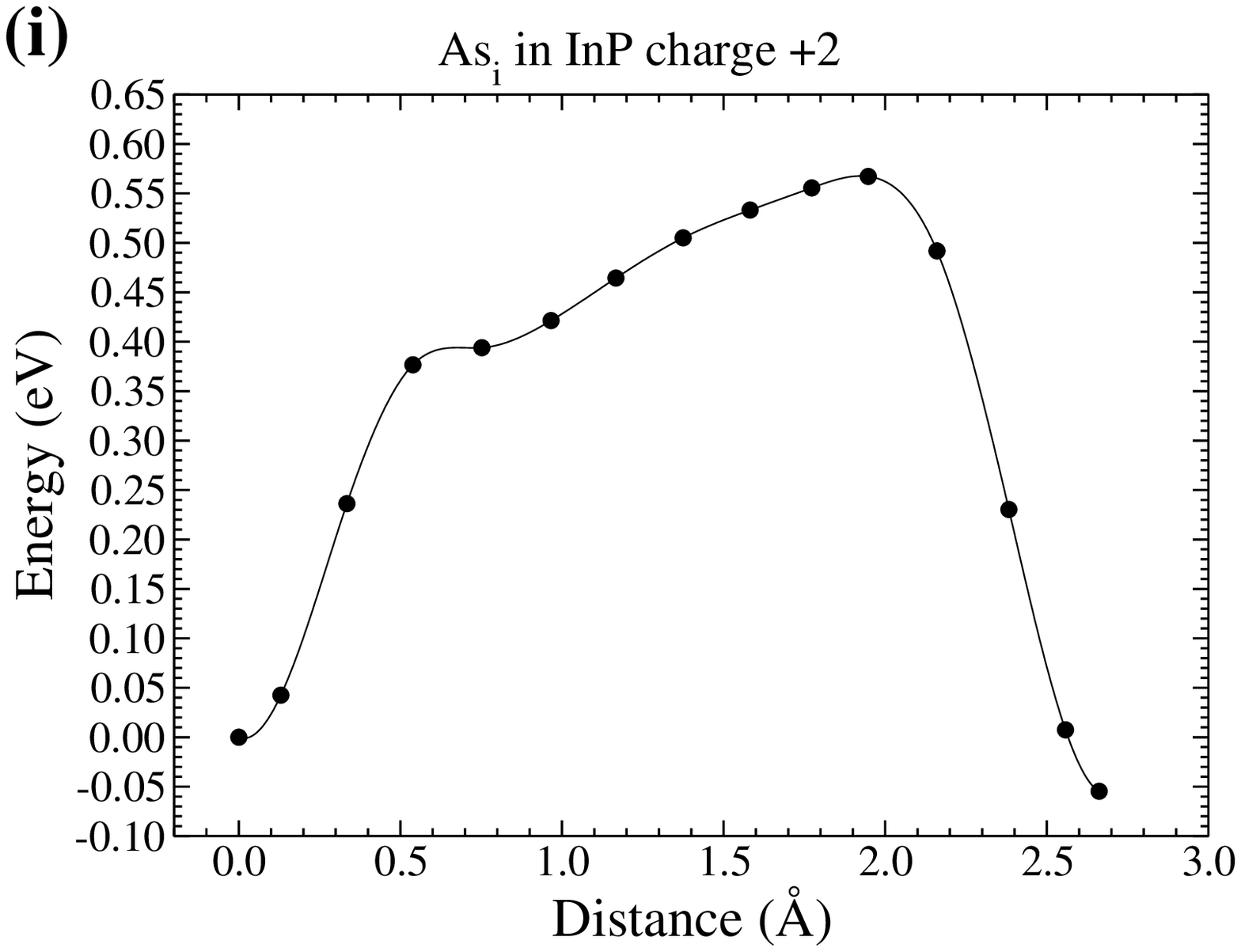}
\includegraphics[width=0.99\columnwidth]{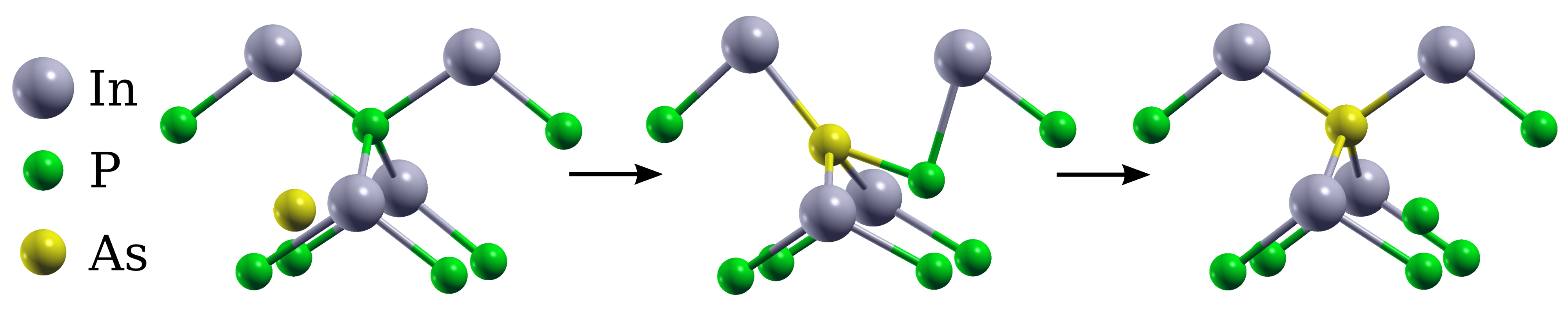}
\end{center}
\end{minipage}
\caption{\label{fig:figure6} Energy profiles and schemes of the transitions with atomic configurations of initial state, saddle point and final state for different jump types of As\textsubscript{i} in InP: (a) -- rotation of the (As-P)\textsubscript{i} split-interstitial in InP in a neutral charge state; (b) -- rotation of the (As-P)\textsubscript{i} split-interstitial in InP in a neutral charge state (type~2); (c) -- movement of an As atom from a neutral split-interstitial with the formation of a split-interstitial near the neighboring site in the [110] direction in the sublattice of P atoms in InP (As-P)\textsubscript{i}$\rightarrow$(As-P)\textsubscript{i}; (d) -- transition of an As atom from a neutral split-interstitial to a P sublattice site with the formation of a split-interstitial consisting of two P atoms (As-P)\textsubscript{i}$\rightarrow$ As\textsubscript{P}+(P-P)\textsubscript{i}; (e) -- transition of an interstitial As atom in InP in the charge state +1 with a change in the direction of the In-As bond through the metastable split-interstitial state (In-As)\textsubscript{i}$\rightarrow$(As-P)\textsubscript{i}$\rightarrow$(In-As)\textsubscript{i}, the atomic configurations of the initial state, the intermediate metastable split-interstitial state, and the final state are shown; (f) --  transition of an interstitial As atom in InP in the +1 charge state in the direction perpendicular to the plane in which the bonds of the In atom with the two nearest P atoms are located through the intermediate metastable split-interstitial state, the atomic configurations of the initial state (In-As)\textsubscript{i}$\rightarrow$(As-P)\textsubscript{i}$\rightarrow$As\textsubscript{P}+(In-P)\textsubscript{i}, the intermediate metastable split-interstitial state, and the final state are shown; (g) -- transition of an interstitial As atom in InP in the +1 charge state from an In-As split-interstitial site to a neighboring In atom with formation of an In-As split-interstitial with different bond direction (In-As)\textsubscript{i}$\rightarrow$(In-As)\textsubscript{i}; (h) -- rotation of the plane, in which the bonds of the In atom with the As atom and one of the two P atoms are located, by an angle of 60$^\circ$ for (In-As)\textsubscript{i} in InP in the +1 charge state; (i) -- transition of an interstitial As atom in InP to a substitution position in the P sublattice with the displacement of the P atom into the interstitial position for the +2 charge state As\textsubscript{i}$\rightarrow$As\textsubscript{P}+P\textsubscript{i}.}
\end{figure*}

We have considered different types of atoms transitions for diffusion via interstitial atoms. The energy barrier for rotation of the axis of a neutral split-interstitial As\textsubscript{i} in InP by an angle of 60$^\circ$ from the [110] direction to the [01$-$1] direction is 0.231~eV (figure~\ref{fig:figure6}a). The second type of rotation by an angle of 60$^\circ$ from the [110] direction to the [01$-$1] direction is shown in the figure~\ref{fig:figure6}b. The energy barrier for this transition is 0.249~eV. Movement of the As atom with formation of a split-interstitial near the neighbor site in the [110] direction in the P atoms sublattice and transition of the P atom from the split-interstitial site to the lattice site occurs with an energy barrier of 0.390 eV (figure~\ref{fig:figure6}c). In this case the C\textsubscript{s} symmetry, that the initial and final states have, is broken for the intermediate states. The As atom leaves the symmetry plane and shifts to one of the two sides.
The transition of an As atom from the split-interstitial to a P sublattice site with the formation of a split-interstitial consisting of two P atoms occurs with an energy barrier of 0.352~eV (figure~\ref{fig:figure6}d), while the energy of the final configuration As\textsubscript{P}P\textsubscript{i} is lower by 0.124~eV than the initial As\textsubscript{i} energy, so the reverse transition has an energy barrier of 0.476~eV. This transition also occurs with C\textsubscript{s} symmetry breaking. For each site in the P atoms sublattice, there are 12 directions of transition to the nearest sites of the P sublattice. The most energy favorable transition paths for the neutral As\textsubscript{i} in directions non-parallel to the axis of the split-interstitial are divided into several stages: a rotation or several rotations of the axis of the split-interstitial by an angle of 60$^\circ$ and transition along the axis of the split-interstitial.

For the atomic configuration of As\textsubscript{i} in charge state +1 shown in figure~\ref{fig:figure3}b, there are 12 symmetry equivalent configurations for one site: 6 possible arrangements of the plane in which the bonds of the In atom with the As atom and the two nearest P atoms are located, for each of which two directions of the In atom bond with the As atom are possible. The transition of the As atom with a change in the direction of the In-As bond (figure~\ref{fig:figure6}e) occurs with an energy barrier of 0.340~eV. In this case, the transition occurs in two stages with the movement of the As atom out of the plane in which the bonds of the In atom with the two nearest P atoms are located, and the formation of an As-P split-interstitial in the intermediate metastable state. The energy barrier for leaving the split-interstitial metastable state is 0.028~eV.

The transition of the As atom in the direction perpendicular to the plane of the bonds of the In atom with the As atom and the two nearest P atoms also occurs through the intermediate metastable split-interstitial state (figure~\ref{fig:figure6}f). In the final state of this transition, the As atom moves to the P site and an interstitial P is formed in the configuration of the In-P split-interstitial, which is close to the shown in the figure~\ref{fig:figure3}b configuration of the In-As split-interstitial. The energy of this state is lower by 0.145~eV than the energy of the initial configuration. The energy barriers are 0.340~eV for the transition from the initial state to the metastable state with the As-P split-interstitial configuration, and 0.115~eV for the transition from the metastable to the final state. The energy barriers for transitions in the reverse direction are 0.572~eV and 0.028~eV, respectively.

In the configuration of the In-As split-interstitial in InP in charge state +1 (figure~\ref{fig:figure3}b), there are 2 out of 12 nearest sites in the In sublattice for which the distance between the In and As atoms is the shortest. The transition in the direction of one of these sites with the formation in it of an In-As split-interstitial of the same configuration, but with a different direction of the In-As bond, occurs with an energy barrier of 0.781~eV (figure~\ref{fig:figure6}g). Two such transitions lead to the transition of the As atom to a neighbor In atom in the plane which contains the bonds of the In atom with the As atom and the two P atoms. If we assume that an As atom makes only this type of jumps, we get that it will move inside the tetrahedron formed by In atoms of the In sublattice. For the macroscopic movement of an As atom along interstitial positions, a combination of this type of transition with the transition of changing the direction of the In-As bond is necessary (figure~\ref{fig:figure6}e). Transitions perpendicular to the plane, in which the bonds of the In atom with the two nearest P atoms are located, through the intermediate metastable split-interstitial state (figure~\ref{fig:figure6}f) have a lower barrier energy and implement the indirect interstitial (interstitialcy) mechanism. Rotation of the plane, in which the bonds of the In atom with the As atom and one of the two P atoms are located, by an angle of 60$^\circ$ for As\textsubscript{i} in InP in the +1 charge state occurs with an energy barrier of 0.487~eV (figure~\ref{fig:figure6}h).

\begin{figure}[t]
\begin{center}
\includegraphics[width=0.75\columnwidth]{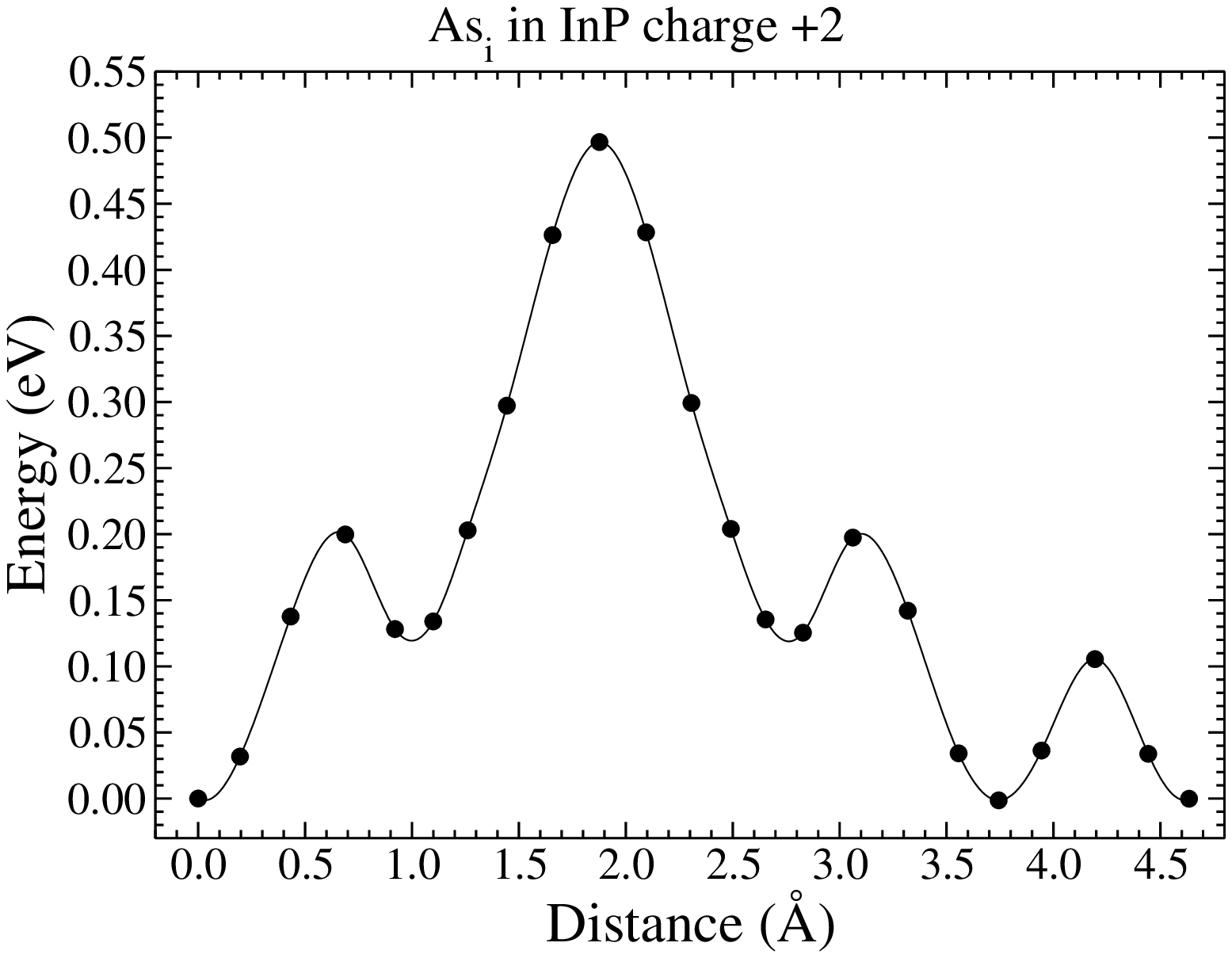}
\includegraphics[width=0.90\columnwidth]{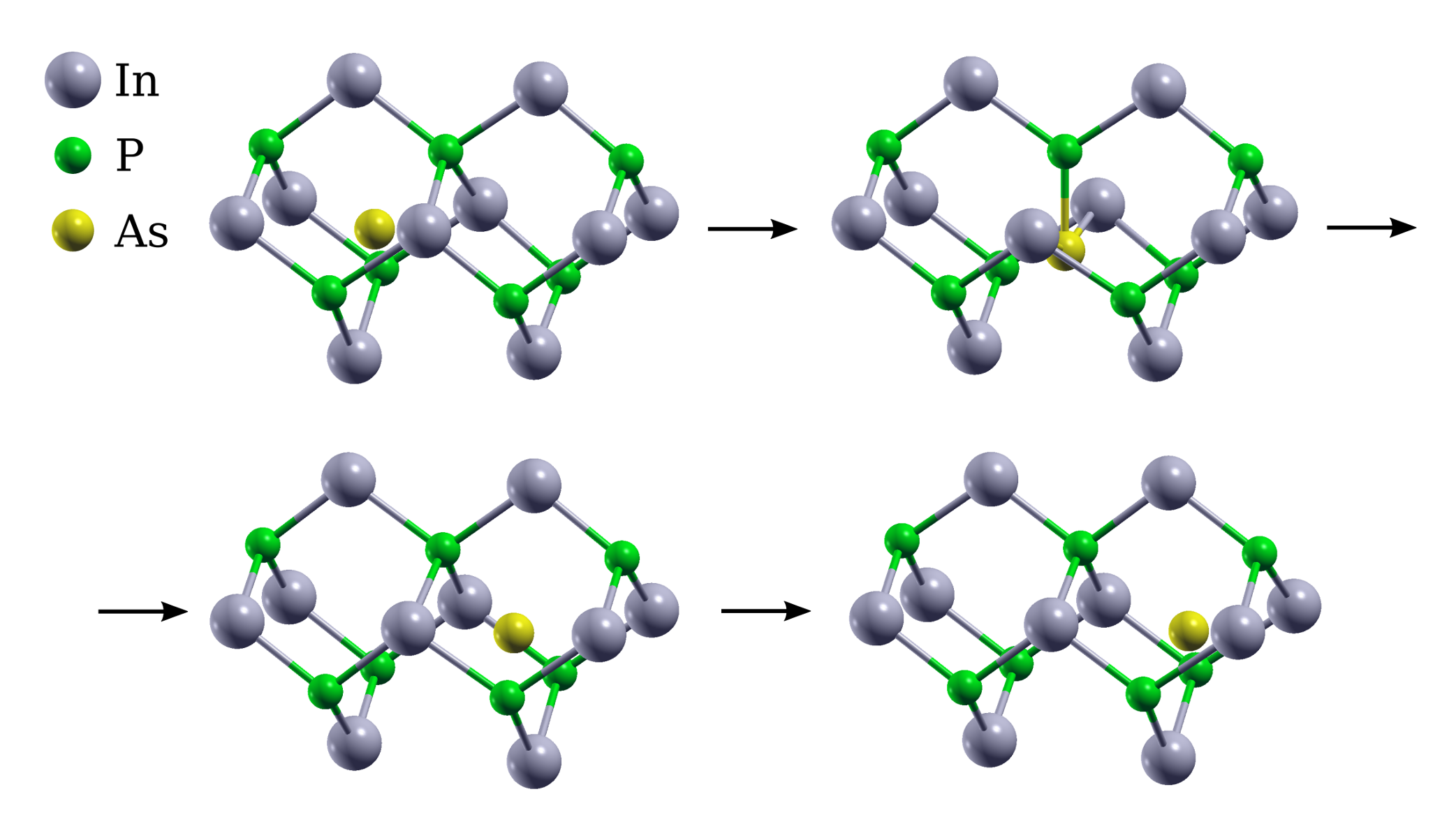}
\caption{\label{fig:figure7} Migration of an interstitial As atom in InP for the +2 charge state: energy profile and scheme of the transition. The atomic configurations of the initial state, the saddle point, the nearest minimum energy C\textsubscript{3v} configuration, and the final state in which the As atom is displaced by one translation vector of the primitive cell are shown.}
\end{center}
\end{figure}

In the +2 charge state of the interstitial As atom in InP, the most energy favorable configuration of atoms has C\textsubscript{3v} point symmetry with the As atom located inside the tetrahedron formed by P atoms and shifted to one of the faces of the tetrahedron (figure~\ref{fig:figure3}c). Accordingly, there are 4 equivalent positions of the interstitial As atom inside each tetrahedron. The transition between such equivalent configurations occurs with an energy barrier of 0.105~eV (figure~\ref{fig:figure7}). The energy profile shown in the figure~\ref{fig:figure7} corresponds to the movement of an interstitial As atom in InP in the +2 charge state by one translation vector of a primitive cell. This profile contains a transition of an As atom to the nearest C\textsubscript{3v} position in the nearest empty tetrahedron formed by P atoms with an energy barrier of 0.497 eV, and a transition between two equivalent C\textsubscript{3v} positions inside one tetrahedron with an energy barrier of 0.105~eV. For the first transition, two local energy minima are also observed. Figure~\ref{fig:figure6}i shows the transition of the interstitial As atom to a substituting position in the P sublattice with the displacement of the P atom into the interstitial position. This transition has an energy barrier of 0.567~eV, and the reverse transition has an energy barrier of 0.622~eV.

Tables~\ref{tab:table6} and~\ref{tab:table7} show the calculation results of migration energy barriers and attempt frequencies for diffusion of interstitial As and P atoms in InP.
\begin{table}
\caption{\label{tab:table6}  Migration barriers and attempt frequencies for the diffusion of interstitial As atoms in InP.}
\begin{tabular}{c p{3.3cm} c c}
\hline
As\textsubscript{i} charge & Jump type & $E_m$, eV & $\nu$, THz\\
\hline
+2 & between tetrahedra & 0.497 & 2.43\\
+2 & between equivalent C\textsubscript{3v} positions & 0.105 & 2.43\\
+2 & As\textsubscript{i}$\rightarrow$As\textsubscript{P}+P\textsubscript{i} & 0.567 & 3.36\\
+2 & As\textsubscript{P}+P\textsubscript{i}$\rightarrow$As\textsubscript{i} & 0.622 & 4.13\\
+1 & (In-As)\textsubscript{i}$\rightarrow$(In-As)\textsubscript{i} & 0.781 & 1.78\\
+1 & rotation of (In-As)\textsubscript{i} & 0.487 & 2.12\\
+1 & (In-As)\textsubscript{i}$\rightarrow$(As-P)\textsubscript{i} & 0.340 & 1.70\\
+1 & As\textsubscript{P}+(In-P)\textsubscript{i}$\rightarrow$(As-P)\textsubscript{i} & 0.572 & 1.76\\
+1 & (As-P)\textsubscript{i}$\rightarrow$As\textsubscript{P}+(In-P)\textsubscript{i} & 0.115 & 1.69\\
+1 & (As-P)\textsubscript{i}$\rightarrow$(In-As)\textsubscript{i} & 0.028 & 1.69\\
0 & 60$^\circ$ rotation of (As-P)\textsubscript{i} & 0.231 & 3.03\\
0 & 60$^\circ$ rotation of (As-P)\textsubscript{i}, type 2 & 0.249 & 1.46\\
0 & (As-P)\textsubscript{i}$\rightarrow$(As-P)\textsubscript{i} & 0.390 & 1.69\\
0 & (As-P)\textsubscript{i}$\rightarrow$As\textsubscript{P}+(P-P)\textsubscript{i} & 0.352 & 1.34\\
0 & As\textsubscript{P}+(P-P)\textsubscript{i}$\rightarrow$(As-P)\textsubscript{i} & 0.476 & 1.17\\
\hline
\end{tabular}
\end{table}
\begin{table}
\caption{\label{tab:table7} Migration barriers and attempt frequencies for the diffusion of interstitial P atoms in InP.}
\begin{tabular}{c p{3.3cm} c c}
\hline
P\textsubscript{i} charge & Jump type & $E_m$, eV & $\nu$, THz\\
\hline
+2 & between tetrahedra & 0.447 & 3.89\\
+2 & between equivalent C\textsubscript{3v} positions & 0.111 & 3.89\\
+2 & P\textsubscript{i}+P\textsubscript{P}$\rightarrow$P\textsubscript{P}+P\textsubscript{i} & 0.483 & 4.91\\
+1 & (In-P)\textsubscript{i}$\rightarrow$(In-P)\textsubscript{i} & 0.775 & 2.04\\
+1 & rotation of (In-P)\textsubscript{i} & 0.622 & 2.41\\
+1 & (In-P)\textsubscript{i}$\rightarrow$(P-P)\textsubscript{i} & 0.425 & 1.65\\
+1 & (P-P)\textsubscript{i}$\rightarrow$(In-P)\textsubscript{i} & 0.035 & 1.65\\
0 & (P-P)\textsubscript{i}$\rightarrow$(P-P)\textsubscript{i} & 0.396 & 1.61\\
0 & 60$^\circ$ rotation of (P-P)\textsubscript{i} & 0.133 & 2.64\\
\hline
\end{tabular}
\end{table}

\subsection{Migration barriers for diffusion of interstitial P and As atoms in InAs}
Tables~\ref{tab:table8} and~\ref{tab:table9} show the calculation results of the migration energy barriers and attempt frequencies for diffusion of interstitial As and P atoms in InAs. Energy profiles for the processes of migration of interstitial As and P atoms in InAs, as well as interstitial P atoms in InP are given in the supplementary material.
\begin{table}
\caption{\label{tab:table8}  Migration barriers and attempt frequencies for the diffusion of interstitial As atoms in InAs.}
\begin{tabular}{c p{3.3cm} c c}
\hline
As\textsubscript{i} charge & Jump type & $E_m$, eV & $\nu$, THz\\
\hline
+2 & between tetrahedra & 0.358 & 2.13\\
+2 & between equivalent C\textsubscript{3v} positions & 0.048 & 2.13\\
+2 & As\textsubscript{i}+As\textsubscript{As}$\rightarrow$As\textsubscript{As}+As\textsubscript{i} & 0.395 & 2.72\\
+1 & (In-As)\textsubscript{i}$\rightarrow$(In-As)\textsubscript{i} & 0.567 & 1.64\\
+1 & rotation of (In-As)\textsubscript{i} & 0.434 & 1.89\\
+1 & (In-As)\textsubscript{i}$\rightarrow$(As-As)\textsubscript{i} & 0.307 & 1.40\\
+1 & (As-As)\textsubscript{i}$\rightarrow$(In-As)\textsubscript{i} & 0.098 & 1.52\\
0 & (As-As)\textsubscript{i}$\rightarrow$(As-As)\textsubscript{i} & 0.327 & 1.32\\
0 & 60$^\circ$ rotation of (As-As)\textsubscript{i} & 0.284 & 1.90\\
\hline
\end{tabular}
\end{table}
\begin{table}
\caption{\label{tab:table9}  Migration barriers and attempt frequencies for the diffusion of interstitial P atoms in InAs.}
\begin{tabular}{c p{3.3cm} c c}
\hline
P\textsubscript{i} charge & Jump type & $E_m$, eV & $\nu$, THz\\
\hline
+2 & between tetrahedra & 0.360 & 3.27\\
+2 & between equivalent C\textsubscript{3v} positions & 0.095 & 3.27\\
+2 & P\textsubscript{i}$\rightarrow$P\textsubscript{As}+As\textsubscript{i} & 0.454 & 3.27\\
+2 & P\textsubscript{As}+As\textsubscript{i}$\rightarrow$P\textsubscript{i} & 0.490 & 2.68\\
+1 & (In-P)\textsubscript{i}$\rightarrow$(In-P)\textsubscript{i} & 0.702 & 1.97\\
+1 & rotation of (In-P)\textsubscript{i} & 0.566 & 2.23\\
+1 & (In-P)\textsubscript{i}$\rightarrow$(P-As)\textsubscript{i} & 0.396 & 1.56\\
+1 & P\textsubscript{As}+(In-As)\textsubscript{i}$\rightarrow$(P-As)\textsubscript{i} & 0.256 & 1.49\\
+1 & (P-As)\textsubscript{i}$\rightarrow$P\textsubscript{As}+(In-As)\textsubscript{i} & 0.024 & 1.41\\
+1 & (P-As)\textsubscript{i}$\rightarrow$(In-P)\textsubscript{i} & 0.117 & 1.41\\
0 & 60$^\circ$ rotation of (P-As)\textsubscript{i} & 0.189 & 1.55\\
0 & 60$^\circ$ rotation of (P-As)\textsubscript{i}, type 2 & 0.166 & 2.50\\
0 & (P-As)\textsubscript{i}$\rightarrow$(P-As)\textsubscript{i} & 0.311 & 1.38\\
0 & (P-As)\textsubscript{i}$\rightarrow$P\textsubscript{As}+(As-As)\textsubscript{i} & 0.415 & 1.13\\
0 & P\textsubscript{As}+(As-As)\textsubscript{i}$\rightarrow$(P-As)\textsubscript{i} & 0.378 & 1.34\\
\hline
\end{tabular}
\end{table}
The migration barrier for the movement of a neutral split-interstitial (As-As)\textsubscript{i} in InAs to a neighbor position in the [110] direction is $E_m$=0.327~eV, which is comparable to the result of Reveil et al.~\cite{Reveil_2017} for this transition $E_m$=0.41~eV. The local density approximation gives slightly higher barrier energies (0.41~eV for a neutral As\textsubscript{i} in InAs and 2.0~eV for a neutral V\textsubscript{As} in InAs~\cite{Reveil_2017}) than the generalized gradient approximation (0.327~eV for a neutral As\textsubscript{i} in InAs and 1.787~eV for a neutral V\textsubscript{As} in InAs). This type of transition, according to our calculations, occurs through the metastable split-interstitial state with breaking of the C\textsubscript{s} symmetry for both P\textsubscript{i} in InP and As\textsubscript{i} in InAs, as well as for As\textsubscript{i} in InP and P\textsubscript{i} in InAs, which is consistent with the calculation of Reveil et al.~\cite{Reveil_2017} for As\textsubscript{i} in InAs. A similar type of transition was also observed for neutral As\textsubscript{i} in GaAs~\cite{Wright_2016}.

\section{Discussion}
\begin{figure}
\includegraphics[width=0.99\columnwidth]{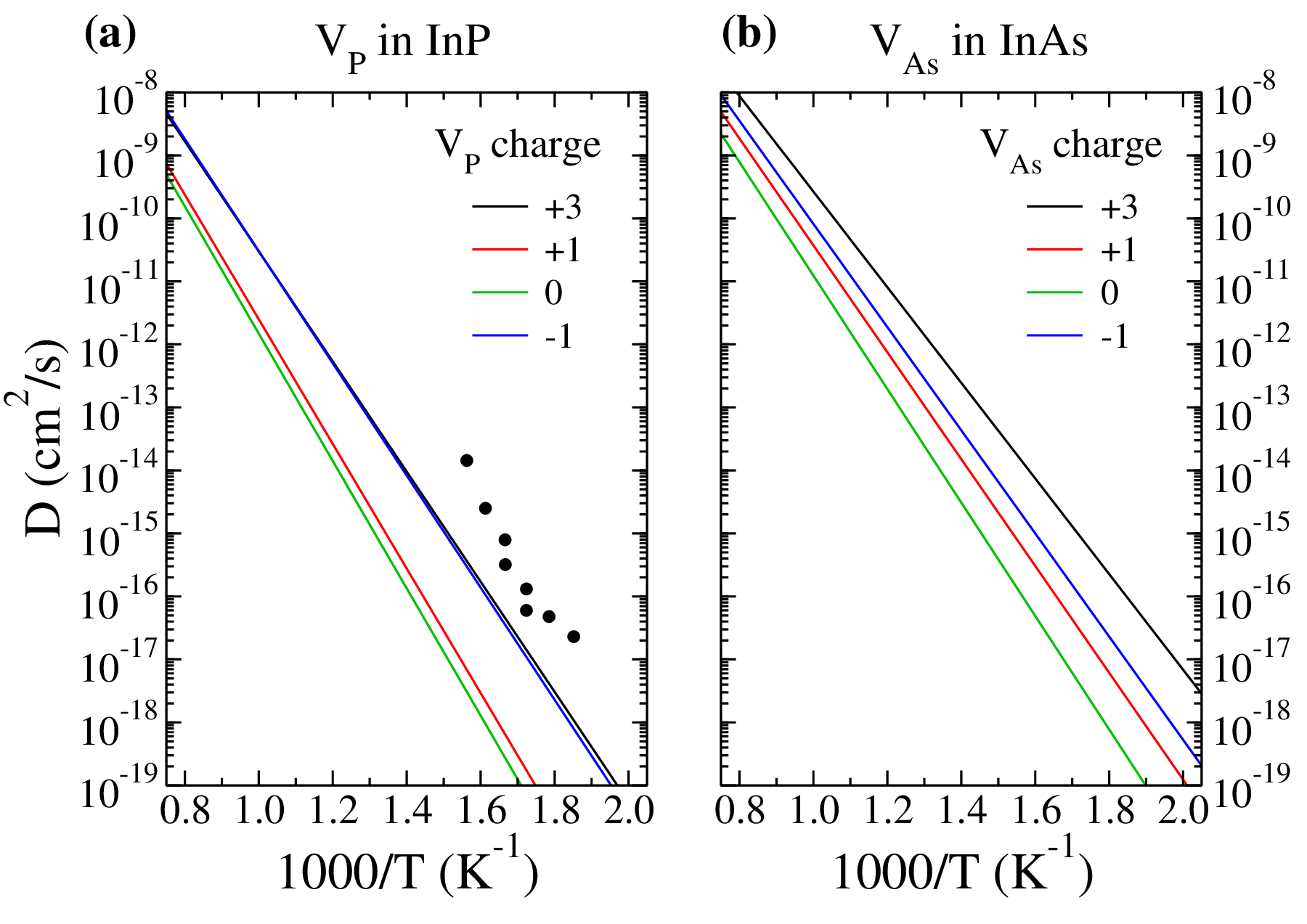}
\caption{\label{fig:figure8} Calculated temperature dependences of the diffusion coefficient of vacancies of group-V elements in InP and InAs for various charge states of the vacancy (solid lines). The dots show the diffusion coefficients of P vacancies in InP calculated using the equation~\ref{eq:eq4} with the jump rate from the experimental results of Slotte et al.~\cite{Slotte_2003} for V\textsubscript{P} in p-type InP.}
\end{figure}

Figure~\ref{fig:figure8} shows the temperature dependences of the diffusion coefficient of P vacancies in InP and As vacancies in InAs calculated using the equation~\ref{eq:eq4} with energy barriers and attempt frequencies from the tables~\ref{tab:table4} and~\ref{tab:table5}. The dots in the figure~\ref{fig:figure8} show the diffusion coefficients of P vacancies in InP calculated using the equation~\ref{eq:eq4} with the jump rate from the experimental work of Slotte et al.~\cite{Slotte_2003}, in which the authors used positron annihilation spectroscopy to study the diffusion of P vacancies in p-type InP. For p-InP, considering that in accordance with the figure~\ref{fig:figure1}a V\textsubscript{P} are in the charge state +3, our calculation gives $E_m$=1.734~eV and $D_0$=1.64$\cdot10^{-2}$ cm\textsuperscript{2}/s. Our results for the energy barrier for vacancy diffusion are in good agreement with the results of Slotte et al.~\cite{Slotte_2003} for the activation energy of diffusion $E_m$=1.8$\pm$0.2~eV. In absolute values, the experimental diffusion coefficient is 4--40 times higher than our calculations, which can be due to the presence of unaccounted migration entropy $\Delta S_m\approx2.6k_B$.

\begin{figure}[h]
\includegraphics[width=0.99\columnwidth]{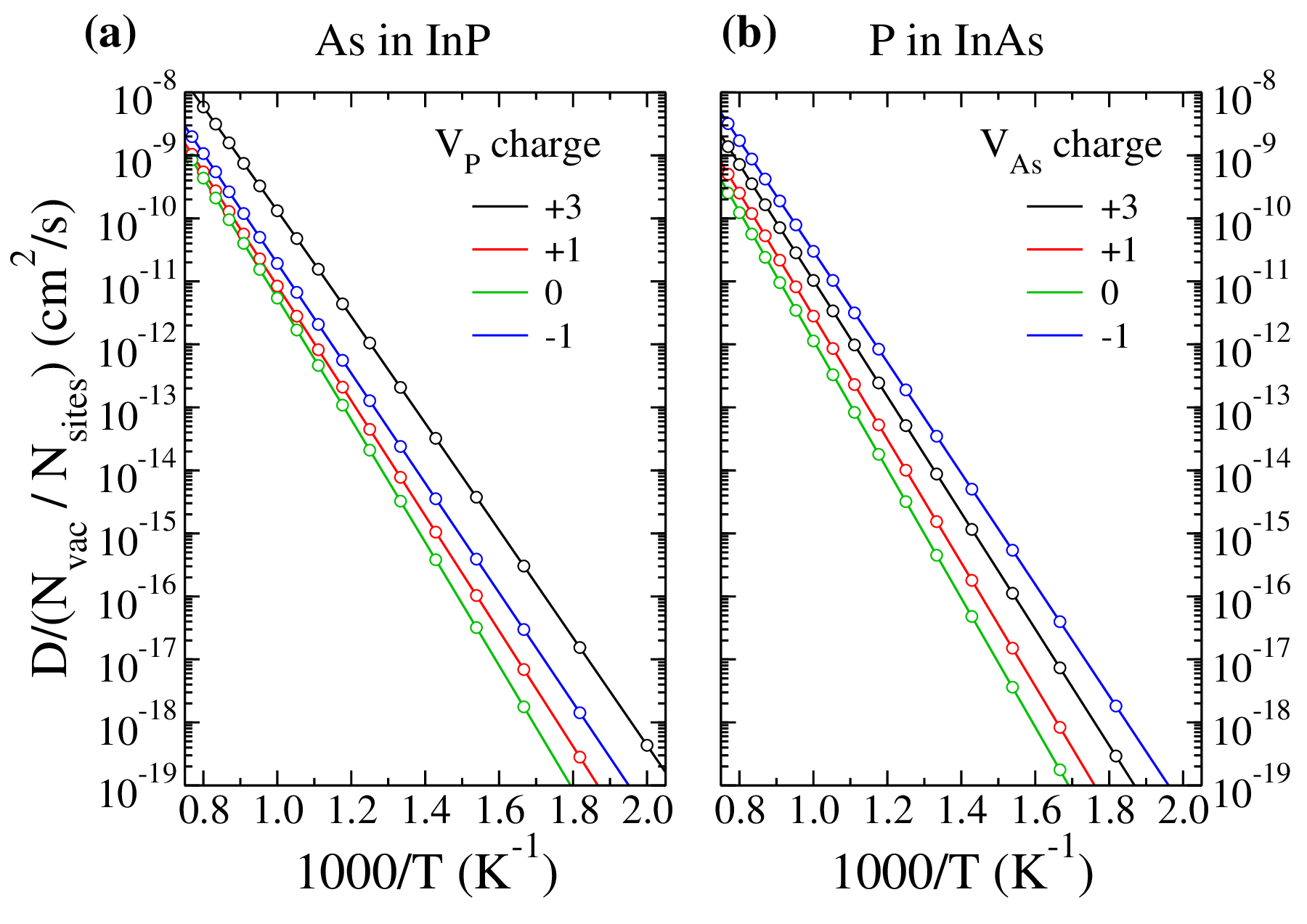}
\caption{\label{fig:figure9} Calculated temperature dependences of the ratio of the diffusion coefficient to the relative concentration of vacancies for diffusion of As atoms in InP and P atoms in InAs by vacancy mechanism for various charge states of the vacancy. Solid lines show calculation with the correlation coefficient from the Manning model; circles show calculation using KineCluE code with energy barriers, attempt frequencies and binding energies from the tables~\ref{tab:table3}--\ref{tab:table5}.}
\end{figure}

Figure~\ref{fig:figure9} shows the calculated temperature dependences of the ratio of the diffusion coefficient to the relative concentration of vacancies N\textsubscript{vac}/N\textsubscript{sites} for diffusion of As atoms in InP and P atoms in InAs by vacancy mechanism. The calculation was carried out according to the equation~\ref{eq:eq7} with energy barriers and attempt frequencies from the tables~\ref{tab:table4} and~\ref{tab:table5}, and with binding energies of the As\textsubscript{P}V\textsubscript{P} complexes in InP and P\textsubscript{As}V\textsubscript{As} complexes in InAs from the table~\ref{tab:table3}. The correlation coefficient was calculated using the Manning model (equations~\ref{eq:eq5},~\ref{eq:eq6}). The circles in the figure~\ref{fig:figure9} show the calculation using the KineCluE code~\cite{Schuler_2020} with the same energy barriers, attempt frequencies and binding energies of the As\textsubscript{P}V\textsubscript{P} and P\textsubscript{As}V\textsubscript{As} complexes. It can be seen from the figure that both methods give very similar results for the diffusion coefficient. Table~\ref{tab:table10} shows the activation energies and pre-exponential factors obtained by fitting the calculation results shown in the figure~\ref{fig:figure9} by thermally activated dependence: $D/(N_{vac}/N_{sites})=D_0\exp(-E_m/kT)$. We note that the $D_0$ factors in the table~\ref{tab:table10} do not include the factors with the formation entropy of the vacancies. 

\begin{table}
\caption{\label{tab:table10}  Migration barriers and pre-exponential factors for the ratio of the diffusion coefficient for the vacancy mechanism to the relative concentration of vacancies, taking into account the temperature dependence of the correlation coefficient and the binding energy of vacancy and impurity atom, obtained by fitting the calculation results shown in the figure~\ref{fig:figure9} by thermally activated dependence: $D/(N_{vac}/N_{sites})=D_0\exp(-E_m/kT)$.}
\begin{tabular}{c c c c c}
\hline
 & \multicolumn{2}{c}{As in InP} & \multicolumn{2}{c}{P in InAs}\\
\hline
Charge  & $E_m$, eV & $D_0$, cm\textsuperscript{2}/s & $E_m$, eV & $D_0$, cm\textsuperscript{2}/s \\
\hline
+3 & 1.595 & 1.59$\cdot10^{-2}$ & 1.820 & 1.57$\cdot10^{-2}$ \\
+1 & 1.793 & 9.33$\cdot10^{-3}$ & 1.932 & 1.55$\cdot10^{-2}$ \\
0 & 1.862 & 1.42$\cdot10^{-2}$ & 2.023 & 1.82$\cdot10^{-2}$ \\
-1 & 1.730 & 1.01$\cdot10^{-2}$ & 1.739 & 1.76$\cdot10^{-2}$ \\
\hline
\end{tabular}
\end{table}

\begin{figure}[t]
\includegraphics[width=0.99\columnwidth]{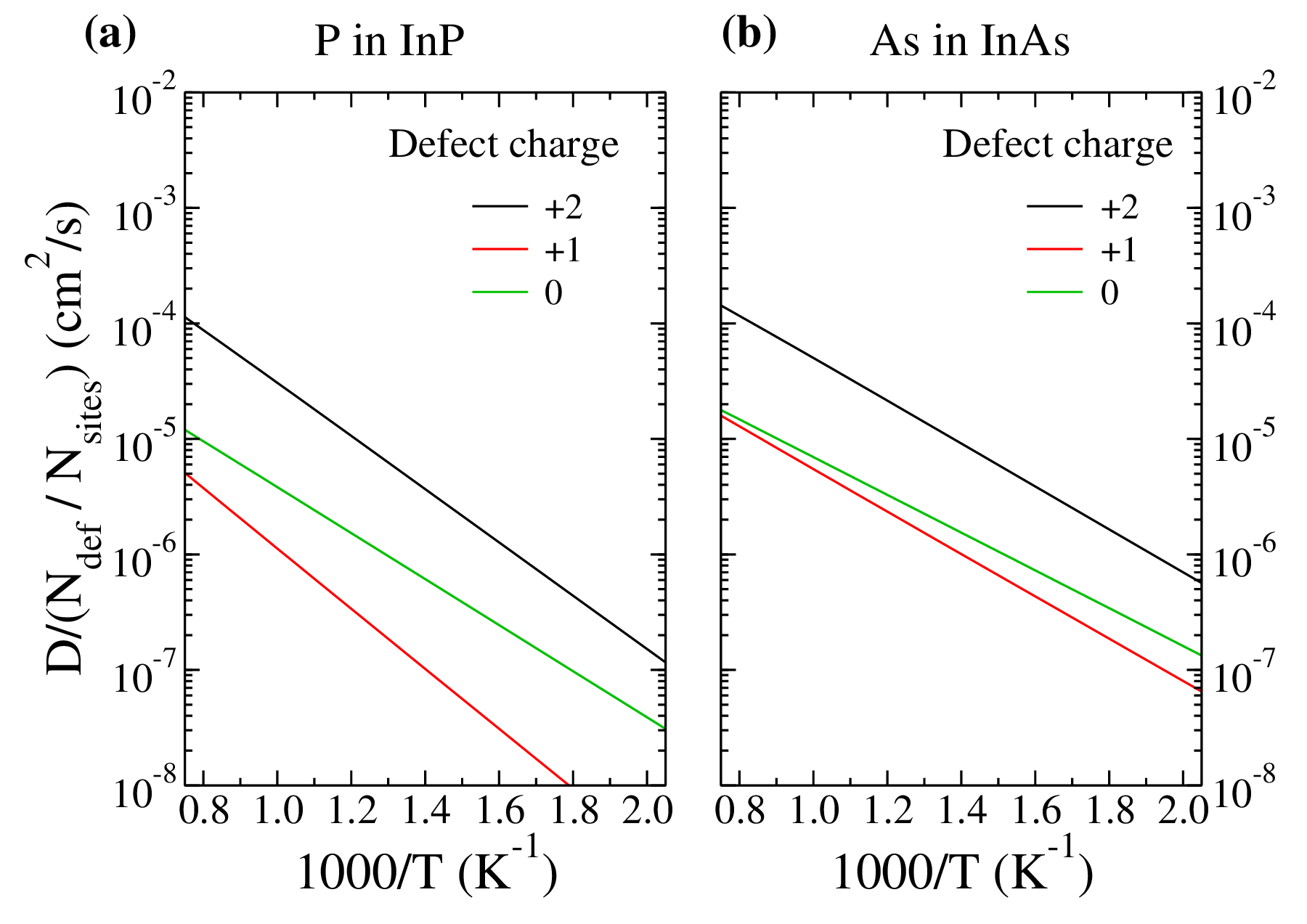}
\includegraphics[width=0.99\columnwidth]{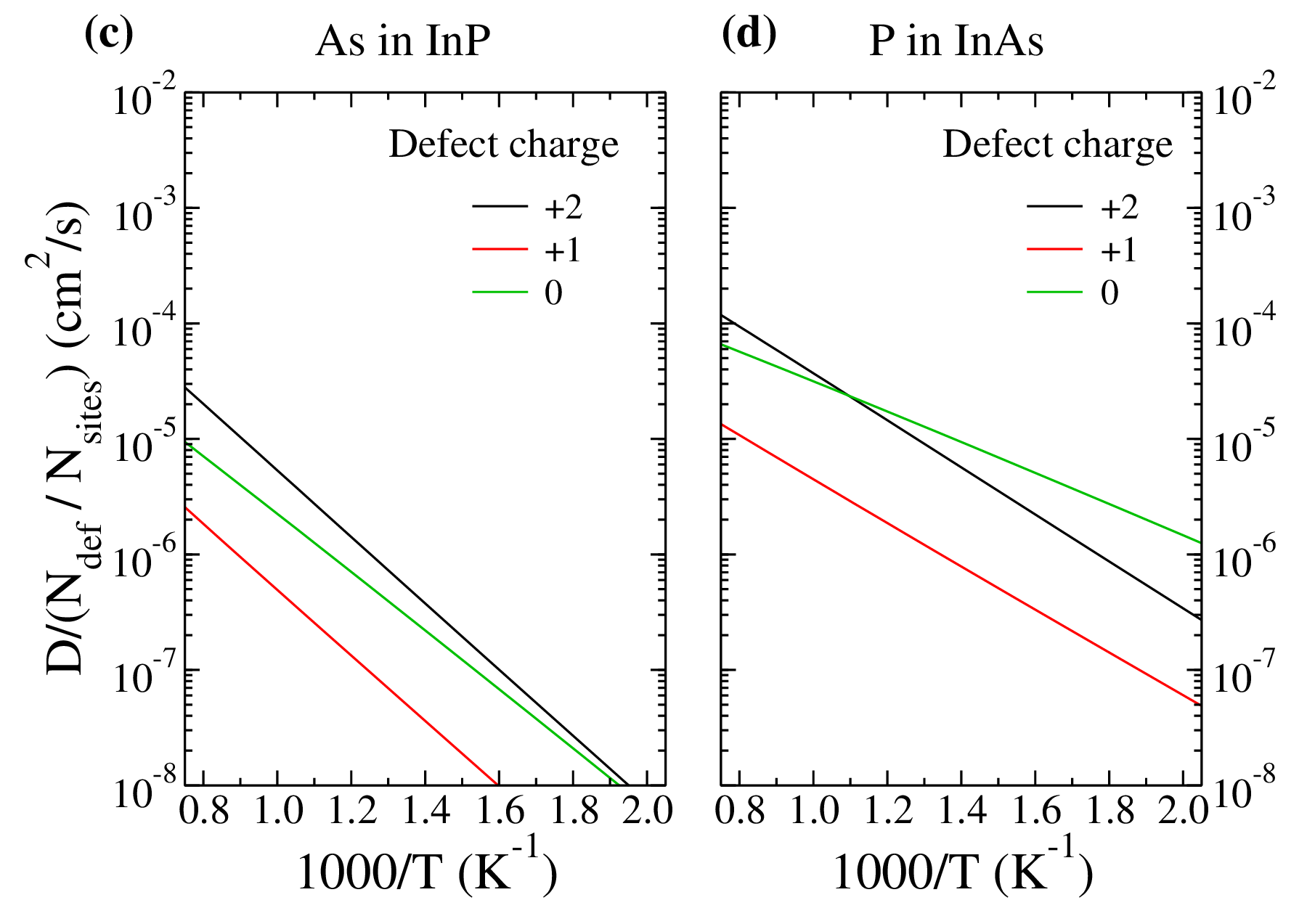}
\caption{\label{fig:figure10} Calculated temperature dependences of the ratio of the diffusion coefficient via interstitial atoms to the relative concentration of interstitial atoms for P atoms in InP (a), As atoms in InAs (b), As atoms in InP (c) and P atoms in InAs (d) for various charge states of the defect. Calculation was carried out using the KineCluE code with energy barriers and attempt frequencies from the tables~\ref{tab:table6}--\ref{tab:table9}.}
\end{figure}

Figure~\ref{fig:figure10} shows the calculated temperature dependences of the ratio of the diffusion coefficient to the relative concentration of interstitial atoms N\textsubscript{def}/N\textsubscript{sites} for diffusion via interstitial atoms for As and P atoms in InP and InAs. These dependencies were calculated using the KineCluE code~\cite{Schuler_2020} with energy barriers and attempt frequencies from the tables~\ref{tab:table6}--\ref{tab:table9}. In the case of asymmetric barriers, the corresponding binding energies were set from the energy profiles of migration for different configurations of the impurity atom and the defect.

Diffusion of the neutral split-interstitials P\textsubscript{i} in InP and As\textsubscript{i} in InAs is determined by the activation energies of the transitions (As-As)\textsubscript{i}$\rightarrow$(As-As)\textsubscript{i} and (P-P)\textsubscript{i}$\rightarrow$(P-P)\textsubscript{i}. Rotations have lower activation energies and they alone do not lead to macroscopic atomic motion. The diffusion coefficient of a split-interstitial as a defect without regard to which atoms form it can be calculated as 
\begin{equation}
\label{eq:eq8}
 D_I=\frac{1}{6}r^2 f Z \nu\exp\left(-\frac{E_m}{k_B T}\right),
\end{equation}
here $f$=1, since the jumps in different directions are independent, $Z$=2, because of at any given time only two directions are available for a jump, $r=\frac{a}{\sqrt{2}}$. The diffusion coefficient of atoms for this mechanism will be less than the diffusion coefficient of split-interstitials, since with each movement of a split interstitial one of the atoms moves to a lattice site. The diffusion coefficient of atoms in this case
\begin{equation}
\label{eq:eq9}
 D_{at}=\frac{1}{6}r^2 f Z \nu \frac{N_I}{N_{sites}}\exp\left(-\frac{E_m}{k_B T}\right),
\end{equation}
where $\frac{N_I}{N_{sites}}$ is the ratio of the number of split-interstitials to the number of sites in the group-V elements sublattice. The correlation coefficient $f$ in this case is less than unity. If the probability of the rotation is much greater than the jump probability, the correlation coefficient is $f=0.4$, as follows from the modeling with the KineCluE code. For slower rotations the correlation coefficient is lower and can be approximately described by the expression
\begin{equation}
\label{eq:eq10}
 f\approx\frac{0.4}{1+A\exp\left(-\frac{\Delta E}{k_B T}\right)},
\end{equation}
where $\Delta E$ is the difference between the barrier energies for the jump of a split-interstitial to a neighboring site and for the rotation of the split-interstitial.

For the charge states $-$2 and $-$1 of P\textsubscript{i} and As\textsubscript{i} in InP and InAs, the probability of a jump of an interstitial atom to a lattice site with moving of another atom from the lattice site to an interstitial position is comparable to or less than the probability of a jump from an interstitial to an interstitial position. In this case, the probability of movement only through interstitial positions decreases as the exponential function $\left(\frac{w_I}{w_I+w_S}\right)^n$ of the number of jumps $n$, where $w_I$ is the rate of jumps between interstitial positions and $w_S$ is the rate of jumps from the interstitial to the substitutional position. At a low concentration of interstitial atoms, the time during which an impurity atom is in a substitution position significantly exceeds the time during which it is in an interstitial position, and the rate of jumps from a substitution to an interstitial position via indirect interstitial mechanism is proportional to the concentration of interstitial atoms. Thus, for all charge states of P\textsubscript{i} and As\textsubscript{i} in InP and InAs, the diffusion process occurs via indirect interstitial mechanism and the diffusion coefficient is proportional to the concentration of interstitial atoms.

\begin{figure}
\begin{center}
\includegraphics[width=0.8\columnwidth]{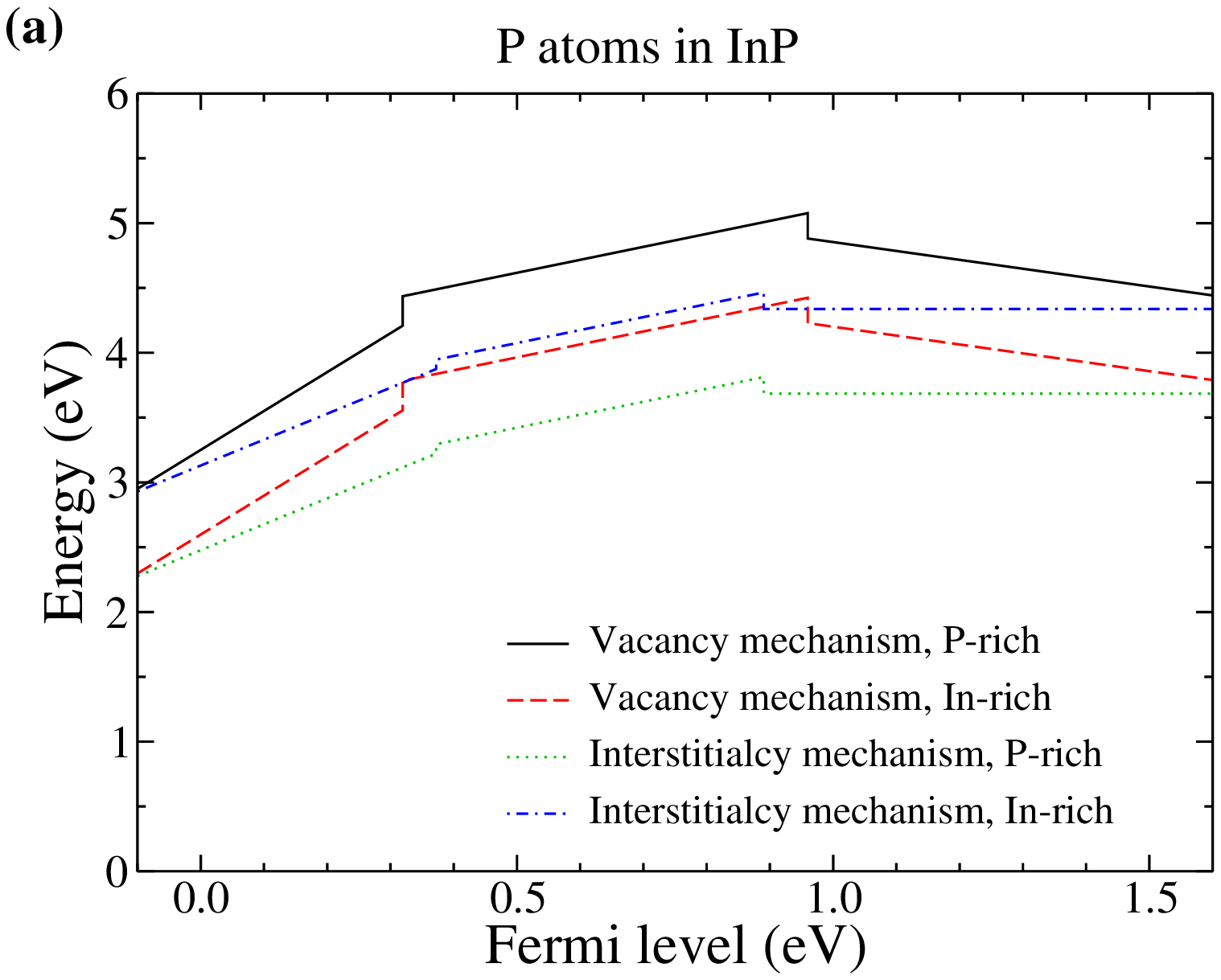}
\includegraphics[width=0.8\columnwidth]{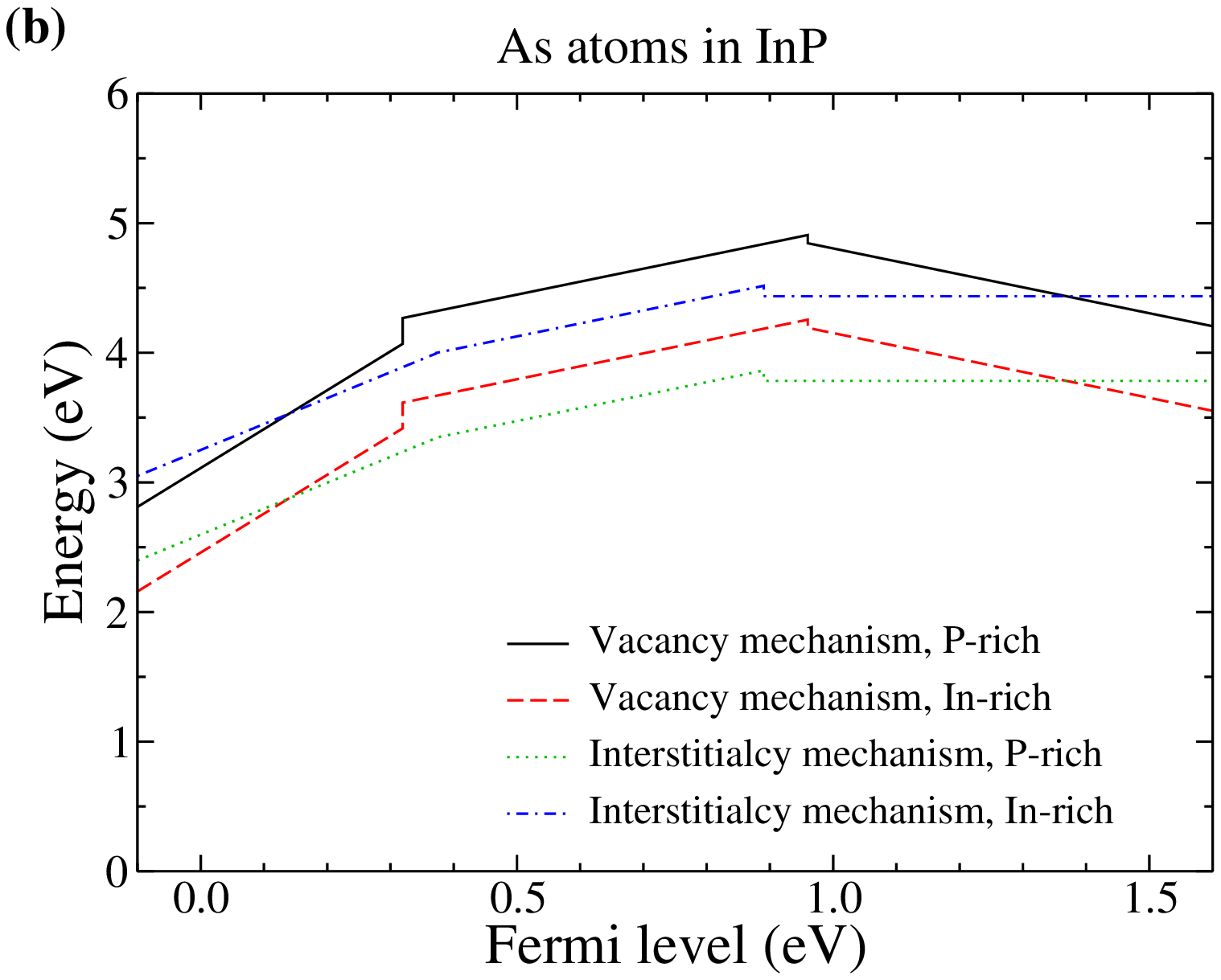}
\includegraphics[width=0.8\columnwidth]{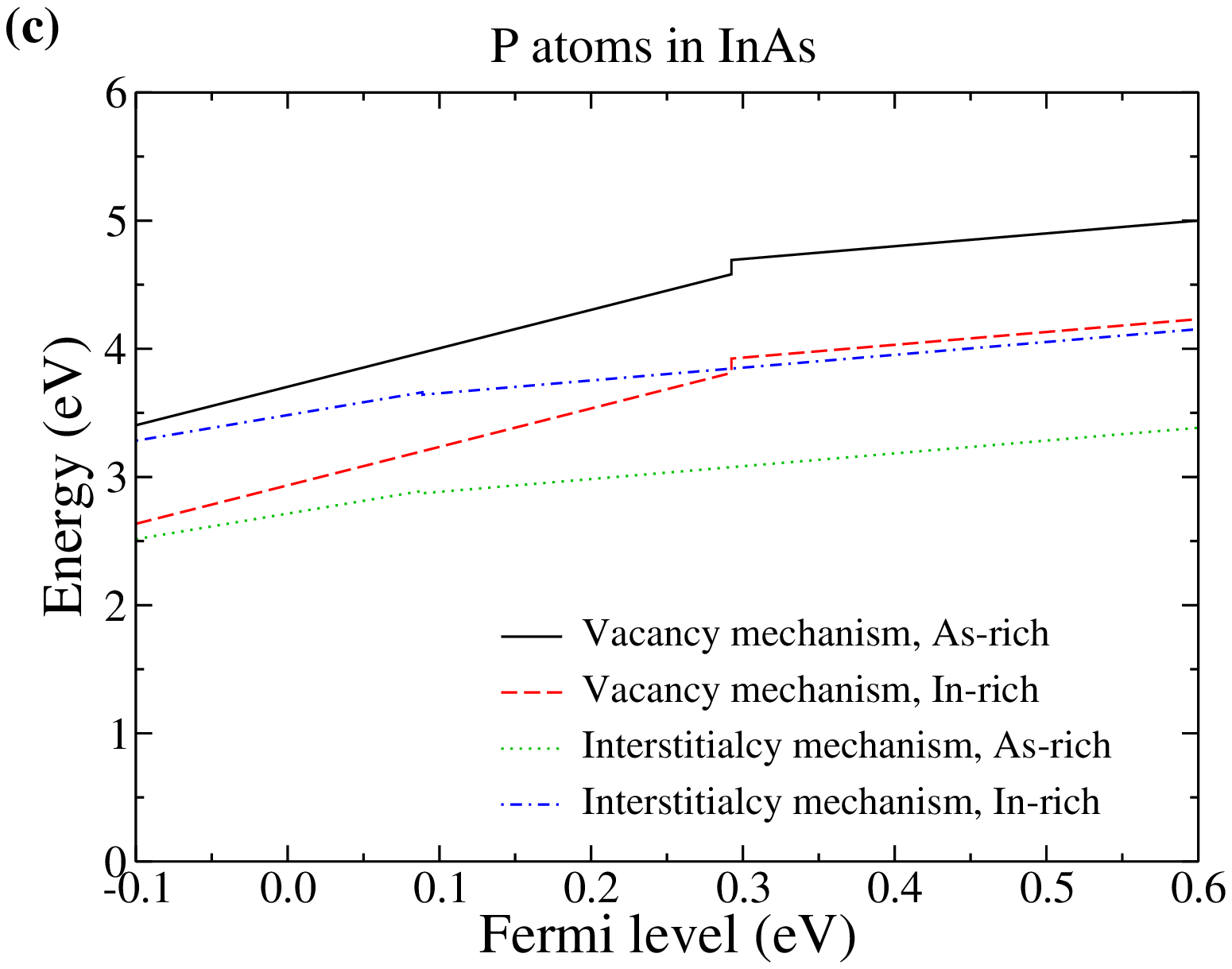}
\includegraphics[width=0.8\columnwidth]{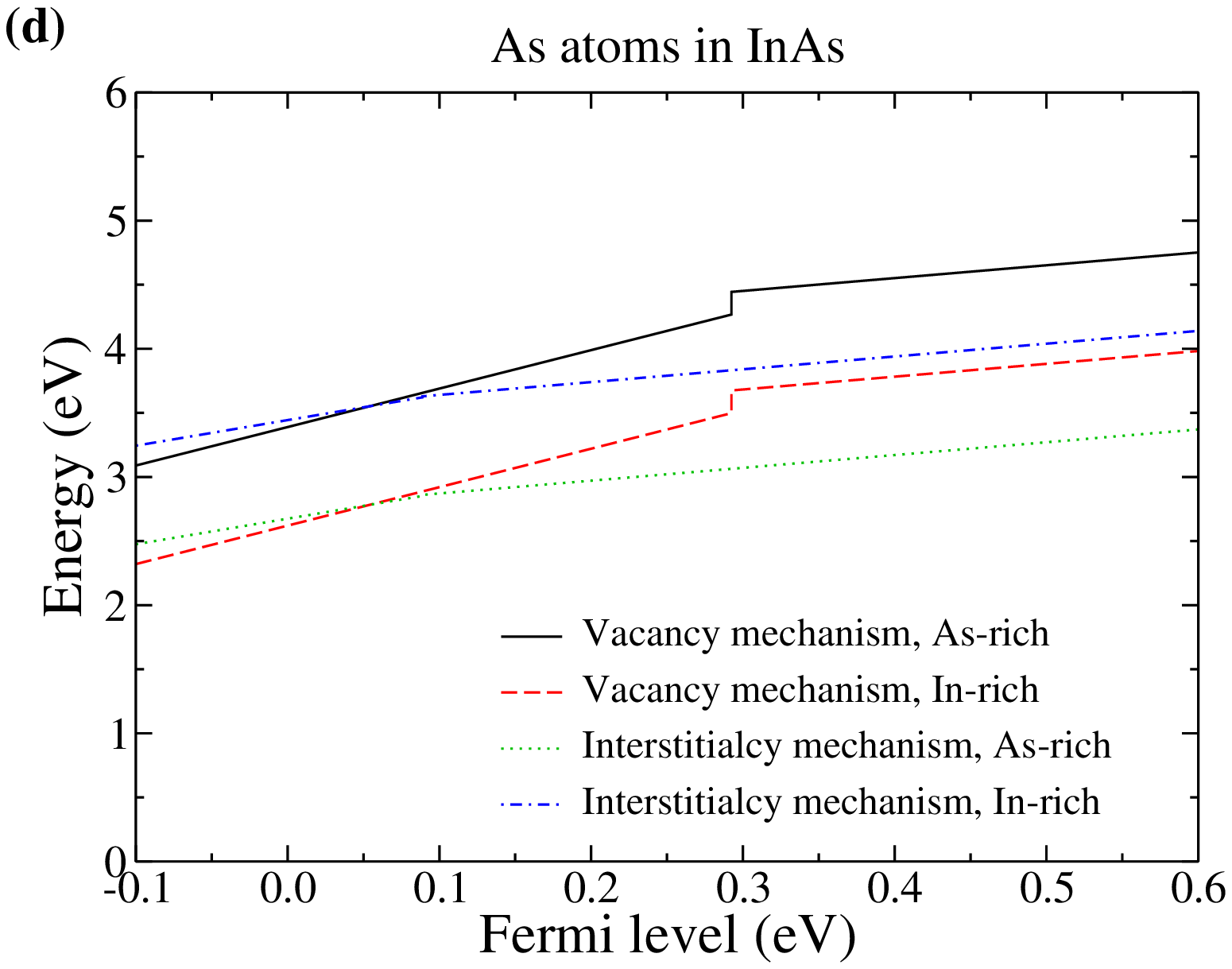}
\caption{\label{fig:figure11} Dependence of the sum of the defect formation energy and the effective migration barrier energy on the Fermi level for the vacancy and indirect interstitial (interstitialcy) mechanisms of diffusion of As and P atoms in InAs and InP.}
\end{center}
\end{figure}

Table~\ref{tab:table11} shows the activation energies obtained by fitting the calculation results shown in the figure~\ref{fig:figure10} by thermally activated dependence: $D/(N_{def}/N_{sites})=D_0\exp(-E_m/kT)$. 
\begin{table*}
\caption{\label{tab:table11}  Migration barriers and pre-exponential factors for the ratio of the diffusion coefficient for the indirect interstitial mechanism to the relative concentration of interstitial atoms, taking into account the temperature dependence of the correlation coefficient and the binding energy of the impurity atom and the defect, obtained by fitting the calculation results shown in the figure~\ref{fig:figure10} by thermally activated dependence: $D/(N_{def}/N_{sites})=D_0\exp(-E_m/kT)$. }
\begin{center}
\begin{tabular*}{0.8\linewidth}{c c c c c c c c c}
\hline
 & \multicolumn{2}{c}{As in InP} & \multicolumn{2}{c}{P in InP} & \multicolumn{2}{c}{As in InAs} & \multicolumn{2}{c}{P in InAs} \\
\hline
Charge  & $E_m$, eV & $D_0$, cm\textsuperscript{2}/s & $E_m$, eV & $D_0$, cm\textsuperscript{2}/s & $E_m$, eV & $D_0$, cm\textsuperscript{2}/s & $E_m$, eV & $D_0$, cm\textsuperscript{2}/s \\
\hline
+2 & 0.568 & 1.6$\cdot10^{-2}$ & 0.449 & 5.6$\cdot10^{-3}$ & 0.359 & 3.3$\cdot10^{-3}$ & 0.399 & 3.8$\cdot10^{-3}$ \\
+1 & 0.572 & 3.7$\cdot10^{-4}$ & 0.521 & 4.7$\cdot10^{-4}$ & 0.367 & 3.9$\cdot10^{-4}$ & 0.380 & 3.7$\cdot10^{-4}$ \\
0 & 0.492 & 6.8$\cdot10^{-4}$ & 0.393 & 3.7$\cdot10^{-4}$ & 0.324 & 3.0$\cdot10^{-4}$ & 0.255 & 6.1$\cdot10^{-4}$ \\
\hline
\end{tabular*}
\end{center}
\end{table*}
Figure~\ref{fig:figure11} shows the dependence of the sum of the defect formation energy and the migration energy barrier, obtained by fitting the calculation results by a dependence of the form $D_0\exp(-E_m/kT)$ with one activation energy (figures~\ref{fig:figure8}--\ref{fig:figure10}, tables~\ref{tab:table4},~\ref{tab:table5},~\ref{tab:table10},~\ref{tab:table11}), for the diffusion of P and As atoms by vacancy and indirect interstitial mechanisms. For the vacancy diffusion mechanism, the migration energy barriers are higher than that for the interstitialcy mechanism, however, higher formation energies of the interstitial atoms lead to the fact that the total activation energies of diffusion are comparable for the vacancy and the interstitialcy mechanisms. The general trend is that for P-rich conditions in InP and As-rich conditions in InAs, the interstitial diffusion mechanism has lower activation energy, and for In-rich conditions, the vacancy diffusion mechanism has lower activation energy. However, for P atoms diffusion in n-type InAs, the diffusion activation energy for the interstitialcy mechanism is lower than that for the vacancy mechanism even under In-rich conditions.

In the work of Sallese et al.~\cite{Sallese_1994}, the diffusion of a thin layer of InAs in InP was studied by photoluminescence measurements, the activation energy of $E_a=3.8\pm0.2$~eV had been obtained, the diffusion coefficient at 830$^\circ$C was 7$\cdot10^{-17}$~cm\textsuperscript{2}/s, which corresponds to $D_0$=16.1~cm\textsuperscript{2}/s. Sallese et al.~\cite{Sallese_1994} performed annealing without additional phosphorus supply, so we assume that the conditions were In-rich and the dominant defects were phosphorus vacancies. Taking into account the residual electron concentration in the InP layers of the order of $10^{15}$~cm\textsuperscript{-3}~\cite{Rudra_1991}, we will assume that the Fermi level is in the upper half of the band gap and take $E_{form}$=2.0~eV for estimation. According to our calculations, taking into account the dependence of the correlation factor on temperature, the energy of the diffusion barrier for As diffusion in InP by the vacancy mechanism is 1.730~eV for the $-$1 charge state of the vacancy. In this case, the pre-exponential coefficient for the diffusion coefficient without taking into account the entropy of defect formation and migration entropy is 1.01$\cdot10^{-2}$~cm\textsuperscript{2}/s. Having estimated the activation energy for the diffusion coefficient as the sum of the formation energy and the migration energy $E_a$=3.73~eV, we find that for agreement of the calculation with the experiment~\cite{Sallese_1994}, it is necessary to assume $D_0$=7.7~cm\textsuperscript{2}/s. The higher pre-exponential factor compared to the calculation is associated with the presence of additional entropy $\Delta S$=6.6$k_B$, which consists of the vacancy formation entropy and the migration entropy. In this case, the entropy of vacancy formation contains a contribution from the presence of several symmetry equivalent configurations, for D\textsubscript{2d} symmetry of the vacancy N\textsubscript{config}=3. The obtained result is consistent with the results of Hurle work~\cite{Hurle_2010}, in which the author had found $\Delta H_f$=2.177~eV and $\Delta S_f$=6.09$k_B$ for neutral phosphorus vacancy in InP by fitting with a thermodynamic model the experimental data on the concentration of point defects~\cite{Morozov_1983} obtained by precision measurements of the lattice constant and density of InP crystals.

\section{Conclusions}
Using density functional theory methods, the atomic structure and formation energies of group-V elements vacancies and interstitial P and As atoms in InP and InAs have been calculated and the symmetry groups of thermodynamically stable configurations of these defects have been determined. According to the calculation with the hybrid HSE functional, the phosphorus vacancy in InP has thermodynamically stable charge states +3, +1 and $-$1 with symmetry groups T\textsubscript{d}, C\textsubscript{2v} and D\textsubscript{2d} in the corresponding charge states, the arsenic vacancy in InAs has thermodynamically stable charge states +3 and +1 with symmetry groups T\textsubscript{d} and C\textsubscript{2v}, respectively. For interstitial As and P atoms in InP and InAs in a neutral charge state, the most energy favorable configuration is the [110] split-interstitial. For As\textsubscript{i} and P\textsubscript{i} in InP and for As\textsubscript{i} and P\textsubscript{i} in InAs in the charge state +1, the most energy favorable configuration is a split-interstitial with an In atom with C\textsubscript{s} symmetry; in the charge state +2 the configuration with C\textsubscript{3v} symmetry is most energy favorable.

The main types of migration jumps for the found configurations have been determined. The most energy favorable migration paths and energy barriers of the migration transitions have been calculated. In the case of diffusion of As and P in InP and InAs via interstitial atoms the diffusion process occurs via indirect interstitial mechanism. The migration energy barriers for the vacancy diffusion mechanism are 1.60--1.86~eV for As in InP, 1.73--2.01~eV for P in InP, 1.51--1.79~eV for As in InAs, 1.74--2.02~eV for P in InAs, for different charge states. For the indirect interstitial diffusion mechanism, the migration energy barriers are lower than for the vacancy mechanism and have values of 0.49--0.57~eV for As in InP, 0.39--0.52~eV for P in InP, 0.32--0.37~eV for As in InAs, 0.26--0.40~eV for P in InAs, for different charge states.

Interstitial As and P atoms in InP and InAs have higher formation energies compared to the group-V vacancies. The total diffusion activation energies are comparable for the vacancy and the interstitialcy mechanisms. The vacancy diffusion mechanism gives lower activation energies for the diffusion coefficient in the case of In-rich conditions for P and As diffusion in InP, As diffusion in InAs, and P diffusion in p-type InAs. The indirect interstitial diffusion mechanism gives lower activation energies for the diffusion coefficient of P and As for P-rich conditions in InP and As-rich conditions in InAs, and also in the case of In-rich conditions for P diffusion in n-type InAs.

The temperature dependences of the diffusion coefficients of the group-V elements vacancies in InP and InAs and substitution atoms As and P in InP and InAs have been estimated for vacancy and indirect interstitial mechanisms. The results of the energy barriers calculations are consistent with experimental data~\cite{Slotte_2003, Sallese_1994}. For agreement of the absolute values of diffusion coefficients with experiment, it is necessary to take into account the entropy of defect formation and migration entropy. The obtained results will be useful for modeling diffusion processes occurring under various experimental conditions in structures based on InP and InAs.
\section{Acknowledgments}
The authors greatly acknowledge support of the Ministry of Science and Higher Education of the Russian Federation within the state task of Rzhanov Institute of Semiconductor Physics SB RAS (FWGW-2022-0005). The Siberian Branch of the Russian Academy of Sciences (SB RAS) Siberian Supercomputer Center and Novosibirsk State University are gratefully acknowledged for providing computational resources.

\includepdf[pages={1-11}]{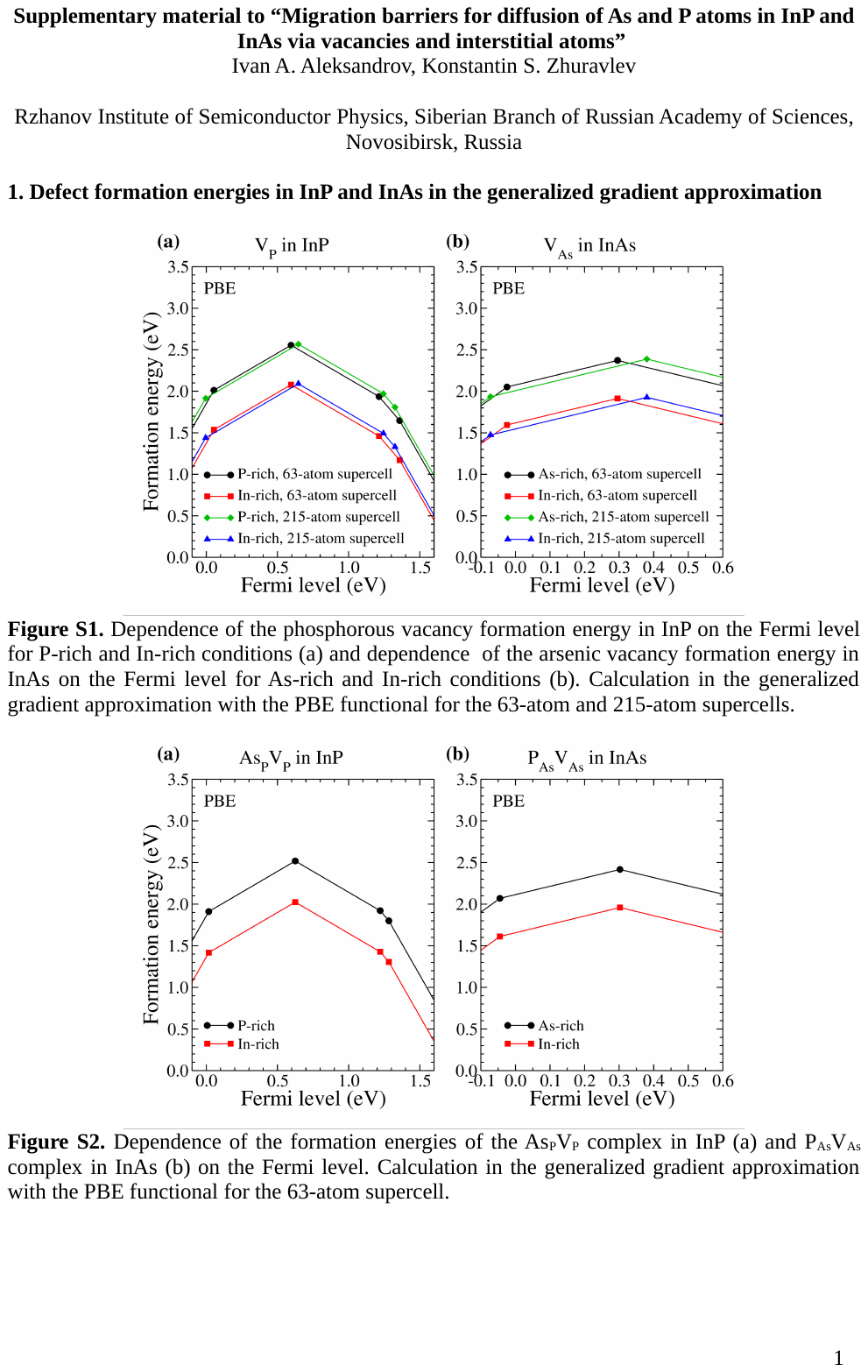}

\end{document}